\documentclass[journal,10pt]{IEEEtran}
\usepackage{todonotes}

\usepackage{amsmath,amssymb,bm}
\usepackage{rotating}
\usepackage{trfsigns}
\usepackage{aligned-overset}
\usepackage[
    backend=biber,
    sorting=none,
    maxbibnames=99,
    defernumbers=true,
    style=ieee,
    giveninits=true,
]{biblatex}
\AtEveryBibitem{
 \clearfield{url}
 \clearfield{issn}
}
\usepackage{siunitx}
\DeclareSIUnit{\ms}{\milli\second}   % Millisekunde
\DeclareSIUnit{\us}{\micro\second}   % Mikrosekunde
\DeclareSIUnit{\ns}{\nano\second}    % Nanosekunde
\DeclareSIUnit{\dBm}{dBm}  

\usepackage{comment}

\usepackage{trfsigns}
\usepackage[export]{adjustbox}

\usepackage{graphicx}
\usepackage{xcolor}
  
\usepackage{tikz}
\usetikzlibrary{shapes.geometric, arrows.meta, positioning,calc,backgrounds,shapes.multipart}
\usepackage{pgfplots}
\pgfplotsset{compat=1.18}

\usepackage{microtype}
\usepackage{placeins}

\usepackage[acronyms]{glossaries}
\newacronym{3gpp}{3GPP}{Third Generation Partnership Project}

\newacronym{admm}{ADMM}{Alternating Directions of Multipliers Method}
\newacronym{agv}{AGV}{Automated Guided Vehicels}
\newacronym{anm}{ANM}{Atomic Norm Minimization}
\newacronym{adc}{ADC}{Analog-to-Digital Converter}
\newacronym{acf}{ACF}{Auto-Correlation Function}
\newacronym{awgn}{AWGN}{Additive White Gaussian Noise}
\newacronym{asic}{ASIC}{Application Specific Integrated Circuit}
\newacronym{arpack}{ARPACK}{ARnoldi PACKage}
\newacronym{api}{API}{Application Programmable Interface}
\newacronym{aut}{AUT}{Antenna Under Test}
\newacronym[plural=AOI, firstplural=Areas of Interest (AOI)]{aoi}{AOI}{Area of Interest}
\newacronym{ai}{AI}{Artifical Intelligence}
\newacronym{aic}{AIC}{Akaike Information Criterion}
\newacronym{agc}{AGC}{Automatic Gain Control}
\newacronym{af}{AF}{Ambiguity Function}

\newacronym{blos}{BLOS}{Beyond Line Of Sight}
\newacronym{bp}{BP}{Basis Pursuit}
\newacronym{bs}{BS}{Base Station}
\newacronym{bpdn}{BPDN}{Basis Pursuit Denoising}
\newacronym{blue}{BLUE}{Best Linear Unbiased Estimator}
\newacronym{bic}{BIC}{Bayesian Information Criterion}
\newacronym{bibo}{BIBO}{Bounded-Input-Bounded-Output}
\newacronym{btp}{BTP}{Bistatic Target Path}
\newacronym{bt}{BT}{bandwidth-time}

\newacronym{cnn}{CNN}{Convolutional Neural Network}
\newacronym{crb}{CRB}{Cramér-Rao Bound}
\newacronym{cs}{CS}{Compressed Sensing}
\newacronym{cf}{CF}{Crest factor}
\newacronym{cr}{CR}{Compression Ratio}
\newacronym{cu}{CU}{Central Unit}
\newacronym{cmos}{CMOS}{Complementary Metal Oxide Semiconductor}
\newacronym{cots}{COTS}{Commercial-Off-The-Shelf}
\newacronym{cpu}{CPU}{Central Processing Unit}
\newacronym{cfar}{CFAR}{Constant False Alarm Rate}
\newacronym{cvx}{CVX}{Convex Optimization toolboX}
\newacronym{comp}{CoMP}{Cooperative Multi-Point}
\newacronym{cpcl}{CPCL}{Cooperative Passive Coherent Location}
\newacronym{csi}{CSI}{Channel State Information}
\newacronym{cir}{CIR}{Channel Impulse Response}
\newacronym{cdf}{CDF}{Cumulative Distribution Function}
\newacronym{cam}{CAM}{Cooperative Awareness Messages}
\newacronym{cpx}{CPX}{Cyclic Prefix}

\newacronym{doa}{DoA}{Direction of Arrival}
\newacronym{dod}{DoD}{Direction of Departure}
\newacronym[plural=DMC,firstplural=Diffuse Multipath Components (DMC)]{dmc}{DMC}{Diffuse Multipath Components}
\newacronym{das}{DAS}{Delay-and-Sum}
\newacronym{dft}{DFT}{Discrete Fourier Transform}
\newacronym{idft}{iDFT}{Inverse Discrete Fourier Transform}
\newacronym{dtft}{DTFT}{Discrete Time Fourier Transform}
\newacronym{idtft}{iDTFT}{Inverse Discrete Time Fourier Transform}
\newacronym{dct}{DCT}{Discrete Cosine Transform}
\newacronym{dsp}{DSP}{Digital Signal Processor}
\newacronym{dnn}{DNN}{Deep Neural Network}
\newacronym{dlr}{DL}{Deep Learning}
\newacronym{dbs}{DBS}{Distributed Base Station}
\newacronym{dmrs}{dmrs}{DeModulation Reference Signal}

\newacronym{eadf}{EADF}{Effective Aperture Distribution Function}
\newacronym{etadf}{ETADF}{Effective Time-Aperture Distribution Function}
\newacronym{esprit}{ESPRIT}{Estimation of Signal Parameters via Rotational Invariance Techniques}
\newacronym{ett}{ETT}{Eigenvalue Threshold Test}
\newacronym{edc}{EDC}{Efficient Detection Criterion}
\newacronym{eft}{EFT}{Exponential Fitting Test}
\newacronym{expm}{EM}{Expectation Maximization}
\newacronym{ekf}{EKF}{Extended Kalman Filter}
\newacronym{ems}{FG EMS}{Fachgebiet Elektronische Messtechnik und Signalverarbeitung}
\newacronym{etof}{eToF}{Excess Time of Flight}
\newacronym{etoa}{eToA}{Excess Time of Arrival}
\newacronym{evv}{eVV}{Excess Velocity vector}

\newacronym{fc}{FC}{fully-connected}
\newacronym{ft}{FT}{Fourier Transform}
\newacronym{fht}{FHT}{Fast Hadamard Transform}
\newacronym[longplural={Fast Fourier Transforms}]{fft}{FFT}{Fast Fourier Transform}
\newacronym{fmcw}{FMCW}{Frequency-Modulated Continuous-Wave}
\newacronym{fpga}{FPGA}{Field Programmable Gate Array}
\newacronym{fri}{FRI}{Finite Rate of Innovation}
\newacronym{fir}{FIR}{Finite Impulse Response}
\newacronym{fim}{FIM}{Fisher Information Matrix}
\newacronym{fmc}{FMC}{Full Matrix Capture}
\newacronym{fista}{FISTA}{Fast Iterative Shrinkage-Thresholding Algorithm}
\newacronym{frvm}{FRVM}{Fast Relevance Vector Machine}
\newacronym{flop}{FLOP}{Floating Point Operation}
\newacronym{fl}{FL}{Federated Learning}
\newacronym{frf}{FRF}{Frequency Response Functions}
\newacronym{fdma}{FDMA}{Frequency-Division Multiple Access}

\newacronym{gfcs}{grid-free CS}{grid-free compressive sensing}
\newacronym{gpu}{GPU}{Graphical Processing Unit}
\newacronym{gtd}{GTD}{Geometrical Theory of Diffraction}
\newacronym{gan}{GAN}{Generative Adversarial Network}
\newacronym{gdop}{GDoP}{Geometric Dilution of Precision}
\newacronym{gnb}{gNB}{base station}
\newacronym{gpsdo}{GPSDO}{GPS disciplined oscillators}

\newacronym{hrpe}{HRPE}{High Resolution Parameter Estimator}
\newacronym{hfss}{Ansys HFSS}{Ansys High Frequency Electromagnetic Simulation Software}

\newacronym[longplural={Inverse Fast Fourier Transforms}]{ifft}{IFFT}{Inverse Fast Fourier Transform}
\newacronym{ir}{IR}{Impulse Response}
\newacronym{iid}{iid}{independent and identically distributed}
\newacronym{iir}{IIR}{Infinite Impulse Response}
\newacronym{irf}{IRF}{Impulse Response Function}
\newacronym{icas}{ICAS}{Integrated Communications and Sensing}
\newacronym{isac}{ISAC}{Integrated Sensing and Communications}
\newacronym{ista}{ISTA}{Iterative Shrinkage-Thresholding Algorithm}
\newacronym{ici}{ICI}{Inter Carrier Interference}
\newacronym{iot}{IoT}{Internet of Things}

\newacronym{jt}{JT}{Joint Transmission}
\newacronym{jpda}{JPDA}{Joint Probabilistic Data Association}

\newacronym{kld}{KLD}{Kullback-Leibler Divergence}
\newacronym{kpi}{KPI}{Key Performance Indicator}

\newacronym{ls}{LS}{Least Squares}
\newacronym{lasso}{LASSO}{Least Absolute Shrinkage and Selection Operator}
\newacronym{lse}{LSE}{Line Spectral Estimation}
\newacronym{lfsr}{LFSR}{Linear Feedback Shift Register}
\newacronym{lo}{LO}{Local Oscillator}
\newacronym{los}{LOS}{Line of Sight}
\newacronym{lti}{LTI}{Linear Time-Invariant}
\newacronym{ltv}{LTV}{Linear Time-Variant}
\newacronym{lam}{LAM}{Large Area Monitoring}

\newacronym{mimo}{MIMO}{Multiple Input Multiple Output}
\newacronym{mmv}{MMV}{multiple measurement vectors}
\newacronym{mmse}{MMSE}{misspecified mean squared error}
\newacronym{mse}{MSE}{mean squared error}
\newacronym{bce}{BCE}{Binary Crossentropy}
\newacronym{mlbs}{MLBS}{Maximum Length Binary Sequence}
\newacronym{mno}{MNO}{Mobile Network Operator}
\newacronym{mpc}{MPC}{Multipath Component}
\newacronym{msm}{MSM}{M-Sequence Method}
\newacronym{mwc}{MWC}{Modulated Wideband Converter}
\newacronym{mpm}{MPM}{Matrix Pencil Method}
\newacronym{mpu}{MPU}{Microprocessor Unit}
\newacronym{mumimo}{MU MIMO}{multi-user MIMO}
\newacronym{mu}{MU}{multi-user}
\newacronym{ms}{MS}{multi-sensor}
\newacronym{ml}{ML}{Machine Learning}
\newacronym{mri}{MRI}{Magnetic resonance imaging}
\newacronym{music}{MUSIC}{Multiple Signal Classification}
\newacronym{mkl}{MKL}{Math Kernel Library}
\newacronym{mcrb}{MCRB}{Misspecified Cramér-Rao Bound}
\newacronym{mmle}{MMLE}{Misspecified Maximum-Likelihood Estimator}
\newacronym{mbpe}{MBPE}{Model-Based Propagation Parameter Estimation}
\newacronym{mrf}{MRF}{Multistatic Reflectivity Function}
\newacronym{mec}{MEC}{Mobile Edge Cloud}
\newacronym{msisac}{MS-ISAC}{Multi-Sensor ISAC}
\newacronym{mht}{MHT}{Multi-Hypotheses Tracker}
\newacronym{mi}{MI}{Mutual Information}

\newacronym{ndt}{NDT}{Nondestructive Testing}
\newacronym{nde}{NDE}{Nondestructive Evaluation}
\newacronym{nn}{NN}{Neural Net}
\newacronym{nist}{NIST}{National Institute of Standards and Technology}
\newacronym{npn}{NPN}{Non-Public Networks}
\newacronym{nr}{NR}{New Radio}
\newacronym{nlos}{NLOS}{Non Line of Sight}
 
\newacronym{omp}{OMP}{Orthogonal Matching Pursuit}
\newacronym{oop}{OOP}{Object Oriented Programming}
\newacronym{ota}{OTA}{Over The Air}
\newacronym{ofdm}{OFDM}{Orthogonal Frequency-Division Multiplexing}
\newacronym{ofdma}{OFDMA}{Orthogonal Frequency-Division Multiple Access}
\newacronym{afdm}{AFDM}{Affine Frequency Division Multiplexing}
\newacronym{otfs}{OTFS}{Orthogonal Time Frequency Space}

\newacronym{pdp}{PDP}{Power Delay Profile}
\newacronym{pap}{PAP}{Power Angular Profile}
\newacronym{pn}{PN}{Pseudo-Noise}
\newacronym{pwc}{PWC}{Plane Wave Compounding}
\newacronym{pcl}{PCL}{Passive Coherent Location}
\newacronym{pwi}{PWI}{Plane Wave Imaging}
\newacronym{pura}{PURA}{Patch Uniform Rectangular Array}
\newacronym{pymax}{PyMAX}{Python Maximization Approach}
\newacronym{pts}{PTS}{Pseudo-True Solution}
\newacronym{pdf}{pdf}{probability density function}
\newacronym{pi}{PI}{Principal Investigator}
\newacronym{pidl}{PIDL}{Physics Informed Deep Learning}
\newacronym{pinn}{PINN}{Physics Informed Neural Network}
\newacronym{pbdl}{PBDL}{Physics Based Deep Learning}
\newacronym{prs}{PRS}{Position Reference Signal}

\newacronym{qam}{QAM}{quadrature amplitude modulation}

\newacronym{ran}{RAN}{Radio Access Network}
\newacronym{ranic}{RIC}{RAN Intelligent Controller}
\newacronym{relu}{ReLU}{Rectified Linear Unit}
\newacronym{resnet}{ResNet}{Residual Neural Network}
\newacronym{ram}{RAM}{Random Access Memory}
\newacronym{rcs}{RCS}{Radar Cross Section}
\newacronym{rd}{RD}{Random Demodulator}
\newacronym{rx}{Rx}{receiver}
\newacronym{rem}{REM}{Reconstruction Error Metric}
\newacronym{rmse}{RMSE}{Root Mean Squared Error}
\newacronym{rms}{RMS}{root mean squared}
\newacronym{ric}{RIC}{Restricted Isometry Constant}
\newacronym{rip}{RIP}{Restricted Isometry Property}
\newacronym{ris}{RIS}{Reconfigurable Intelligent Surface}
\newacronym{rc}{RC}{Raised Cosine}
\newacronym{roi}{ROI}{Region of Interest}
\newacronym{roc}{ROC}{Region of Convergence}
\newacronym{rt}{RT}{Raytracing}
\newacronym{rimax}{RIMAX}{Richter Maximization Approach}
\glsunset{rimax}
\newacronym{rvm}{RVM}{Relevance Vector Machine}
\newacronym{rss}{RSS}{Received Signal Strength}
\newacronym{rfid}{RFID}{Radio Frequency Identification}
\newacronym{refodat}{REFODAT}{Repositorium für Forschungsdaten in Thüringen}
\newacronym{re}{RE}{Resource Element}
\newacronym{rb}{RB}{Resource Block}
\newacronym{rrh}{RRH}{Remote Radio Head}
\newacronym{rru}{RRU}{Remote Radio Unit}

\newacronym{samurai}{SAMURAI}{Synthetic Aperture Measurements of Uncertainty in Angle of Incidence}
\newacronym[plural=SC,firstplural=Specular Components (SC)]{sc}{SC}{Specular Components}
\newacronym{sdp}{SDP}{semi-definite program}
\newacronym{sdr}{SDR}{Signal to Diffuse Ratio}
\newacronym{simd}{SIMD}{Single Instruction Multiple Data}
\newacronym{svd}{SVD}{singular value decomposition}
\newacronym{svm}{SVM}{Support Vector Machine}
\newacronym{soe}{SOE}{Sparsity Order Estimation}
\newacronym{sgd}{SGD}{Stochastic Gradient Descent}
\newacronym{stuca}{StUCA}{Stacked Uniform Circular Array}
\newacronym{spucpa}{SPUCPA}{Stacked Polarimetric Uniform Circular Patch Array}
\newacronym{suca}{SUCA}{Stacked Uniform Circular Array}
\newacronym{saft}{SAFT}{Synthetic Aperture Focusing Technique}
\newacronym{sota}{SOTA}{State of the Art}
\newacronym{ssd}{SSD}{Solid State Device}
\newacronym{ssr}{SSR}{Sparse Signal Recovery}
\newacronym{sa}{SA}{Synthetic Aperture}
\newacronym{sh}{SH}{Spherical Harmonics}
\newacronym{spw}{SPW}{Single Plane Wave}
\newacronym{shm}{SHM}{Structural Health Monitoring}
\newacronym{snr}{SNR}{Signal-to-Noise Ratio}
\newacronym{stela}{STELA}{Soft-Thresholding with Exact Line Search Algorithm}
\newacronym{siso}{SISO}{Single Input Single Output}
\newacronym{simo}{SIMO}{Single Input Multiple Output}
\newacronym{swe}{SWE}{Spherical Wave Expansion}
\newacronym{sme}{SME}{Spherical Mode Expansion}
\newacronym{sage}{SAGE}{Space-Alternating Generalized Expectation-Maximization}
\newacronym{stft}{STFT}{Short Time Fourier Transformation}
\newacronym{sl}{SL}{Side Link}
\newacronym{slc}{SLC}{Sensor Level Cooperation}
\newacronym{scf}{ScF}{Scattering Function}
\newacronym{sf}{SF}{Spreading Function}
\newacronym{sar}{SAR}{Synthetic Aperture Radar}

\newacronym{th}{T\&H}{Track and Hold}
\newacronym{tf}{TF}{Transfer Function}
\newacronym{tx}{Tx}{transmitter}
\newacronym{twista}{TWISTA}{Two-step Iterative Shrinkage-Thresholding Algorithm}
\newacronym{tof}{ToF}{Time of Flight}
\newacronym{tdoa}{TDoA}{Time Difference of Arrival}
\newacronym{toa}{ToA}{Time of Arrival}
\newacronym{tdd}{TDD}{Time Division Duplex}
\newacronym{tdma}{TDMA}{Time-division Multiple Access}
\newacronym{trf}{TRF}{Time Reversal Focusing}
\newacronym{tr}{TR}{Time Reversal}

\newacronym{uca}{UCA}{uniform circular array}
\newacronym{ura}{URA}{Uniform Rectangular Array}
\newacronym{ula}{ULA}{Uniform Linear Array}
\newacronym{uwb}{UWB}{Ultra-Wideband}
\newacronym{usndt}{US-NDT}{Ultrasonic Non-destructive Testing}
\newacronym{ue}{UE}{User Equipment}
\newacronym{ul}{UL}{Uplink}
\newacronym{dl}{DL}{Downlink}
\newacronym{uav}{UAV}{Unmanned Aerial Vehicles}
\newacronym{udc}{UDC}{Up/Down Converter}
\newacronym{utm}{UTM}{Uncrewed Aircraft Systems Traffic Management}
\newacronym{usrp}{USRP}{Universal Software Radio Peripheral}

\newacronym{vna}{VNA}{Vector Network Analyser}
\newacronym{vsh}{VSH}{Vector Spherical Harmonics}
\usepackage[
  hidelinks,         % keine bunten Boxen um Links
  pdfencoding=auto,  % Unicode für PDF-Strings automatisch
  psdextra           % erlaubt mehr Makros in PDF-Strings
]{hyperref}

\usepackage{orcidlink}
\usepackage{cleveref}

\DeclareMathOperator*{\Rect}{rect}
\DeclareMathOperator*{\Sinc}{sinc}

\newcommand{\ScPr}[2]{{\left\langle #1,#2 \right\rangle}}
\newcommand{\Norm}[1]{{\left\Vert #1\right\Vert}}

\title{Distributed Multisensor ISAC}

\author{
    \IEEEauthorblockN{%\input{dist_ms_isac}
        Reiner Thom\"a\IEEEauthorrefmark{3}\IEEEauthorrefmark{1}\,\orcidlink{0000-0002-9254-814X},
    	Carsten Andrich\IEEEauthorrefmark{1}\,\orcidlink{0000-0002-4795-3517},
    	Michael D\"obereiner\IEEEauthorrefmark{2}\,\orcidlink{0000-0001-9675-9860},
        Reza Faramarzahangari\IEEEauthorrefmark{1}\,\orcidlink{0009-0008-5300-8088},
        Jonas Gedschold\IEEEauthorrefmark{1}\,\orcidlink{0000-0002-0251-887X},
        Marc Francisco Colaco Miranda\IEEEauthorrefmark{1}\,\orcidlink{0009-0006-1256-629X},
        Saw James Myint\IEEEauthorrefmark{1}\,\orcidlink{0009-0007-3788-7126},
        Steffen Schieler\IEEEauthorrefmark{1}\,\orcidlink{0000-0003-4480-234X},
        Christian Schneider\IEEEauthorrefmark{1}\IEEEauthorrefmark{2}\,\orcidlink{0000-0003-1833-4562},
        Sebastian Semper\IEEEauthorrefmark{1}\IEEEauthorrefmark{2}\,\orcidlink{0000-0002-2610-7389},
        Carsten Smeenk\IEEEauthorrefmark{2}\,\orcidlink{0009-0007-9062-0025},
	    Gerd Sommerkorn\IEEEauthorrefmark{1}\,\orcidlink{0009-0003-1111-322X},
	    Zhixiang Zhao\IEEEauthorrefmark{1}\,\orcidlink{0009-0001-4733-9226}
	}                                     
\\
	\IEEEauthorblockA{
		\IEEEauthorrefmark{1}Technische Universit\"at Ilmenau, Institute for Information Technology, Ilmenau, Germany\\
		\IEEEauthorrefmark{2}Fraunhofer Institute of Integrated Circuits, Dep. EMS, Ilmenau, Germany\\
        \IEEEauthorrefmark{3}Corresponding author via \url{reiner.thomae@tu-ilmenau.de}
	}
}

\date{March 2025}

\begin{document}
\maketitle

\begin{abstract}
Integrated Sensing and Communications (ISAC) will become a service in future mobile communication networks. It enables the detection and recognition of passive objects and environments using radar-like sensing. The ultimate advantage is the reuse of the mobile network and radio access resources for scene illumination, sensing, data transportation, computation, and fusion. It enables building a distributed, ubiquitous sensing network that can be adapted for a variety of radio sensing tasks and services.

In this article, we develop the principles of multi-sensor ISAC (MS-ISAC). MS-ISAC corresponds to multi-user MIMO communication, which in radar terminology is known as distributed MIMO radar. 
First, we develop basic architectural principles for MS-ISAC and link them to example use cases. We then propose a generic MS-ISAC architecture. After a brief reference to multipath propagation and multistatic target reflectivity issues, we outline multilink access, coordination, precoding and link adaptation schemes for MS-ISAC. Moreover, we review model-based estimation and tracking of delay~/~Doppler from sparse OFDMA~/~TDMA frames. We emphasize Cooperative Passive Coherent Location (CPCL) for bistatic correlation and synchronization. Finally, issues of multisensor node synchronization and distributed data fusion are addressed. 

% kein latex code im abstract oder den keywords


\end{abstract}

\section{Introduction and Motivation}
\label{sec:introduction}
\gls{isac} is considered one of the key features of future 6G mobile communication~\cite{ref28,A1,A2, A3}. 
Despite different interpretations, we understand \gls{isac} as a means of radar detection and localization of passive objects (``targets'') that are not equipped with a radio tag (``not connected'').
We therefore use the term ``sensing'' as a synonym for ``radar sensing''. 
The targets reveal their existence and position by radio wave reflection when properly illuminated. 
In contrast to well-known dedicated radar systems, \gls{isac} exploits the inherent resources of the mobile communication system on the radio access and network level. 
Simple proposals assume to share the radio access resources, e.g. base station sites and hardware, for orthogonal transmission of communication and sensing waveforms. 
However, there are more advanced versions for integrating communication and sensing functions. 
In this paper, we solely assume that the primary system is the mobile communication network, which is extended by new sensing functionalities while taking maximum advantage from the existing communication functions. 
Sometimes this approach is referred to as ``communication centered''. 
In its most resource efficient operational mode, \gls{isac} reuses the radio resources foreseen for communication purposes also for target illumination. 
Obviously, the concurrent use of the same limited frequency resources for two different services is most ``bandwidth efficient''. 
This allows to utilize the entire and constantly growing frequency spectrum, which is allocated for communication, also for the sensing task. 
This way, radar sensing can utilize the very different propagation conditions at frequencies ranging from FR1 to millimeter waves (mmWave) at FR2 and, in the future, even higher. 
Conversely, mobile radio can take advantage of the knowledge of the propagation environment gained from radar sensing for situation aware link adaption. 
Summarizing, the tight integration of communication and sensing functions in one system offers enormous synergy potential. 

Optimum design and full understanding of the possibilities and trade-offs of such fully integrated communication and sensing systems is still in its infancy. 
From an academic point of view, it is a challenge to determine to what extent the two communities for mobile communications and radar can benefit and learn from each other, despite or perhaps because of the different progress and developments they have made in recent decades. 

The advantage of a fully integrated design of \gls{isac} systems (often also called Integrated Communication and Sensing, ICAS) is multifaceted %complex
and goes far beyond frequency reuse. 
We strongly believe that \gls{isac} concepts will benefit enormously from \gls{mumimo} communication concepts, as positioning is an inherently cooperative task that requires the coordination of multiple spatially distributed sensor links. Therefore, the sophisticated \gls{mumimo} access schemes that have revolutionized the mobile radio access in communications have a similar potential for cooperative multi-link \gls{isac}. 
The idea of a distributed \gls{ms} \gls{mimo} \gls{isac} network which can be considered equivalent to \gls{mumimo} and was first introduced in~\cite{EUMW, ref12, ref3}. 
On the physical access level, the diverse functions of the \gls{mu} \gls{ran}, which are subjected to international standardization by 3GPP, already can create benefits for radar sensing. 
This includes the full range of broadcast and orthogonal multiuser access schemes in time, frequency and space, as well as signaling techniques for channel state estimation and subsequent link adaptation, predistortion, and resource allocation. 
Moreover, the \gls{mumimo} paradigm offers a perspective for distributed sensor cooperation between sensors at different levels, ranging from radio access coordination to the establishment of an adaptive radar sensing network in which the same network is used for target detection, data transport, and data fusion. 
Further, with distributed computing facilities such as the \gls{mec} we have all resources at our disposal which we may need to apply \gls{ml} and \gls{ai} for adaptive resource allocation, target parameter estimation, and scene recognition. 
This way, \gls{isac} can become a ubiquitous and cognitive radar sensing network~\cite{ref9}. 
Public and private cellular networks will offer an interesting perspective for \gls{isac} since they are controlled and administrated by \glspl{mno}, which now have all tools at hand to offer an administrated radar sensing service with well-defined sensing quality of service. 
On the other hand, the sensing functionality can be more decentralized and organized ad-hoc if the radio network is a meshed network of direct device-to-device communication of cars or drones.

Concerning the term \gls{mimo} we have to distinguish between collocated and distributed \gls{mimo}. While collocated \gls{mimo} in radar has a very specific meaning.
It includes a pair of a sparse and a smaller dense \gls{tx} and \gls{rx} array at the same platform to build a synthetic (quasi monostatic) radar array.
Comparably, a distributed \gls{mimo} radar, is based on transmitters and receivers that illuminate and observe (``sense'') the target from distinct locations that are far away from each other. 
This is also referred to as bistatic or multistatic radar if several \gls{tx}-\gls{rx} pairs are involved. 
The advantage of the distributed \gls{mimo} architecture in \gls{isac} is the main topic of this paper. 
In communications, \gls{mimo} was at first understood as a combination of antenna arrays at the \gls{bs} and at the remote \gls{ue}. 
The purpose was to establish a single data link between \gls{tx} and \gls{rx} with increased data throughput (\gls{mimo} switching or beamforming) or increased robustness (diversity). 
Only later multi-user \gls{mimo} was considered, which comprises several distributed \glspl{ue}, usually with multiple antennas on the \gls{rx} side~\cite{ref0}.

The main contributions of this paper are as follows: 
At first we discuss basic options for distributed \gls{isac} system concepts in conjunction with respective use cases. 
Then we introduce the generic architecture of a distributed \gls{ms} \gls{mimo} \gls{isac} network. 
We develop the basic multilink sensor access schemes and further elaborate on a reference signal reconstruction and synchronization scheme called \gls{cpcl}, which we have already proposed for the first time in~\cite{ref1}. 
We will further address the 3D target state vector estimation in dynamic multilink scenarios and give a comprehensive overview on multiple sensor link adaptation, coordination, adaptation and precoding in multipath environments. 
Finally, we offer a perspective on distributed \gls{msisac} resource allocation, data fusion and tracking. Since all the principles we propose are highly dependent on multipath propagation, we will we shortly refer to propagation aspects without referring to propagation simulation or measurement in detail.

\section{MS ISAC Architectures and Use Cases}
\label{sec:architectures}
In this introductory section, we discuss basic system architectures for distributed \gls{isac} networks using examples in close connection with promising use cases. 
To this end, we selected three network paradigms that differ in terms of access to network resources, involvement of the \gls{ue} and mobility of sensor nodes. 
We further underline and motivate the \gls{msisac} architecture by referencing respective use cases that were discussed in the German 6G research initiative and coordinated by the 6G Platform~\cite{ref24}. Another white paper on use cases was published by the European standardization organization ETSI~\cite{ref24a}. 
The use cases selected serve as examples that benefit from distributed \gls{msisac} systems. 
% Of course, this overview cannot be complete. 

In addition to applications for car and drone traffic, we focus on supporting public safety and protection of critical infrastructures, because these have not been discussed much in mobile communications so far, although they are urgently needed. 
We believe that these use cases directly benefit from \gls{isac} and can initiate pilot applications, which is crucial when it comes to new services to be offered by mobile operators. 
In the following discussion, we limit ourselves to radar detection and localization. 
While sensor data in other physical domains such as video, infrared, LiDAR, acoustic, etc., can be used together with radio sensing the discussion of these is beyond the scope of this paper, as we are focusing on the direct reuse of the communication radio interface for sensing. 
In addition, we note the radio interface can also be used to receive radio signals that are actively emitted by target objects. A malicious drone, for instance can be detected by its active radio emissions but as well by radar reflections even if it stops transmitting. 
Although this can be very important for some of the use cases below, we do not take it into account here.
\subsection{Infrastructure-only Sensing}
This variant uses only the infrastructure for target illumination, reflection detection, and data processing, which offers technical and economic advantages.
%This variant, consisting of target illumination, reflection detection, and data processing, is completely taken over by the infrastructure, which has technical and economic advantages. 
The sensor nodes are stationary and integrated in a well-defined environment. 
This includes power supply, synchronization, access to local computational resources, real-time data exchange between nodes and to the fusion center (perhaps via a fixed front haul network) and low latency, stable internet access. 
Since no \gls{ue} is included in the sensing loop, commercial issues with mobile communication users are not directly affected. 
This may also relax privacy issues, since no user data are shared to external parties/devices.   

\begin{figure}
    \centering
    \includegraphics[width=0.57\linewidth,bgcolor=gray!10,rndcorners=5,rndframe={color=gray!50, width=\fboxrule, sep=\fboxsep}{5}]{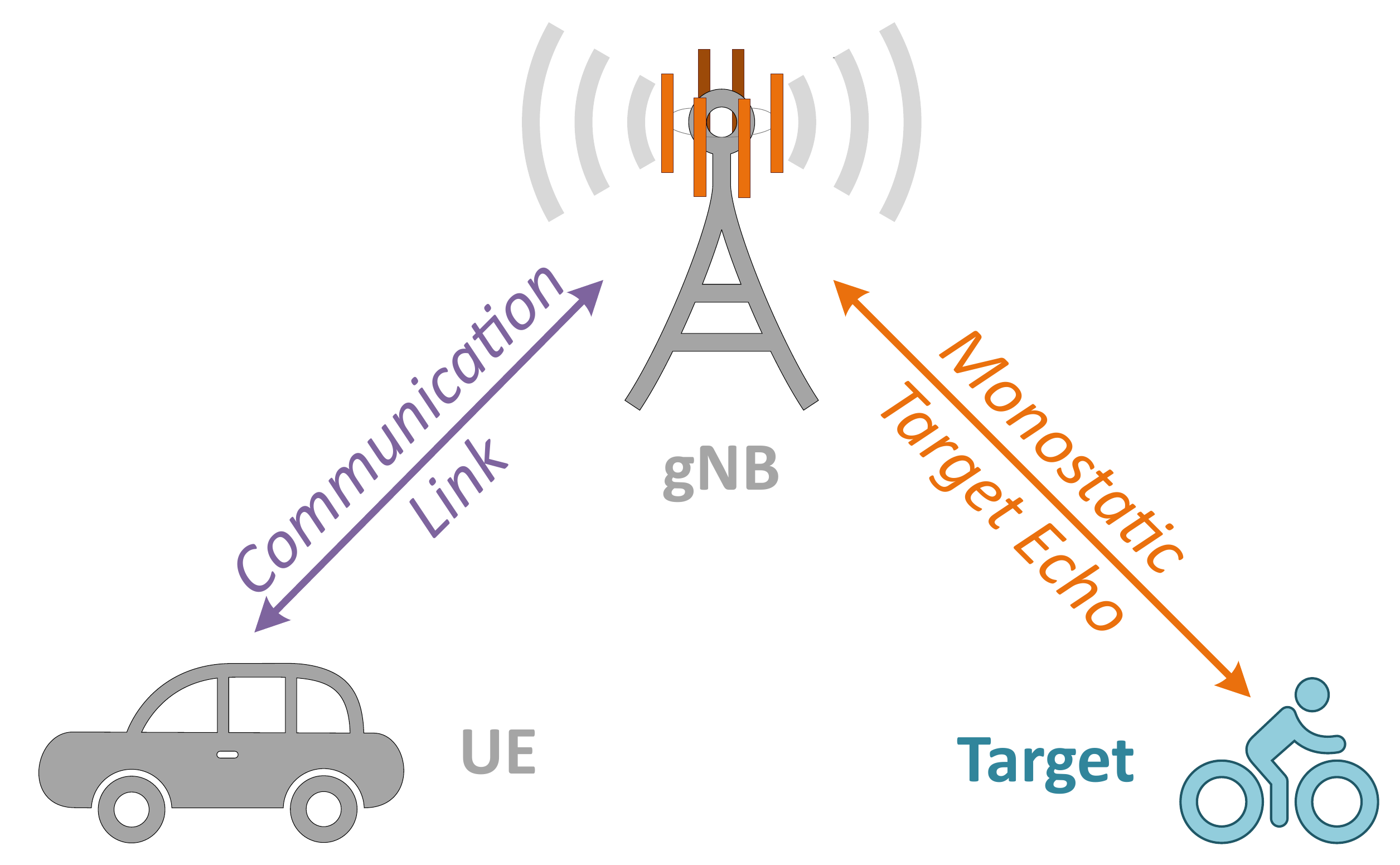}
    \caption{Infrastucture based sensing (Solitary \gls{gnb} - monostatic). The solitary \gls{gnb} communicates with the \gls{ue} and at the same time receives radar target returns.}
    \label{fig:infrasensMono}
\end{figure}

In terms of technical aspects, \gls{isac} was originally considered mainly as a monostatic single site concept, consisting of a \gls{gnb} equipped with an antenna array, \Cref{fig:infrasensMono}. 
This has the logistical advantage of being a standalone single-station sensing system, where the array is used for \gls{doa} estimation. 
Together with \gls{tof} estimates from detected target returns (target range) one can estimate the target location. 
However, cross range spatial target resolution degrades with distance and the underlying monostatic radio sensing geometry would require full duplex radio access or at least collocated \gls{tx}~/~\gls{rx} radio interfaces (quasi monostatic).
But despite of several research activities towards full duplex access, it is not yet standard in mobile radio since established communication waveforms (such as \gls{ofdm}) are not well suited for full duplex. 
Direct \gls{tx} to \gls{rx} feed through would be a serious problem for quasi-monostatic configurations and for lower frequencies an antenna array suitable for \gls{doa} estimation is not always available. 
Moreover, for higher frequencies (FR2 and beyond), analog phased arrays, or respectively hybrid arrays, will be used which are not directly applicable for high resolution \gls{doa} estimation. 
A related and unconventional usage of spatial precoding for \gls{doa}~/~\gls{dod} estimation is described in \Cref{sec:access} of this paper.

\begin{figure}
    \centering
    \includegraphics[width=\linewidth,bgcolor=gray!10,rndcorners=5,rndframe={color=gray!50, width=\fboxrule, sep=\fboxsep}{5}]{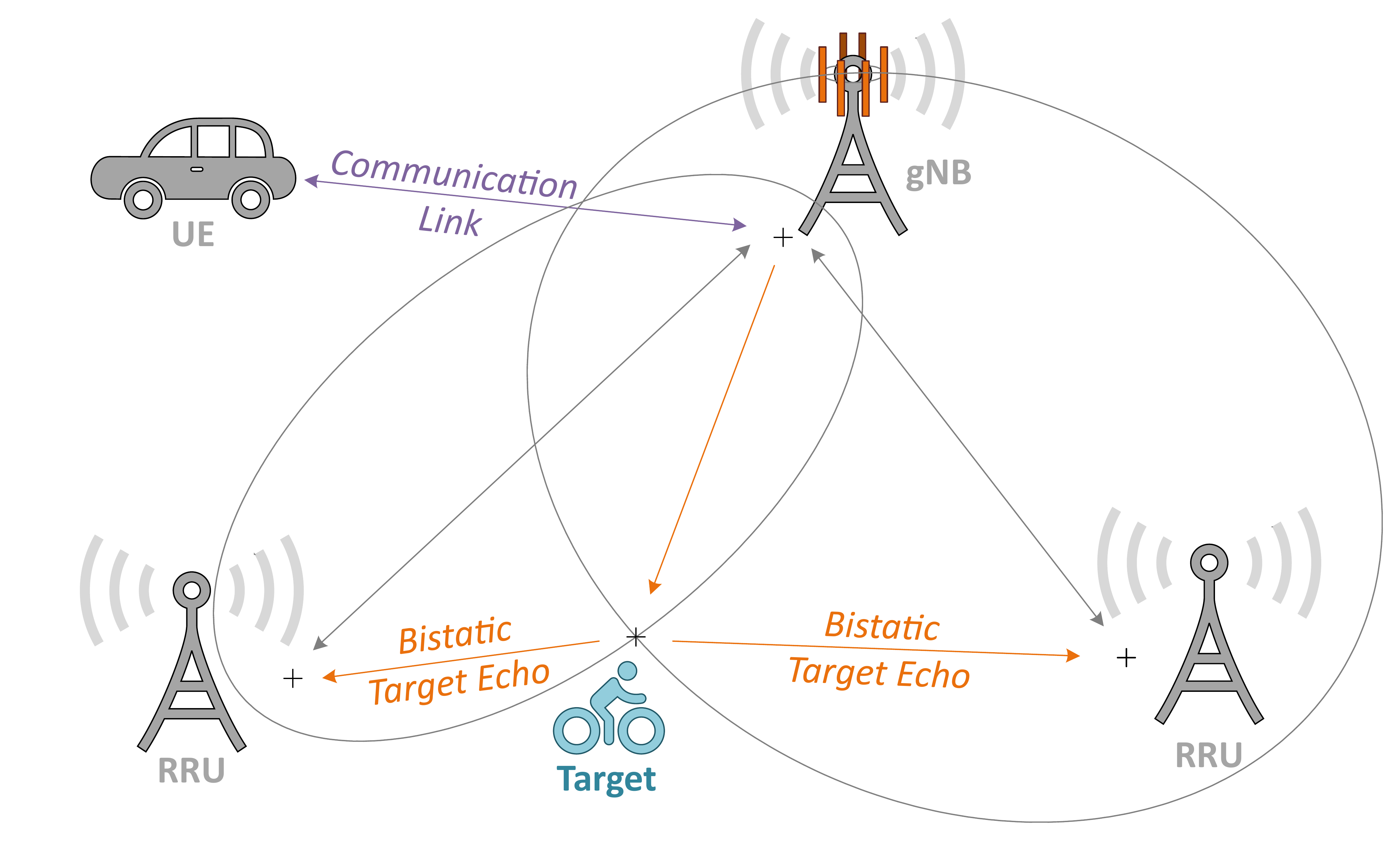}
    \caption{Infrastucture based sensing (Distributed \gls{gnb} - multistatic). Multiple \glspl{rru} connected to a cloud-controlled baseband radio pool form a distributed base station.  
    One \gls{rru} communicates with the \gls{ue} and at the same time illuminates the scenario, while other \glspl{rru} receive bistatic radar returns.}
    \label{fig:infrasensMulti}
\end{figure}

Many of the inherent problems associated with the monostatic single station architecture can be solved with the \gls{dbs} concept depicted in \Cref{fig:infrasensMulti}, which changes the radar geometry from monostatic towards bistatic radio access~\cite{EUMW, ref3, ref12}. Each \gls{rru} receives the backscattered waves transmitted from another \gls{rru} for communication purposes in its respective downlink. 
For sensing, this has several advantages. 
%First of all,
To begin with, it does not require a full duplex radio interface since the direct \gls{tx}\,2\,\gls{rx} feed through (also called self-interference) can be controlled by the distance, position and isolation of the radio nodes and, thus, can be kept small enough. 
However, the sensing \glspl{rru} must be silent while one of the other \glspl{rru} is transmitting. 
This reduces the efficiency of communication. 
Moreover, we must make sure that the different \glspl{rru} have overlapping radio coverage. 

As for the localization paradigm, we move from single station joint distance and \gls{doa} estimation to multiple bistatic (or multistatic) range estimation.  Moreover, joint range/Doppler estimation for position/speed-related measurements is considered throughout this paper. Multilateration requires that there is sufficient distance between the transmitting and receiving nodes. With the density and spatial spread of the radio nodes deployed, we can control localization performance and coverage. \Cref{fig:infrasensMulti} already indicates that target position estimation uses the  \gls{etof}, which results from a comparison of the \gls{los} \gls{tof} vs. \gls{tof} of the path that is routed via the target (called \gls{btp}), which ends up in an ellipse indicating the target position. 
The focal points of the ellipse are the \gls{tx} and \gls{rx} locations, respectively. 
To obtain the target position, we need multiple (at least three) intersecting ellipses, where the accuracy depends on the length of the \mbox{\gls{tx}-\gls{rx}~base~line}. 
Since this multilateration localization does not require antenna arrays, it is also applicable at lower frequencies where suitable arrays may not be available. 
In addition, it saves measurement time, since beam search is not required, but beamforming can be additionally applied for \gls{snr} gain and clutter filtering. 
Further advantages of spatially distributed sensing result from the target-related spatial diversity gain. More details will be discussed \cref{sec:propagation:diversity}. 

The distributed base station concept in \Cref{fig:infrasensMulti} includes synchronized \glspl{rru} (also called \glspl{rrh}), connected to a central, software-defined and cloud-controlled baseband radio pool (CRAN) via fixed wireless or fiber-optic fronthaul links. There are several reasons for this architecture, e.g. flexibility and easy reconfigurability, better spatial coverage, increased capacity for long-range access~\cite{ref21,ref22}, and enhanced positioning services~\cite{ref20}. 
For \gls{isac}, a modification of the distributed \gls{rru} concept was proposed in~\cite{ref19}. It is based on \glspl{rru} dedicated for receiving only. These ``sniffer'' nodes were originally proposed to submit another, well-separated radio interface for quasi-monostatic sensing (replacing full duplex radio interface). 
However, if the sniffer is deployed at a bigger distance from the \gls{rru} (e.g. several hundreds of meters away or more in a free space cellular environment), we end up with a clearly bistatic architecture or better to say a multistatic architecture if the sniffer receives also \gls{dl} signals from other \glspl{rru} or if there are multiple sniffers deployed. The proprietary link to the CRAN makes the sensor part of the infrastructure. The distributed, modular \gls{isac} architecture is further leveraged by the Open RAN concept that supports use of multivendor radio modules and increases functional flexibility (even out of standard) based on software-centric architectures~\cite{ref23}. Dedicated computational facilities (e.g. based on high-performance graphical processing units) for most computationally demanding radar data processing, data fusion and resource allocation fall in the scope of this concept, as well as \gls{npn} that are installed in wide area industrial campuses are promising for distributed infrastructure-centric \gls{isac} ecosystems.

\paragraph*{Use Cases} Infrastructure-only sensing seems to be a promising use case if there is a well-defined (perhaps spacious) region that needs ubiquitous and around-the-clock surveillance even if no \gls{ue} is around. 
This can be large outdoor industrial areas like power plants, logistic centers, railroad marshaling yards, harbors, etc., which need to be protected against intruders, spying, and terror attacks. 
While some high sensitive premises are already well protected by special security systems, others are not, mostly because of economic reasons. 
Some elements of the critical infrastructure are located in remote areas, just not observed and weakly secured. 
Examples are base station towers, water supply stations and energy infrastructure facilities. 
The latter is becoming increasingly important with the more decentralized power networks (power supply nodes or other de-central facilities, wind turbines, fields of photovoltaic collectors, etc.). 
Other examples are linear spaciously extended elements of infrastructure, such as power lines, waterways, or railway tracks. 
Also road traffic scenarios need more attention. 
While car manufacturers follow the paradigm of the so-called ``ego-view'' which is related to the perception of the sensors installed in the respective cars to support autonomous driving, the ``regulatory view'' would deserve more attention. The latter relies on a comprehensive perception of the large-scale road situations at junctions or major intersections, accident hot spots or other critical positions for traffic supervision by regulation authorities. 
Since the regulatory view cannot be generated from aggregated ego-views of cars, an independent sensor network would be necessary. 
The regulatory view would include recognizing dangerous situations that could lead to accidents and predicting traffic jams or blockages that could be avoided by diverting traffic flows. 
We can also detect illegal activities such as illegal car racing or other traffic offenses as well as wrong-way drivers on motorway slip roads. 
Another specific application is securing highway parking lots against organized cargo theft, which causes immense economic damage on German autobahns.

However, fraudulent or otherwise harmful intruders can also be \glspl{uav} respectively drones. 
The problem is well known from air traffic disruptions at public airports by hobby drones and will become much bigger with the increase of the commercial drone traffic. 
\Glspl{mno} have already started to support \gls{blos} drone connectivity and offer \gls{utm} systems. 
A special airspace has been established called ``U-Space'', in which rules for safe traffic are defined (see e.g. EU Drone policy 2.0).
% https://transport.ec.europa.eu/document/download/1cb5fb4f-4252-4f97-abf4-c4a167b1c7d2_en?filename=COM_2022_652_drone_strategy_2.0.pdf
However, these \gls{utm} systems do not sufficiently consider the chance of a violation of the \gls{utm} rules, may it be intentionally or unintentionally. 
Therefore, besides of cooperative traffic monitoring we would need an independent surveillance system to detect exceptional and threatening situations (``trust but verify''). 
However, the existing air traffic monitoring systems cannot be used for U-space surveillance because they are neither technically designed for this purpose nor can they be operated on an area-wide basis in order to economically secure the U-space level. 
We would need a scaled-down system being available to public and commercial drone operators everywhere. 
U-space surveillance is also necessary for public events against terror attacks carried out by drones and to avoid illegal observation of private and public premises by camera drones. 
An infrastructure-only \gls{msisac} system will have the technical and economic surveillance capability with ubiquitous availability. 
The advantages of \gls{isac} follow from the fact that the independent surveillance capability is enabled by the same mobile communication system that also hosts the cooperative \gls{utm} system with minor infrastructure extensions. 
In addition, the superior range of dedicated radar is compensated for by a denser network of sensors and illuminators---another benefit of network densification. 

We also do not want to conceal the fact that a comprehensive solution to the problem of U-Space surveillance would require additional sensor functions.
For example, radio communications to or from the drones should be monitored.
An active communication link is most easy to detect and already reveals a lot about a target's mission.
Furthermore, no transmit resources are required for target illumination.
On the other hand, radar sensing has the advantage that the drone can be detected even when there is no active telemetry or data connection.
It therefore makes sense to combine both methods of radio sensing, namely localization of actively transmitting targets and radar localization of passive targets.
Finally, receiver interfaces of mobile networks can be enabled for the detection and localization of transmitters.
This is as another advantage of integrating radio sensing functionalities into the mobile communication network.
However, a more detailed discussion of the field of radio surveillance which includes identification, recognition, localization and tracking of active radio emitters is far beyond the scope of this paper.         

\subsection{Uplink~/~Downlink Sensing}

\begin{figure}
    \centering
    \includegraphics[width=\linewidth,trim={0 -30 -20 -30},bgcolor=gray!10,rndcorners=5,rndframe={color=gray!50, width=\fboxrule, sep=\fboxsep}{5}]{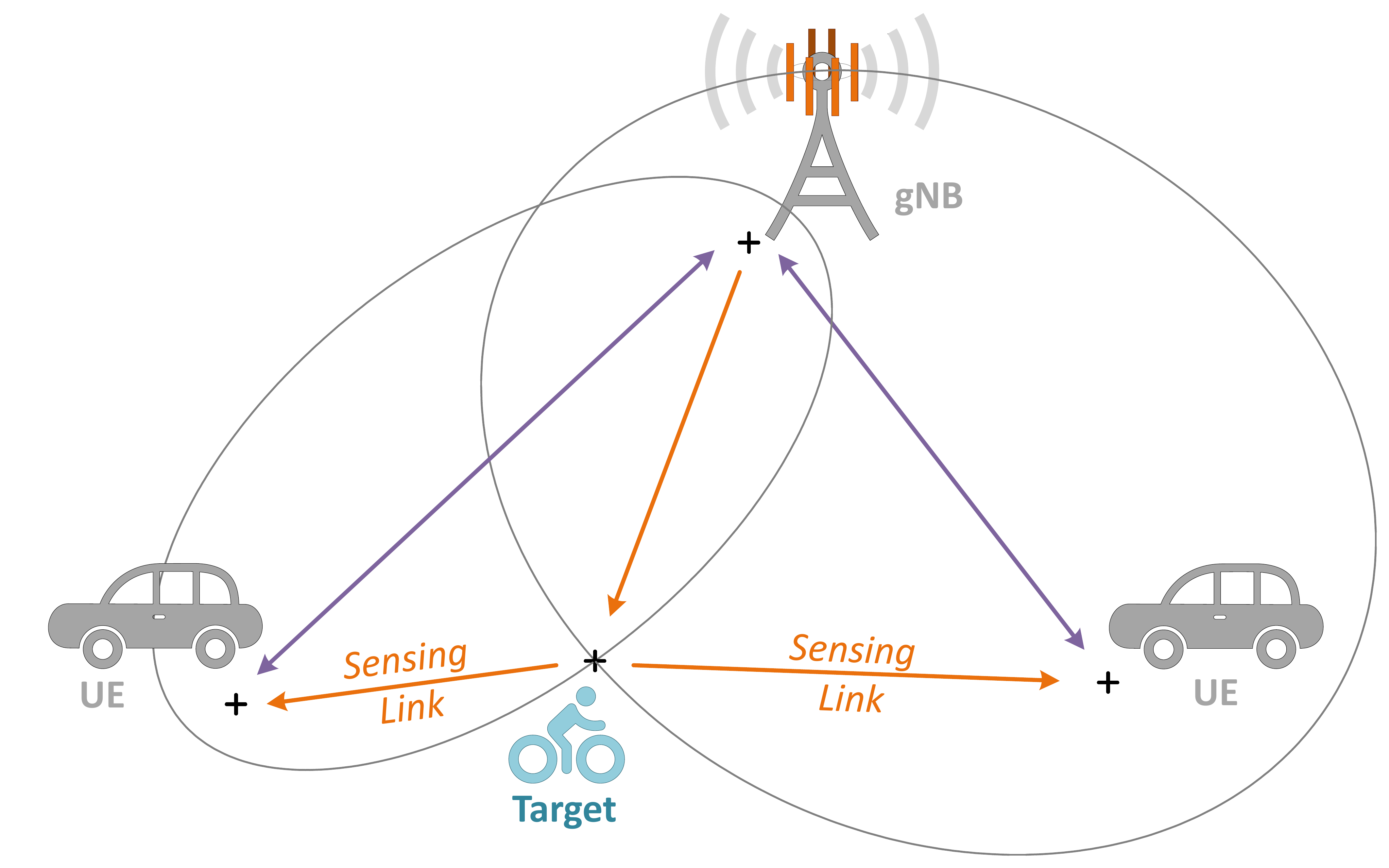}
    \caption{\gls{msisac} with mobile \gls{ue} included in the downlink or uplink.}
    \label{fig:UEincluded}
\end{figure}

Compared to infrastructure-only sensing, uplink~/~downlink sensing utilizes a \gls{ue} directly as an illuminator or a sensor in the sensing cycle, where the \gls{bs} acts as complementary sensor or illuminator, see \Cref{fig:UEincluded}.
Now, the sensing result depends on the availability, position and dynamics of the \glspl{ue}, e.g. cars, drones or sniffing sensor nodes that are deployed only on demand or moving. 
In any case, the sensors and the illuminator are not directly connected by a fixed link as we have postulated in cases of the distributed infrastructure-only sensing. 
However, we assume the \gls{ue} is booked into the \gls{ul} or \gls{dl} of the base station. 
Therefore, the sensing geometry is bistatic in general and corresponds to the ``natural'' arrangement of \gls{ue} vs. infrastructure in mobile communication. 
This situation has some striking similarities to the passive radar principle, which is also called \gls{pcl} and well known for decades~\cite{ref29}. 
\Gls{pcl} relies on a so called transmitter of opportunity, e.g. a digital broadcasting DVB-T or analog FM transmitter. 
The \gls{pcl} sensor tries to receive the clean transmit signal which is used as a correlation reference for the surveillance signal. 
Therefore, a conventional \gls{pcl} sensor typically has two receiver channels, a highly directive channel delivering the multipath-free copy of the \gls{tx} signal and an omnidirectional surveillance channel containing the target response. 
Besides of the target related diversity gain of the bistatic sensing geometry, \gls{pcl} offers the advantage that \gls{tx}\,-\,\gls{rx} synchronization is no longer an issue as we only need to know the \gls{tof} difference between the reference and the surveillance channel at the receiver, which we call \gls{etof}. 
Of course, we have to make sure that the transmit reference is received via \gls{los} and we need to know the position of both, the \gls{bs} and the (mobile) \gls{ue}. 

In continuation of the \gls{pcl} concept we have proposed \gls{cpcl} in~\cite{ref1}. The discussion about \gls{cpcl} will run through the entire paper. It turns out that the advantages of \gls{cpcl} are striking and manifold. 
\Gls{cpcl} is a way to fully integrate sensing functionality into the radio access scheme for communication. 
It allows to use the pilot and reference signals for \gls{tx}\,/\,\gls{rx} frame synchronization and cooperation based on feedback signaling. 
It also paves the way to utilize the complete communications data payload for radar sensing, while taking full advantage of the flexibility of 5G~\gls{nr} access schemes for signal adaption. 

A mobile \gls{isac} sensor in the \gls{dl} can already obtain benefits even if the sensor is only passively sniffing and does not act as a fully authorized communication user. 
In this reduced \gls{cpcl}-mode, a frame synchronized receiver can take full advantage of \gls{fft} based frequency domain system identification methods~\cite{ref30,ref96} over classical \gls{pcl}. 
This includes cyclic prefix removal and thus lower variance since intercarrier leakage almost disappears. 
Moreover, the received signal can be equalized which supports transmit reference signal recovery without requiring a dedicated (authorized) receiver channel. 
In the sniffer mode, the \gls{dl} plays the role as broadcast sensing channel even if it is not a broadcast channel for communication. 
However, as there is no active interaction with the illuminating node, the network resources cannot be used for data fusion. 
If the mobile \gls{isac} sensor is a fully authorized \gls{ue}, the communication functions on the network level can be used for reporting of sensing results and subsequent data fusion (with access to computational resources), predistortion at transmitter side and resource allocation which makes \gls{cpcl} most attractive if mobile or deployable \gls{ue} devices are included. 

Actually, there are considerable differences between \gls{ul} and \gls{dl} \gls{isac} in terms of multisensor access and reasonable application scenarios. 
While a mobile sensor, being a fully connected \gls{ue} or just a sniffing node, can use the complete multiuser communication payload transmitted by the \gls{gnb} in the \gls{dl} for broadcast \gls{isac} illumination, the situation in the \gls{ul} is different. 
Access in \gls{ul} has to be designed orthogonally since the target illumination is facilitated by the multiple transmitting \glspl{ue} which are at different positions. 
However, this immediately gives the \gls{gnb} multiple concurrent measurements, hence multiple ellipses, for target localization (assuming that the target is in the sensing coverage area of both illuminating \glspl{ue}). 
But what about localization in the \gls{dl} case? 
From where do we get the multiple results? 
There are two options: In case of multiple \glspl{ue} (respectively mobile sniffing sensors) receiving sensing signals, they would have to share their sensing data for data fusion, which would need dedicated communication channels. 
The other option is that the mobile sensing node should look out for another illuminating \gls{gnb}.
If this auxiliary network node is used for \gls{cpcl} in its reduced form, just as a sniffer node without communication connectivity to the illuminator, it does not need to decode data and no communication resources are consumed. 
However, the respective illumination signal has to be taken as it is and no cooperation with the transmitter (e.g., no adaptive modification of the transmit signal) is possible. 
Required metadata can be transferred by the primary link. 
Obviously, the auxiliary illumination source can be different in its radio parameters from the primary \gls{isac} link, e.g. located at a different frequency or with different bandwidth and modulation. 
This would open interesting perspectives for radar sensing. 
Further drawbacks of \gls{ul} sensing result from the lower transmit power of the mobile \gls{ue} and from lower duty cycle in \gls{tdd} for target illumination.  

\paragraph*{Use Cases} \gls{isac} in down link has the potential to complement the already very well established auto motive radar and LiDAR sensors in cars.
We do not assume that \gls{isac} can replace these sensors, but \gls{isac} can support the ``ego view'' that only emanates from the illuminators and sensors installed on the cars. 
This renders the benefit immediately visible for the road user. 
In \gls{dl} \gls{isac} access, the additional illumination by infrastructure nodes can help to better recognize the local environment of the car. 
The bistatic radar architecture already provides an advantage with its inherent diversity advantage that supports an extended field of view and more comprehensive perception. 
Of course, the biggest gain will arise if all sensor data available in the car, including ego radar and LiDAR, can be fused. 
\gls{isac} can provide additional road safety as it is an independent system, which can be an advantage if the density of automotive radars undergoes some saturation because of mutual interference as an automotive radar does not include an orthogonal multisensor access scheme. 
Other areas of application can be comprehensive perception, navigation and collision avoidance for \gls{agv} in warehouses, extended manufacturing facilities, large logistic facilities, reloading sides, etc. 
As the mobile sensors can rely on the more powerful base station to illuminate the scene, promising applications are conceivable in which the mobile sensor explores its surroundings. 
There is a similar advantage in detecting low-flying drones that might be hidden behind buildings or other terrain structures. 
They become visible when illuminated by base stations on the ground and observed by high-flying sensing drones. 
Also in this scenario, the high transmission power and ubiquitous availability of base stations allow better illumination of the scenario, while the sensing drone only needs to carry a receiver when observing the scenario.

The \gls{ul} scenario seems to be more appropriate to support the operator respectively regulatory view described above for infrastructure only sensing. 
The mobile, distributed node can help to get a better illuminated scenario. 
In special critical situations, the mobile node can be assigned a specific illumination/sensing mission. 

\subsection{Sidelink Based Sensing in Mesh Networks}

Already in 5G, 3GPP has specified the \gls{sl} direct device to device communication~\cite{ref26,ref27}. 
The \gls{sl} it is mainly intended for mobile users, mostly cars.
But also \glspl{uav} can benefit.
The \gls{sl} technology is also called Cellular Vehicle-to-Everything (C-V2X) radio access technology since it is deemed as an extension of the cellular network. 
The \gls{sl} supports both, autonomous operation and resource allocation for direct \gls{ue}$\to$\gls{ue} communication or resource allocation controlled by the \gls{gnb}.
This allows dedicated low latency communication between mobile units but also independent operation if no infrastructure is in reach.
First thoughts about integration of sensing and sidelink communications have been published already~\cite{ref25}. 
The advantage of sidelink-based sensing is that we get a meshed network of mobile nodes that can act both as illuminators and sensors. 
The mobile sensor cloud can be configured and controlled by a supervising instance within the mobile network or it can act ad hoc and self-organized once it has been set in motion. 
Since the weight and energy consumptions of the payload will be limited, cooperative acting and task sharing will become important. 
This relates to sensing (e.g. distributed \gls{toa} localization schemes), computation (distributed fusion and off-loading of preprocessed data to a fusion center) and also data transfer. 
Therefore, communication and sensing radio interfaces can be heterogeneous. 
A master drone can collect data, deliver data fusion capabilities at the edge and transfer data to the cellular network. 
This way, a swarm of drones can accomplish its mission autonomously and finally report the condensed result.

\paragraph*{Use cases} 
Using the \gls{sl} for integrated sensing in dense traffic scenarios can substantially extend the situational awareness compared to the established ego-view oriented automotive radars even if no infrastructure is in reach. It results from the spatial diversity of the meshed, distributed \gls{mimo} radar network, which is spanned by the cluster of neighboring vehicles. Compared to automotive radar, the lower frequencies (FR1) offer wider coverage due to lower path loss and less shadowing, and, to some extent, enables the visibility of targets in non-\gls{los} regions. Hence, meshed \gls{sl} \gls{isac} has the potential to expand the ego-view. It will be a step towards cooperative sensing. An advantage over the pure exchange of ego data can be that \gls{isac} data result from using a common, standardized radio interface. Therefore, car manufactures maybe less reluctant to contribute to this type of a cooperation since they do not need to share their proprietary ego-sensing data. \gls{isac} radar sensing in combination with \gls{sl} communication can also be used to increase trustworthiness for communication since it can help to identify fake transmitters. For instance, a malign \gls{sl} node trying to fool the other connected road users by sending wrong \gls{cam}, can be identified by comparing the pretended \gls{cam} information with the radar ground truth measured by \gls{isac} (relative distance or speed). 

\begin{figure}
    \centering
    \includegraphics[width=0.85\linewidth,trim={-60 -40 -60 -35},bgcolor=gray!10,rndcorners=5,rndframe={color=gray!50, width=\fboxrule, sep=\fboxsep}{5}]{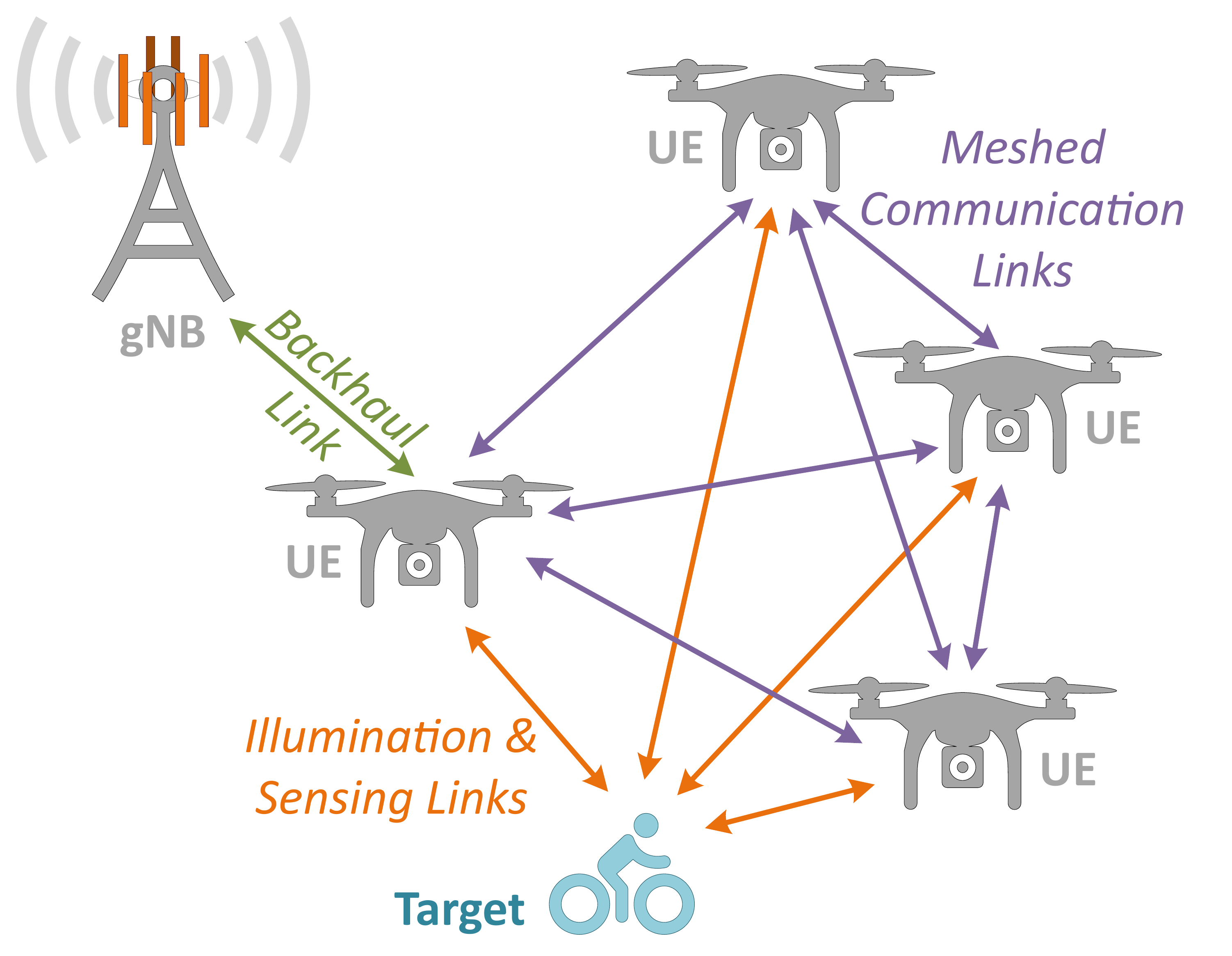}
    \caption{Cooperative \gls{isac} in sidelink (\gls{sl}) communication between drones}
    \label{fig:coopdrones}
\end{figure}

Besides meshed \gls{isac} sensing using the \gls{sl} in public road traffic, other applications with mobile clouds of connected vehicles are on the horizon. Examples are autonomously acting \glspl{agv} or swarms of drones. Application of drones for real-time monitoring of ground activities, for or the purpose of law enforcement, search and rescue, technical inspections, environmental monitoring, etc. is already widespread. For instance, drones can help to increase safety of big public events by early recognizing mass panics and critical streams of movement, detecting blocked or closed exits or even guiding people to open exit routes. Auxiliary \gls{isac} sensing can help to extend the perceptive capabilities by the radio domain. Radar sensing can help when optical sensing fails because of bad whether condition or missing natural optical illumination. With radar sensing we may get an overview picture more easily and measure relative velocities. If, in addition to radar, \gls{isac} also detects intentional or unintentional radio emissions from the ground, we can get a more comprehensive picture of the situation on the ground. Moreover, besides sensing of ground activities, \gls{sl} based sensing can help to localize relative positions of drones towards the swarm for collision avoidance (see \Cref{fig:coopdrones}). 

\section{Generic MS MIMO ISAC Architecture}
\label{sec:generic}
\subsection{Distributed \gls{ms} \gls{mimo} \gls{isac}}\label{sec:generic:msmimoisac}
The three basic architectures of \gls{msisac} systems and the corresponding use cases discussed in the previous chapter have already demonstrated the advantages of distributed \gls{mimo} \gls{isac} concepts. This already included several bistatic, i.e. multistatic, sensing links. It also became clear that the multisensor radio access procedures, the allocation and scheduling of radio resources and synchronization issues will be diverse and extensive. Ultimately, we need a comprehensive architecture for accessing and coordinating multiple sensors. The purpose of this chapter is to outline a generic \gls{ms} \gls{mimo} \gls{isac} architecture and to highlight the potential basic system design. In terms of the radar community, we obtain a distributed \gls{mimo} radar architecture~\cite{ref10,ref10b}, which differs well from a \gls{mimo} radar with collocated antennas~\cite{ref10a}. The latter is a favored solution for quasi-monostatic radio access.
It combines the advantage of separated \gls{tx} and \gls{rx} antennas arrays on the same platform for reduced \gls{tx} to \gls{rx} interference with a reduced hardware overhead for forming a big virtual array from smaller, respectively sparse Tx/Rx subarrays at the expense of some transmit multiplexing scheme. For a more comprehensive discussion of collocated \gls{mimo} radar readers are referred to~\cite{ref10b} and the references given therein. However, distributed \gls{mimo} radar corresponds much better to \gls{ms} \gls{mimo} \gls{isac} as discussed already in the section above. Another striking equivalence becomes obvious when \gls{ms}-\gls{mimo} \gls{isac} is compared to \gls{mumimo}~\cite{ref31}, which was a big step forward in mobile communications compared to single-user \gls{mimo} by adding multiuser capability in the wireless realm. Therefore, it seems worthwhile to consider advanced \gls{mu} access schemes in mobile communication for application to \gls{msisac}. 

\begin{figure}
    \centering
    \includegraphics[width=0.85\linewidth,trim={-50 -50 -50 -50},bgcolor=gray!10,rndcorners=5,rndframe={color=gray!50, width=\fboxrule, sep=\fboxsep}{5}]{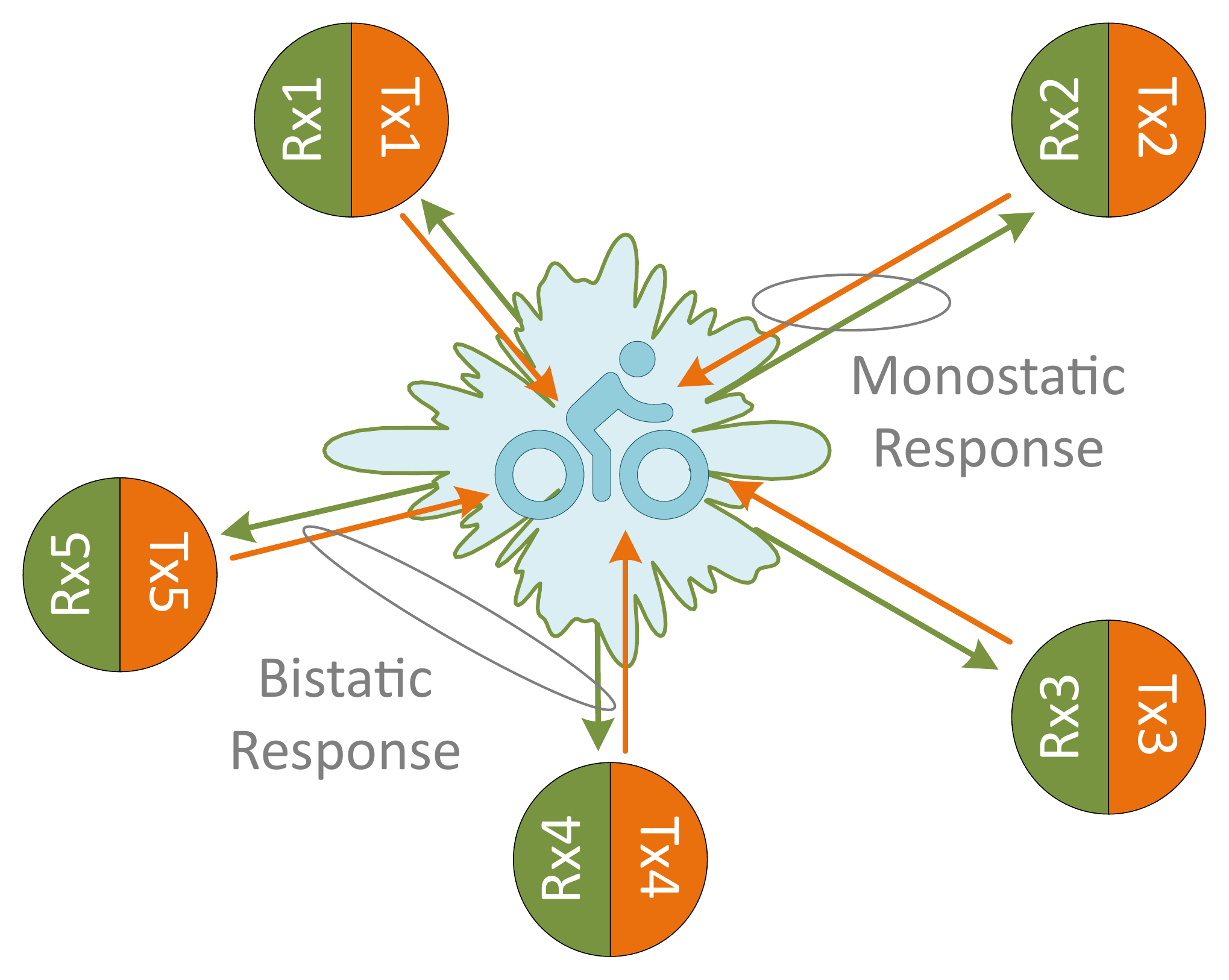}
    \caption{Generic distributed \gls{ms}-\gls{mimo} \gls{isac} architecture consisting of multiple transmit-receive links. In general, the radio nodes can operate in monostatic and/or bistatic access mode.}
    \label{fig:generic_mimo_arch}
\end{figure}

From \Cref{fig:generic_mimo_arch} it becomes clear that collecting the full distributed $N \times N$ \gls{mimo} matrix with the $N$-node \gls{ms}-\gls{isac} network requires $N$ monostatic and $N^2 - N$ bistatic measurements. Each measurement consists of a normalized cross-correlation function of the respective received signal $y_n(t)$ with the transmitted excitation signal $x_m(t)$ (the correlation reference) between nodes $m$ and $n$, which is regarded as an estimate of the corresponding Tx-to-Rx \gls{cir} $\bm h_{m,n}(t)$, respectively \gls{frf} $\bm H_{m,n}(f)$. \Cref{eq:mimo_matrix} depicts the full distributed I/O \gls{mimo} matrix according to \cref{fig:generic_mimo_arch}. The elements in the main diagonal result from the monostatic measurements whereas the entries in the upper and the lower triangular part corresponds to bistatic measurements in the Forward (FW) and Backward (BW) link (\gls{ul} and \gls{dl}, respectively). If FW and BW connections are reciprocal, one half of the bistatic part can be regarded as redundant. However, since radio nodes can be heterogeneous in the FW, BW and access mode (hence non-reciprocal) we can get up to $N^2$ independent measurements. This already shows the difference of multistatic vs. multiple monostatic measurements. Why do we need so many measurements? The advantages are the resulting number of spatial degrees of freedom and diversity. While the former is well defined by the sensor-target geometry, the latter has a statistical meaning. We can select the measurements with the highest geometric relevance for localization and omit the others (e.g., the monostatic links if there is a lack of full duplex radio interface), or we can use the seemingly redundant measurements for further location variance and bias reduction measures and improve target recognition. Surplus measurements can also be used to detect targets in cases of missing \gls{los}, respectively target shadowing. Another advantage of distributed \gls{msisac} comes from increased spatial coverage and robustness since the sensor nodes act independently and can complement and substitute each other. Access coordination between distributed links and collaboration may become more important for \gls{ms}-\gls{mimo} \gls{isac} than for \gls{mumimo} communication, as target localization and tracking is bound to a clear multidimensional geometry and subjected to real-time conditions.

The so called \gls{mimo} matrix $\bm H: \mathbb{R} \rightarrow \mathbb{C}^{N \times N}$ collects the respective \gls{frf} for each of the $N^2$ possible measurements via
\begin{equation}\label{eq:mimo_matrix}
    \bm H(f) = \left(\begin{array}{ccc}
        H_{1,1}(f) & \ldots & H_{1,N}(f) \\
        \vdots         & \ddots & \vdots         \\
        H_{N,1}(f) & \ldots & H_{N,N}(f) \\
    \end{array}\right),
\end{equation}
where each element of $\bm H$ is defined as
\begin{equation}\label{eq:mimo_matrix_elements}
    H_{m,n}(f) = \frac{Y_m(f)}{X_n(f)}, \quad 1 \leqslant n,m \leqslant N,
\end{equation}
and $X_n$ denotes the transmit signal of node $n$ and $Y_m$ the receive signal at node $m$.
Hence, \eqref{eq:mimo_matrix} essentially denotes the estimated \glspl{frf} between nodes $m$ and $n$, corresponding to \Cref{fig:generic_mimo_arch}.
\subsection{Heterogenous Radio Nodes}
The radio nodes of the distributed \gls{ms}-\gls{mimo} \gls{isac} network in \Cref{fig:generic_mimo_arch} can be heterogeneous in many aspects. Heterogeneity can occur on the level of radio parameters or on the level of network parameters. In terms of the radio parameters, we may see differences w.r.t. frequency range, bandwidth, transmit and receive access schemes, antennas, polarimetric access, etc. These parameters are related to wave propagation aspects and to design issues of the radio interface. Since mobile radio frequencies range from well below \SI{1}{\giga\hertz} up to millimeter wave frequencies (perhaps more than \SI{100}{\giga\hertz} in the future), we can take advantage of the different scattering, diffraction and penetration characteristics in this wide span of frequencies. Roughly speaking, the higher the frequency, the more optical-like the propagation mechanisms become. At lower frequencies, we observe more diffraction and diffuse scattering, also material penetration becomes possible. At higher frequencies, objects are more likely to act as reflectors, although surface roughness may become more important and obstructing objects may cause more blockage, and polarization selectivity of target details may become stronger. Finally, the same relative speed results in respectively higher Doppler shift. As far as the radio interface is concerned, the antennas play the most important role. Since the directivity depends to a certain extent on the size of the antenna in relation to the wavelength, the antennas tend to be smaller for the same directivity (or gain). However, link attenuation increases because of the smaller effective antenna area. This leads to the use of larger and therefore more directional antennas, which must be controllable in mobile applications. 

From a technical point of view, the combination of different frequencies offers many advantages, but also poses challenges. While only smaller frequency distances can still be handled by a single \gls{udc} radio interface, carrier aggregation and wider frequency spacing (e.g. FR1 and FR2) may require multiple RF chains or non-heterogeneous radio nodes. While widely separated frequency blocks can increase delay resolution a lot, coherent processing would be necessary which is possible if the frequency blocks belong to the same \gls{udc} channel (interband processing). If the frequency blocks belong to different \gls{udc} channels, non-coherent combining of distant frequency bands is easier to achieve and can still help to increase target resolution, may make detection more robust, and reveal different features for target recognition by frequency diversity. Another advantage of frequency diversity results from the interplay between coverage and resolution. Lower frequencies enable wider coverage with less directional antennas. Higher frequencies with their wider bandwidth channels support higher delay (resp. range) resolution. At the same time, beamforming becomes possible which allows spatial filtering (to suppress clutter) and increase \gls{snr}. However, any beamforming also requires beam search, resp. scanning procedures, which may be seen as too tedious and cumbersome for fast target acquisition. Obviously, a cooperation of low frequency nodes for target acquisition and high frequency nodes for precise localization, recognition and tracking seems to be an advantage. This may include beam search instruction at higher frequencies from measurements at lower frequencies.  

Concerning the antennas and antenna arrays used, a wide variety of differences can occur. Here we only mention some. The difference between linear/planar vs. circular/cylindrical array, e.g., decides about full azimuth angle coverage which seems be more important for sensing than for communications. Moreover, elevation coverage matters if targets are expected in the air (e.g. drones). Important is also the difference between analog (phased array) and digital beamforming. While the former is capable only for beamforming, the latter has the potential of direction estimation. The reason is that estimation of direction is based on mutual correlation processing of array outputs. Analog beamforming (which seems dominating for higher frequencies) embodies a linear weight and sum processor that does not allow access to the individual antenna outputs for correlation processing. Full polarimetric processing requires polarization selective antennas and two corresponding \gls{udc} channels on the \gls{tx} and \gls{rx} side.

The sniffer nodes discussed in \cref{sec:generic } may fall out of the ordinary even more. 
At first, a sniffer will act as receive only. 
But even if it is hosted by a \gls{gnb}, it can cover extended frequency bands outside the host's assigned bands. 
Receiving off-system signals would help to collect additional \gls{isac} data in the reduced \gls{cpcl} mode (like a synchronized passive radar) relative to other coexisting mobile radio networks that produce a valuable illuminating signal (could also include DVB-T broadcast). At the same time, the sniffer can be used to detect and localize alien emitters. This feature opens up other wide fields of \gls{isac} applications and can be an interesting add-on for some use cases. 

If the sniffer is integrated into a distributed base station architecture by using a fixed wireless, RFoF, or optical fronthaul link, it can be directly incorporated into the internal \gls{gnb} base-band processing regime and enjoy all advantages of \gls{cpcl} processing. However, if it is equipped with its own subscriber identity (SIM card), it can act as normal \gls{ue} and can be deployed more easily at preferred locations. Now it can interact with the hosting \gls{gnb} via the normal \gls{ue} \gls{ul}/\gls{dl} radio interface. Addionally, the \gls{cpcl} advantages in terms of carrier and clock reference synchronization still apply. 

Another aspect of heterogeneity on the level of network parameters is access to computational facilities at the network edge. While \gls{mec} in a cellular base station meanwhile is a standard feature, is poses questions in mobile ad hoc networks. In a meshed \gls{isac} network consisting of drones, for example, we are confronted with limited communication and payload capacities in terms of weight and energy supply. On the other hand, the ad-hoc nature of such an \gls{isac} swarm of radio nodes may allow for dynamic configuration of drones with different and specialized sensing capabilities. Also a master drone with advanced computational and long range communication capabilities for centralized data fusion can be added.

\section{Multipath Propagation}
\label{sec:propagation}
\subsection{Multipath Propagation Channel and Multipath Exploitation}
Understanding of multipath propagation issues is crucial to the performance of mobile radio and are critical to system design. 
Multipath is responsible for channel dispersion in the delay, Doppler, and angular domains. 
It leads to temporal variability, fast and slow fading. 
It has been considered the most limiting feature of system performance that needs to be mitigated by equalization, precoding and link adaptation in different domains. 
Despite a certain diversity gain advantage, multipath originally was seen as harmful for high data rates and reliable communication. 
The situation has dramatically changed with the advent of \gls{mimo} radio access, which has shown that multipath can improve communication and enhance capacity in complicated, multipath-rich environments. 
Also in radio localization, multipath may turn from a foe to a friend if correctly exploited~\cite{ref55}. Therefore, measurement and model based propagation studies in rich scattering environments are of outstanding importance also for \gls{isac} system design~\cite{ref3}. Although the topic would deserve more attention, we can give only a concise overview on the most relevant propagation effects to support understanding of basic \gls{msisac} design issues.        

\begin{figure}
    \centering
    \includegraphics[width=0.85\linewidth,trim={-30 -30 -30 -30},bgcolor=gray!10,rndcorners=5,rndframe={color=gray!50, width=\fboxrule, sep=\fboxsep}{5}]{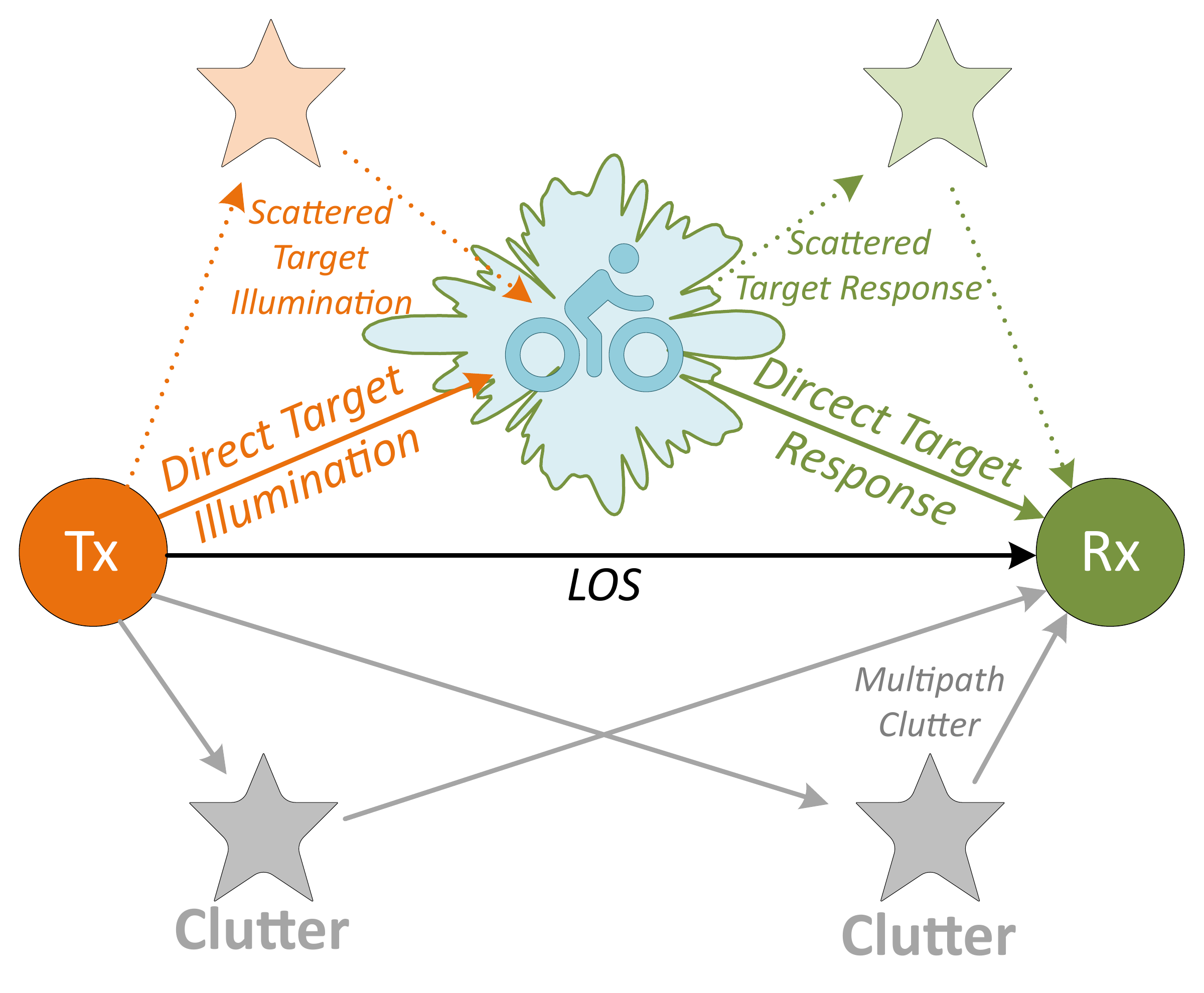}
    \caption{Multipath geometry of the bistatic \gls{isac} propagation channel.}
    \label{fig:multipath_geom}
\end{figure}

The consideration of the propagation channel for radar localization differs remarkably from the situation in communication. Generally speaking, in radar the desired information is contained in the geometric structure of the propagation, while in communication the information is encoded in the unknown transmitted symbols, which have to be received correctly. With \gls{isac}, where communication and sensing functions are integrated, we even have a mixture of both which casts different views on the influence of propagation. \Cref{fig:multipath_geom} depicts the geometric structure of a single bistatic multipath propagation channel. In communication, the total energy per symbol sent and received is used for symbol recognition, hence for maximizing mutual information. In \gls{isac} only those paths that are routed via the target can contribute to target detection and localization. However, the \gls{los} path between \gls{tx} and \gls{rx} (which is often misleadingly referred to simply as “interference”) is also important. In addition to its contribution to communication, it helps for Tx2Rx frame synchronization, submits a reference for differential \gls{etof} and eDoppler estimation in bistatic setups and correlation reference signal recovery, which is necessary for \gls{cpcl} target estimation as described \cref{sec:generic}. In contrast, all other propagation paths that are not routed via the target are useless for radar sensing, perhaps disturbing and therefore referred to as clutter. The role of the target channel, which consists of all multiple paths that are routed via the target, can be thought of as a relay channel as highlighted in \Cref{fig:multipath_geom}. 

The target is most clearly identified by the two \gls{los} paths (Tx-to-target and target-to-Rx in \Cref{fig:multipath_geom}), which we call the \glsreset{btp}\gls{btp}. However, the simple triangular \gls{los}-related model would be invalid if one or more of the three \gls{los} paths are blocked. While the blocked direct Tx2Rx \gls{los} must be given special consideration since it delivers the correlation reference, there are several possibilities to exploit target related multipath. In general, the associated multipath links that are also routed via the target additionally contribute to targets illumination and visibility and may reveal relevant geometric information for localization even if the respective \gls{los} is missing. Spatial precoding and dedicated estimation measures would be necessary to make use of this multipath interaction (“multipath exploitation”), may it be enhanced visibility through spatial multipath focusing (increased illumination power) on the target or enabling target visibility even when the target \gls{los} to \gls{rx} is obstructed. Here we can only give a very concise overview highlighting effects and indicating respective measures to mitigate or exploit multipath. 

The first step to dealing with multipath is to mitigate non-target related multipath, referred to as clutter in \Cref{fig:multipath_geom}. The problem is that clutter can be much stronger than the target reflection. Therefore clutter can mask the targets contribution by the with and sidelobes of its point spread functions. Even if we have somehow managed to resolve the multipath, we still need to make sure that the multipath belongs to a relevant target and not just some other object. \Cref{fig:multipath_geom} also underlines that multidimensional estimation is a powerful means to enhance resolution. For instance, if clutter cannot be resolved in (bistatic) range, directional estimation or filtering can help. Doppler can have a significant effect, as moving targets are revealed by a non-zero Doppler shift if the \gls{isac} nodes are stationary. Doppler can be explicitly measured or implicitly exploited by background subtraction. When the \gls{isac} nodes move, the situation becomes more complicated, as static clutter now also leads to a Doppler shift. However, since we know the position of the \gls{isac} nodes and perhaps also their dynamics (we often can estimate the direct \gls{los}), we can remove this additional Doppler shift or at least take it into account. 

Proactive exploitation of multipath goes even further. While precoding and associated resource allocation measures are shortly referred to in the next section, a detailed discussion about multipath exploiting procedures is beyond the scope of this paper. Roughly speaking, there seem to be two classes of multipath exploiting estimation methods. One is called “finger printing”. It is based on calculating some position related features from observed \gls{isac} data. In the most simple case, this can be received power which undergoes spatial selective fading. However, differential delays, Doppler or directions that can be better attributed to relative target position and movement can be also used. While explicit geometrical calculation of position from these parameters may not be possible in multipath rich environment, machine learning can be applied to identify targets position from measured fingerprints. In addition, reference information calculated from a ray tracing simulation of the scenario (a “digital twin”) can serve as a learning data base~\cite{ref56,ref57}.  

Another approach is to explicitly embeddi a multipath propagation data model based on ray tracing into the position estimation loop. Assuming we know the positions of the dominant interaction points of the environment, the associated reflections can be used as further (or auxiliary) spatially orthogonal measurements like additional illuminations generated by virtual anchor nodes. This inverse model-based estimator effectively traces back the observed specular data to the unknown target position~\cite{ref51}. The advantage relative to finger printing is less learning effort which is required (the explicit knowledge about the structure of the environment is directly applied) and we have more control about which interacting objects are included. This may be also important if “artificial interactions” generated by \glspl{ris} or by amplify-and-forward relays are used. 

Yet another approach that enables target localization with obstructed \gls{los} from \gls{tx} to or from the target to \gls{rx} (e.g. if the target gets lost when tracking) makes use of knowledge about the effect of obstruction and especially if we explicitly know the position of the obstructing object~\cite{ref58,ref59}. Even when the target itself is obscuring the Tx2Rx \gls{los} we can detected the presence of the target when it moves through the \gls{los} between \gls{tx} and \gls{rx} (forward scattering radar~\cite{ref60}).

The \gls{frf} corresponding to the scenario depicted in \cref{fig:multipath_geom} can be separated as 
\begin{align}\label{eq:multipath_geom}
    H_{m,n} = H_{\rm los} + H_{\rm sti} H_{\rm dtr} + H_{\rm dti} H_{\rm str} + H_{\rm dti} H_{\rm dtr} + H_{\rm clt},
\end{align}
where $H_{\rm los}$, $H_{\rm dti}$, $H_{\rm dtr}$, $H_{\rm sti}$, $H_{\rm str}$, and $H_{\rm clt}$ denote the \glspl{frf} for the \gls{los}, direct target illumination, direct target response, scattered target illumination, scattered target response, and clutter, respectively. Clutter summarizes all contributions that are not related to the target (except \gls{los}). Whether the scattered target illumination and response should be considered a clutter depends on whether it is exploited or not, i.e., if it constitutes unwanted interference.
\subsection{Multistatic Target Reflectivity and Diversity Gain}\label{sec:propagation:diversity}
The target's contribution to the end-to-end multipath routing from transmitter to receiver is described by its \gls{mrf}, which is the bistatic transfer function of all pairs of incoming and outgoing \gls{btp} waves including target-related \gls{los} and multipath. Therefore, the target deserves special attention in \gls{isac} propagation measurement and modeling. A complete \gls{mrf} description would be based on 2x2 full polarimetric bistatic angular scanning. The \gls{mrf} measurement range described in~\cite{ref3} uses wideband illumination in order to characterize a target’s inherent time variability (e.g. because of rotating wheels or propellers in case of cars or drones or moving arms and legs in case of walking persons), which will cause characteristic micro Doppler patterns~\cite{ref70}. Together with the bistatic spread in delay and angle, we obtain a full \gls{mrf} description that enables advanced propagation studies, recognition of extended targets and high performance ray optical modeling in multipath rich environments. Note that these characteristics are distance dependent as they are influenced by spherical wavefronts.

At greater distances and lower bandwidths (which does not allow to resolve the shape of the target), the advantage of distributed \gls{mimo} radar shifts to spatial diversity, which affects the probability of target detection. The use of spatial diversity for mitigation of fading is already well known from \gls{mimo} communication where it is related to the signal at the receiver antenna position. In distributed \gls{mimo} radar it is related the target. Since radar targets are usually larger than several wavelengths of the radio carrier (“electrically large targets”) and complex in terms of many structural details of small size, even cavities, the scattered field is composed of contributions from a great number of seemingly distinct scattering centers, with the resulting superposition being a complex circular symmetric Gaussian variable (iid in real and imaginary part). Therefore, the reflected power of the target (related to its \gls{rcs}) follows a $\chi^2$ distribution with two degrees of freedom. This applies as long as the multiple reflections make a similar contribution (in terms of power) to the resulting target backscattering. In case we have a dominating (specular) contribution, the distribution will change to a non-central chi-square distribution (a squared Rician distribution). Moreover, the reflected power maybe subject to a slow variability over time. These scintillations are responsible for signal fading, which has influence to target detection statistics. In radar technology, these statistical variations are combined in so called Swerling’s target models~\cite{ref32}. In mobile radio communications, the same phenomenon is known as small scale and large scale fading. 

Diversity gain increases the probability of detection in case of multiple observations~\cite{ref10c}. It depends on the decorrelation of the combined observations and can be characterized as cooperative diversity or macro-diversity. Therefore, distributed \gls{mimo} radar is often referred to as statistical \gls{mimo} radar to distinguish it from \gls{mimo} radar with coherent signal combining. The possible diversity gain depends of course on the number of independent observations, which in the best case can be up to $N^2$ (see \cref{sec:generic:msmimoisac}). A welcome side effect is that the claimed distributed spatial structure is also advantageous for estimating the 3D dynamic state vector which is described in \cref{sec:access}. The influence of the aspect angle is also evident when we consider special target shapes, such as those targets that attempt to conceal themselves by minimizing their monostatic \gls{rcs}, also known as stealth targets. At the same time, we have to emphasize a very specific spatial constellation of \gls{tx} and \gls{rx}, the so-called forward scattering case where \gls{tx} and \gls{rx} face each other from opposite directions and the target is in between. Surprisingly enough, this situation has many advantages for target detection as the received power may undergo a clear increase because of the creeping wave along the targets surface and the resulting lens effect. For higher frequencies (mmWave) and depending on the distance between target and antennas, also shadowing may occur. The latter case may become relevant for short distance sensing in car-to-car scenarios. 

While the combined $\chi^2$ distributions apply to multiple observations of electrically large targets, we need to distinguish the case of extended targets, which are resolved in the delay domain and are therefore represented by a small number of neighboring range bins. For a given target, this situation depends on the bandwidth of the \gls{isac} waveform. The more bandwidth we have, the more range bins we can resolve -- for the same target. Since the observed reflectivity response of the neighboring range bins is not independent because it originates from the same target seen from the same direction, we can try to maximize received power, respectively the detection probability, by matching the transmit signal to the target response (target matched filter).

\section{MS ISAC Access and Estimation}
\label{sec:access}
In \Cref{sec:generic}, we describe a very general architecture of \gls{ms} \gls{mimo} \gls{isac} systems.
According to the multipath structure of propagation described in \Cref{sec:propagation}, the challenge is to coordinate the radio access, allocate the available radio resources, adjust the radio link parameters, and predistort the transmitted waveform where appropriate to achieve a well-defined \gls{isac} quality of service.
This applies to both radar target localization performance and communication performance, although the two performance measures will not be independent of each other.
Although we aim to reuse the available resources (e.g., use the same waveform for communication and sensing) in order to maximize resource efficiency, the goals for performance optimization may be competitive.

Since we assume that the \gls{isac} system is primarily a mobile communication system, we do not discuss adopting and designing specific waveforms to be used for sensing in parallel to communication.
Instead, we are taking the approach of using the powerful and flexible link access coordination and adaptation schemes that are already included in 5G NR. 

Of course, the way and the criteria to adapt and coordinate may be somewhat different for \gls{isac}, but with 5G NR the ``toolbox'' to implement the variety of adaptation schemes is already there and will be further developed with 6G using powerful methods of \gls{ai} leveraged by high-performant computational instances distributed in the access network. 

The \gls{isac} sensing \glspl{kpi} are not yet well defined.
On the one hand, as with any radar, the fundamental \glspl{kpi} are related to localization accuracy and resolution in terms of the number of dynamic targets that can be resolved and tracked in a certain area.
More specifically, this also includes resolvable parameter dimensions such as range, Doppler and direction, target acquisition time, and coverage.
Many of the \gls{kpi} parameters are statistical in nature, such as detection rate, probability of false alarm or missing targets, probability of true tracks and false tracks.
Others refer to the detection of target types and shapes.
Still others refer to image quality or the ability to detect, analyze, and evaluate complex scenarios.
Due to the largely ubiquitous availability of wireless sensors and the wide variety of assignable radio resources on the radio and network level, \gls{isac} has the potential to be a highly customizable radar sensor system that can meet various performance requirements.
A comprehensive discussion of \gls{isac} \glspl{kpi} is beyond the scope of this paper.
For parameter variance and resolution assessment, the classical parameter is the \gls{crb}, which we also apply for waveform design.
Some recent works~\cite{ref19,ref50} suggest \gls{mi} as a common performance metric for the joint radar and communications system.
However, there is still a lack of effort in defining new metrics for \gls{mimo} radar and \gls{mumimo} communications co-design problems.
Joint optimization, which takes into account the balance between communication and sensing quality of service within the constraints of limited availability and economy of resources, appears to be uncharted territory—but is of great importance for resource scheduling at the network level.

\subsection{Multistatic Dynamic Target State Vector Estimation}

\begin{figure}
    \centering
    \includegraphics[width=\linewidth,bgcolor=gray!10,trim={20 150 10 10},rndcorners=5,rndframe={color=gray!50, width=\fboxrule, sep=\fboxsep}{5}]{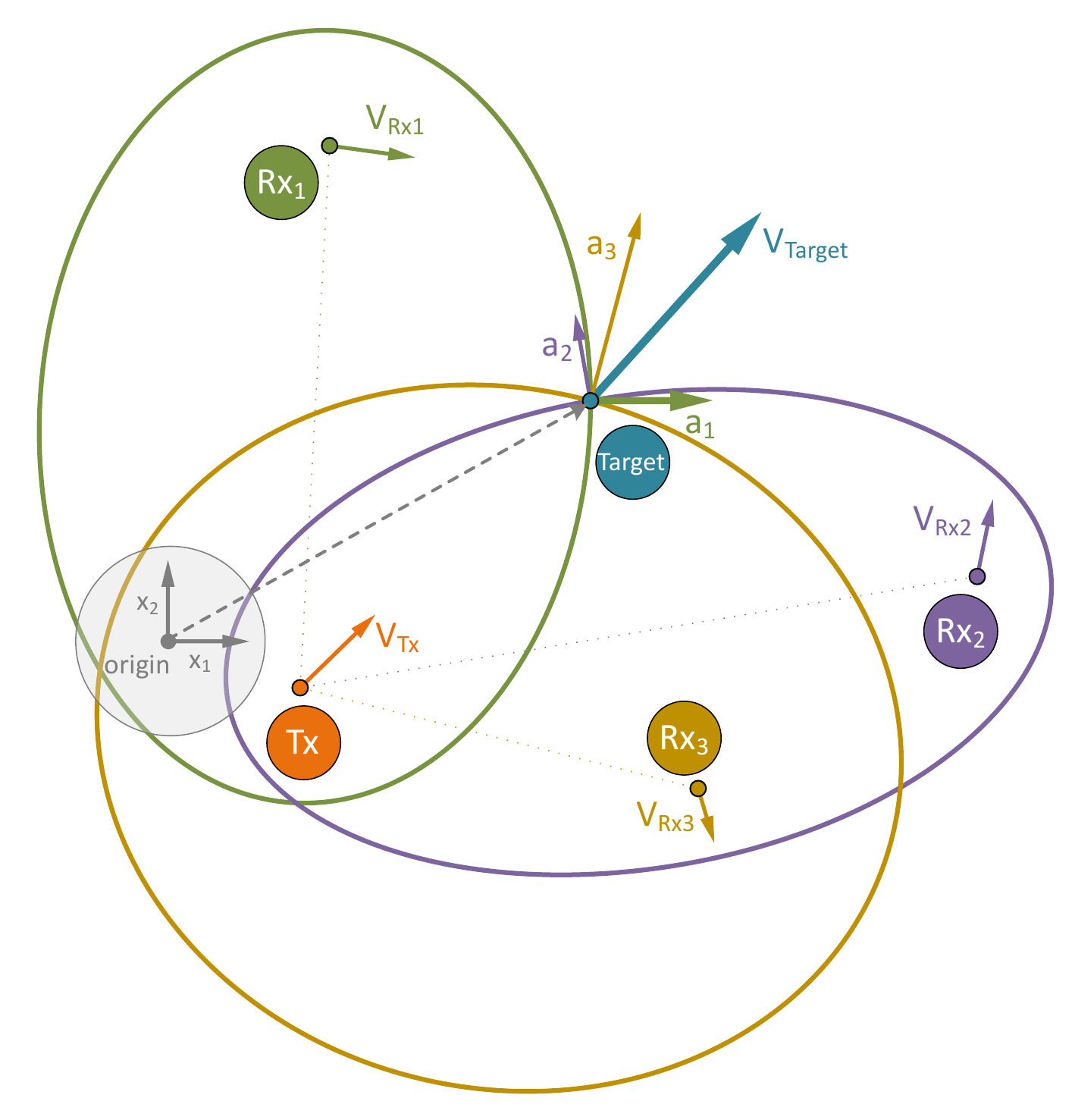}
    \caption{3D \gls{etof} and eDoppler ellipsoids and 3D target state vector. The target is characterized by its instantaneous position vector (here in 2D) and by its differential speed vector with its estimated components. The \gls{isac} sensor nodes are defined by the same two components. But for simplicity, only the speed vectors are depicted here. The positions are indicated just by dots.}
    \label{fig:ellipsoids}
\end{figure}

\gls{isac} aims to estimate the dynamic target state vector, which describes the position, orientation (if possible), and movement of the target in three dimensions.
Therefore, we usually need several measurements taken by spatially distributed \gls{isac} radio nodes to determine targets' coordinates and movements.
These nodes can closely cooperate at the intra-cluster level or more loosely at the inter-cluster level.
While in the first case a local, sensor-centric coordinate system is mostly applied, we need in the second case a sensor-independent global coordinate system since target coordinates are exchanged between different sensor clusters.
The sensor-centric coordinate system can be Cartesian or polar.
While for single station monostatic radars, where range and bearing are measured, the polar coordinate system relative to the radar`s position is the natural choice, the situation with \gls{msisac} is somewhat more complicated.
Due to the bistatic nature of radio access, the iso-\gls{etof} surfaces spanned by the Tx\,/\,Rx pairs forming the cluster of closely cooperating \gls{isac} nodes are equipotential ellipsoids, as shown in \Cref{fig:ellipsoids}.
The same figure also depicts the chosen global Cartesian coordinate system of the dynamic state vector, whose static components result from the crossing point of a measured \gls{etof} ellipsoids from several Tx\,/\,Rx pairs.
The dynamic state vector part of moving targets is characterized by their spatial velocity vector and (if available) by higher spatial derivatives, such as acceleration.
The components of the bistatic \gls{evv} target velocity vector (relative to the velocity of the moving pair of sensors) are identified from the measured excess Doppler shift and its projection to the respective dynamic sensor state vectors of the sensor nodes.
With the static part, we may want extended targets to be localized in relation to some single focal point.
However, the targets are revealed by the first approaching reflection from the target surface relative to the sensor nodes.
Multiple bistatic measurements can help to localize the target with respect to its center of gravity by taking advantage of the target-related diversity and proper averaging.
Moreover, the orientation and shape of the target, as well as its inherent movement (micro Doppler) may be of interest, but this would go beyond the scope of this article. 

From a set of $L$ independent \gls{etof} measurements with transmitter and receiver positions $\bm r_{\rm tx_i}$, $\bm r_{\rm rx_i}$, and corresponding \gls{etof} observations $\tau_i$ and speed of light $c$, we can infer that the position $\bm r_{\rm target}$ of a point target has to satisfy
\begin{equation}\label{eq:etof_ellipses}
    \left\Vert \bm r_{\rm tx_i} - \bm r_{\rm target} \right\Vert_2 
    + \left\Vert \bm r_{\rm rx_i} - \bm r_{\rm target} \right\Vert_2 
    = \tau_i c, \quad 1 \leqslant i \leqslant L
\end{equation}
which corresponds to a system of quadratic equations, which always has a unique solution as soon as $L \geqslant 3$ and if all three ellipses correspond to the same target (c.f. discussion about extended targets above).
Similarly, we can derive a quadratic set of equations between the observed excess Doppler (eDoppler) $\alpha_i$ at a pair of moving \gls{isac} nodes that observe a moving target with velocity $\bm v_{\rm target}$.
Here, we assume that both target as well as \gls{tx} and \gls{rx} move with locally constant velocity, which allows us to formulate the relationship between the eDoppler $\alpha_i$ and the desired velocity $\bm v_{\rm target}$ in Euclidean coordinates via

\begin{align}\label{eq:eDoppler_ellipses}
\alpha_i = 
    & + \ScPr{
        \frac{\bm v_{\rm tx_i}}{\Norm{\bm v_{\rm tx_i}}_2}
    }{
        \frac{\bm r_{\rm tx_i} - \bm r_{\rm target}}{\Norm{\bm r_{\rm tx_i} - \bm r_{\rm target}}_2}
    } \notag \\
    & - \ScPr{
        \frac{\bm v_{\rm target}}{\Norm{\bm v_{\rm target}}_2}
    }{
        \frac{\bm r_{\rm tx_i} - \bm r_{\rm target}}{\Norm{\bm r_{\rm tx_i} - \bm r_{\rm target}}_2}
    } \notag \\
    & - \ScPr{
        \frac{\bm v_{\rm target}}{\Norm{\bm v_{\rm target}}_2}
    }{
        \frac{\bm r_{\rm rx_i} - \bm r_{\rm target}}{\Norm{\bm r_{\rm rx_i} - \bm r_{\rm target}}_2}
    } \notag \\
    & + \ScPr{
        \frac{\bm v_{\rm rx_i}}{\Norm{\bm v_{\rm rx_i}}_2}
    }{
        \frac{\bm r_{\rm rx_i} - \bm r_{\rm target}}{\Norm{\bm r_{\rm rx_i} - \bm r_{\rm target}}_2}
    }.
\end{align}

Here, $\ScPr{\cdot}{\cdot}$ denotes the Euclidean standard 3D inner product.
However, this set of equations is only valid if the \gls{etof} $\tau_i$ and eDoppler $\alpha_i$ can be estimated without error, which in the presence of noise is impossible.
Hence, regularization or least-squares approaches need to be considered.

A visualization of the relationship between target position, target velocity, and the sensing nodes` positions and velocities is given in \Cref{fig:ellipsoids}.
For each link $i$ we depict the observed \gls{tof} $\tau_i$ as ellipses, whose foci are the sensing nodes` positions, and the target is a point on the respective ellipse.
Obviously, all the ellipses intersect at the target's location.
Further, at each \gls{isac} link, the target`s current velocity imposes an observed Doppler shift $\alpha_i$ that can be geometrically derived by orthogonally projecting the velocity $\bm v_{\rm target}$ onto the line that is perpendicular to the ellipse at the target`s position. 
Then after scaling by $\cos{\beta_i/2}$, where $\beta_i$ denotes the bistatic angle between \gls{tx}, target and \gls{rx}, we obtain $\alpha_i$.
Note that for correct calculations, all positions must be given in the same global coordinate system.
Further, the projected vectors $\alpha_i$ as depicted in \Cref{fig:ellipsoids} are drawn to scale for static \gls{isac} nodes. 
In the case of relative motion, the full extent of \eqref{eq:eDoppler_ellipses} has to be considered to calculate the $\alpha_i$.

\begin{figure}
    \centering
    \includegraphics[width=\linewidth,bgcolor=gray!10,rndcorners=5,rndframe={color=gray!50, width=\fboxrule, sep=\fboxsep}{5}]{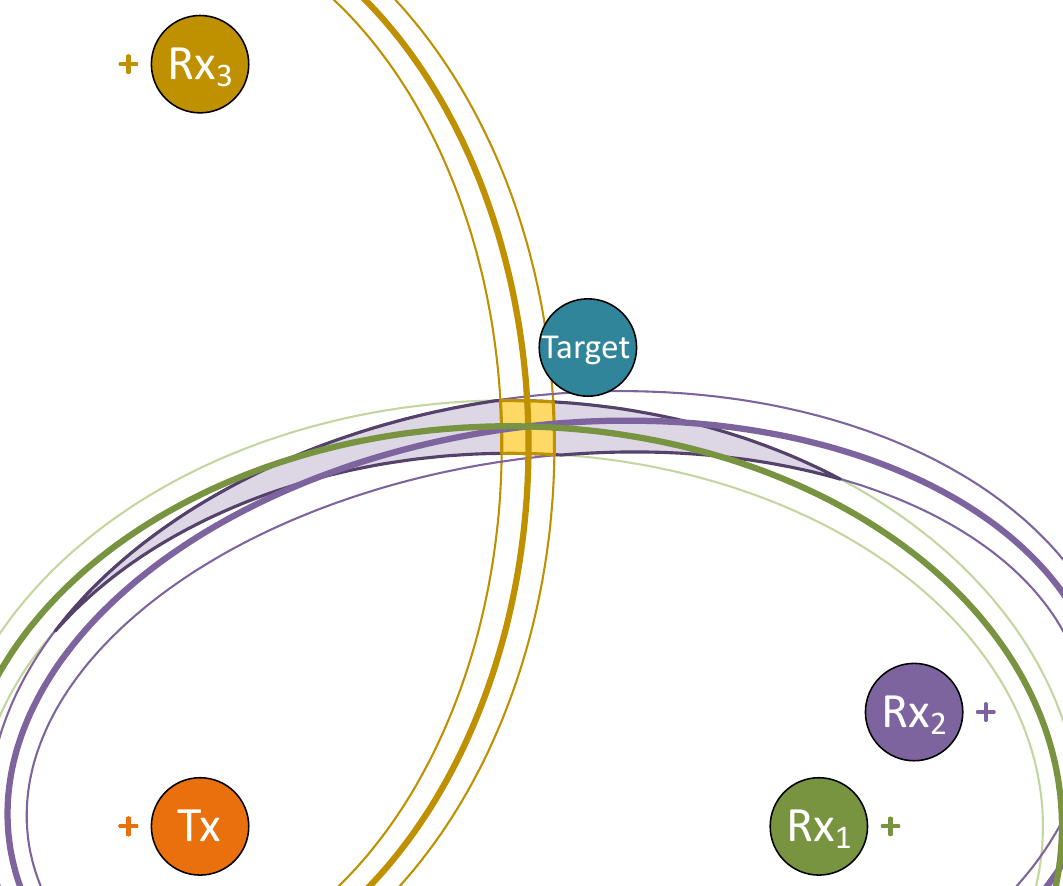}
    \caption{Two crossing ellipses with small and large \gls{gdop}. When the ellipses intersect at a right angle (Tx,Rx$\mathrm{_1}$~\&~Tx,Rx$\mathrm{_3}$), the position error is small (yellow); where they intersect at a flat angle (Tx,Rx$\mathrm{_1}$~\&~Tx,Rx$\mathrm{_2}$), it is large (violet).}
    \label{fig:crossingellipses}
\end{figure}

While a detailed consideration of the estimation of the target parameters from \eqref{eq:etof_ellipses} is beyond the scope of this paper, a general discussion seems worthwhile.
Since the equations are nonlinear, the global solution can be ambiguous.
This is especially true if single measurements are erroneous.
Using surplus measurements for averaging may be desired.
Then a linearization may be helpful, which leads to a local linear Taylor series approximation of \eqref{eq:etof_ellipses}.
This is most useful for differential estimation relative to former verified positions.
This paves the road to achieve a least squares position estimate by applying the Moore-Penrose pseudo inverse to an overdetermined set of linear equations gained from excess measurements.
This procedure also takes advantage of the target-related diversity gain as discussed above.
However, the choice of bistatic Tx\,/\,Rx sensor pairs should be made very carefully, which makes it an important issue of resource allocation.
Not only does the radio access of selected pairs of nodes have to be coordinated and adapted in order to achieve the desired accuracy and resolution of the bistatic measurements.
In addition to reducing the variance by averaging, an excess number of measurements can also be used to identify outliers that may occur, for example, due to an obstructed \gls{los}.
The sensor fusion of the measurement data must also be carried out.
The selection of \gls{isac} nodes also affects data transmission and computation in some fusion centers.
The assignment of the nodes depends on their suitability and relevance, which in turn depends on the geometric constellation of the node pairs in relation to the target and other interacting objects.
This has an influence not only on the \gls{snr} of the estimates.
There is also an error propagation mechanism known from satellite navigation as \gls{gdop} as briefly indicated in \Cref{fig:crossingellipses}.  

\begin{figure}
    % FIGURE 8
    \centering
    \includegraphics[width=\linewidth,trim={-30 -30 -30 -30},bgcolor=gray!10,rndcorners=5,rndframe={color=gray!50, width=\fboxrule, sep=\fboxsep}{5}]{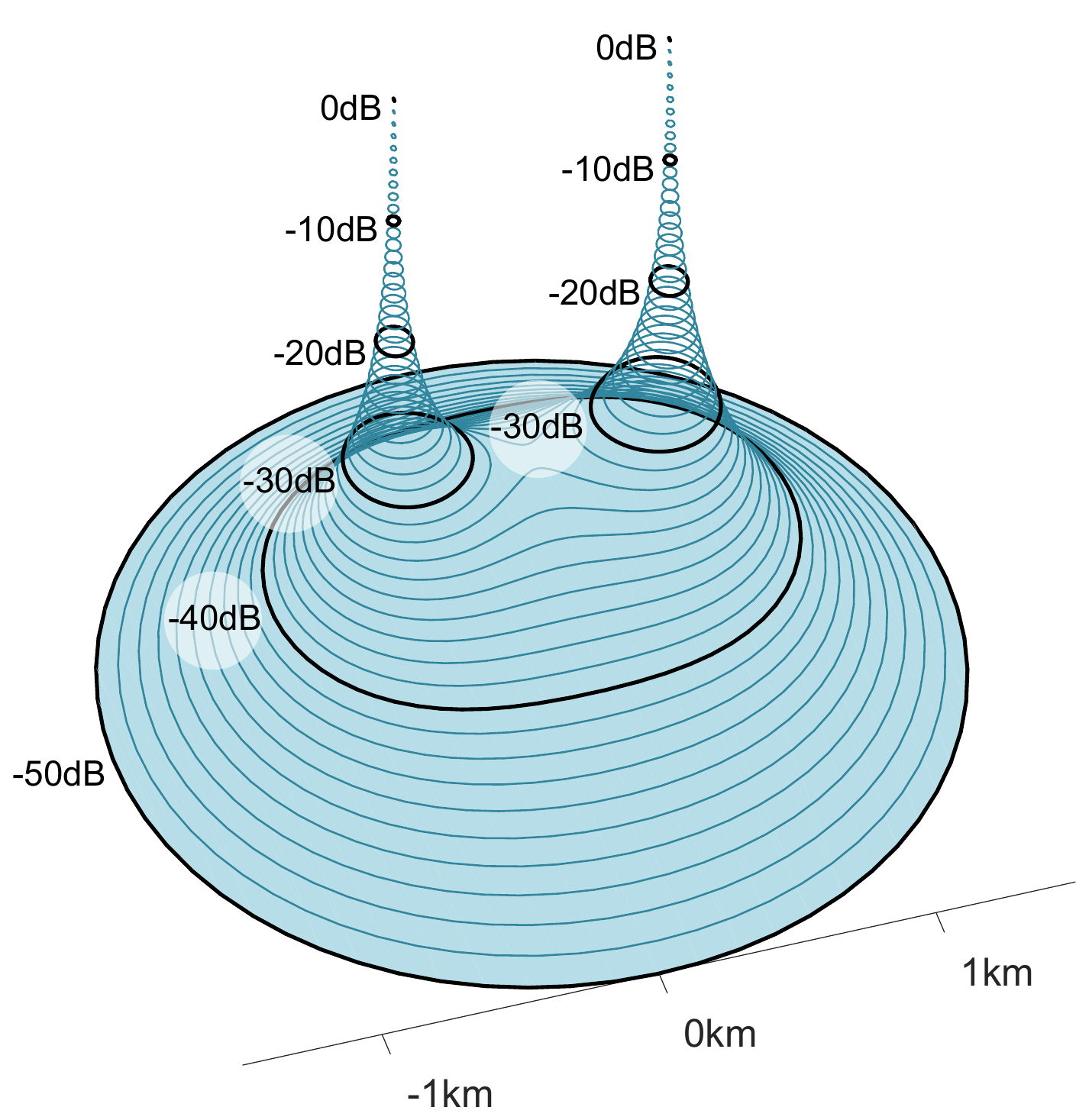}
    \caption{Bistatic radar coverage given by Cassini ovals.}
    \label{fig:cassiniovals}
\end{figure}

The underlying spatially distributed localization paradigm has inherent advantages over the single station range/bearing localization of monostatic \gls{isac} sensors.
Single station monostatic radars suffer from decreasing \gls{snr} and corresponding probability of detection by the fourth power of the radial distance according to the radar equation.
In addition, the cross-range resolution decreases with increasing distance from the target.
While increasing transmit power beyond the regulatory limit is not an appropriate approach for \gls{isac}, more degrees of freedom to balance the spatial distribution of performance arise with distributed \gls{isac}.
The spatial coverage of \gls{msisac} becomes evident from the Cassini ovals, which describe the coverage maps of bistatic radars.
The \SI{0}{\dB} level in \Cref{fig:cassiniovals} refers to direct transmission from Tx to Rx and thus to the Friis equation.
Therefore, the decibels describe the additional attenuation due to excessive range.
We deduce that the best \gls{snr} is achieved when the target is close to one \gls{isac} radio node (Tx or Rx).
It is moderate when the target is located somewhere between the two nodes, and it decreases steeply according to the radar equation when it is far away from both nodes.
Therefore, an advantageous sensor scenario with drones could be that the drone approaches the target while the illuminator is a powerful \gls{gnb} in its fixed position.

As explained in the previous sections, \gls{msisac} performance results from the cooperation of multiple distributed radio nodes, respectively, from the fusion of their measurements.
Therefore, the most obvious approach for enhancing and smoothing area-wide coverage and sensing performance is network densification, as we already know from mobile communications.
By choosing the relative position of the \gls{isac} nodes, we can achieve a smooth coverage and resolution performance by controlling the \gls{snr} and \gls{gdop} parameters.
We can even adjust the performance according to the local needs.
Heterogeneous nodes can help further.
Although multilateration can be seen as an advantage if we do not have a \gls{doa} estimate available, beamforming can be used to increase \gls{snr} and, hence, improve performance in certain situations.
Different frequency bands can be used, e.g., to achieve higher performance at target hot spots with nodes using millimeter wave frequencies or to guarantee minimum sensing performance in less critical situations by auxiliary umbrella nodes at lower frequencies. 

Last but not least, distributed \gls{isac} is better suited for 3D target localization as it has the potential to capture all spatial coordinates. 
For instance, although a single station range\,/\,bearing radar can act as an autonomous localization system, it is ``Doppler blind'' for cross-range moving targets. 

\subsection{\gls{ms} Radio Access and Scheduling}

Obviously, \gls{ms}-\gls{mimo} radio access needs coordination and scheduling.
The multiple radio links in \Cref{fig:generic_mimo_arch} share the same radio environment.
Therefore, we need orthogonal medium access schemes to make sure that the multiple links do not create mutual interference.
The links should permit multiple (quasi) simultaneous measurements because of the dynamic 3D geometric nature of the problem.
These measurements undergo bandwidth and real-time constraints and require adequate temporary coordination because of the dynamic evolution of the whole scenario.
Requirements on range-Doppler resolution and coherent radar integration time place demands on bandwidth, synchronization, and sustained recording time.
Eventually, multi-antenna access focused in the beam domain and matched in polarization at \gls{tx} and/or \gls{rx} side would be necessary.
With the limited resources of the shared medium, we are therefore subject to fundamental performance limits that would indicate how many dynamic targets in a given environment we can reliably detect, resolve, and track.
Finally, yet importantly, in addition to the sensing performance of the physical radio interface, we may have to find a performance balance of \gls{isac} with communication services about radio and other network resources, which also includes demands for sensing data transport and fusion.

\begin{figure}
    \centering
    \includegraphics[width=.75\linewidth,trim={-70 -50 -70 -50},bgcolor=gray!10,rndcorners=5,rndframe={color=gray!50, width=\fboxrule, sep=\fboxsep}{5}]{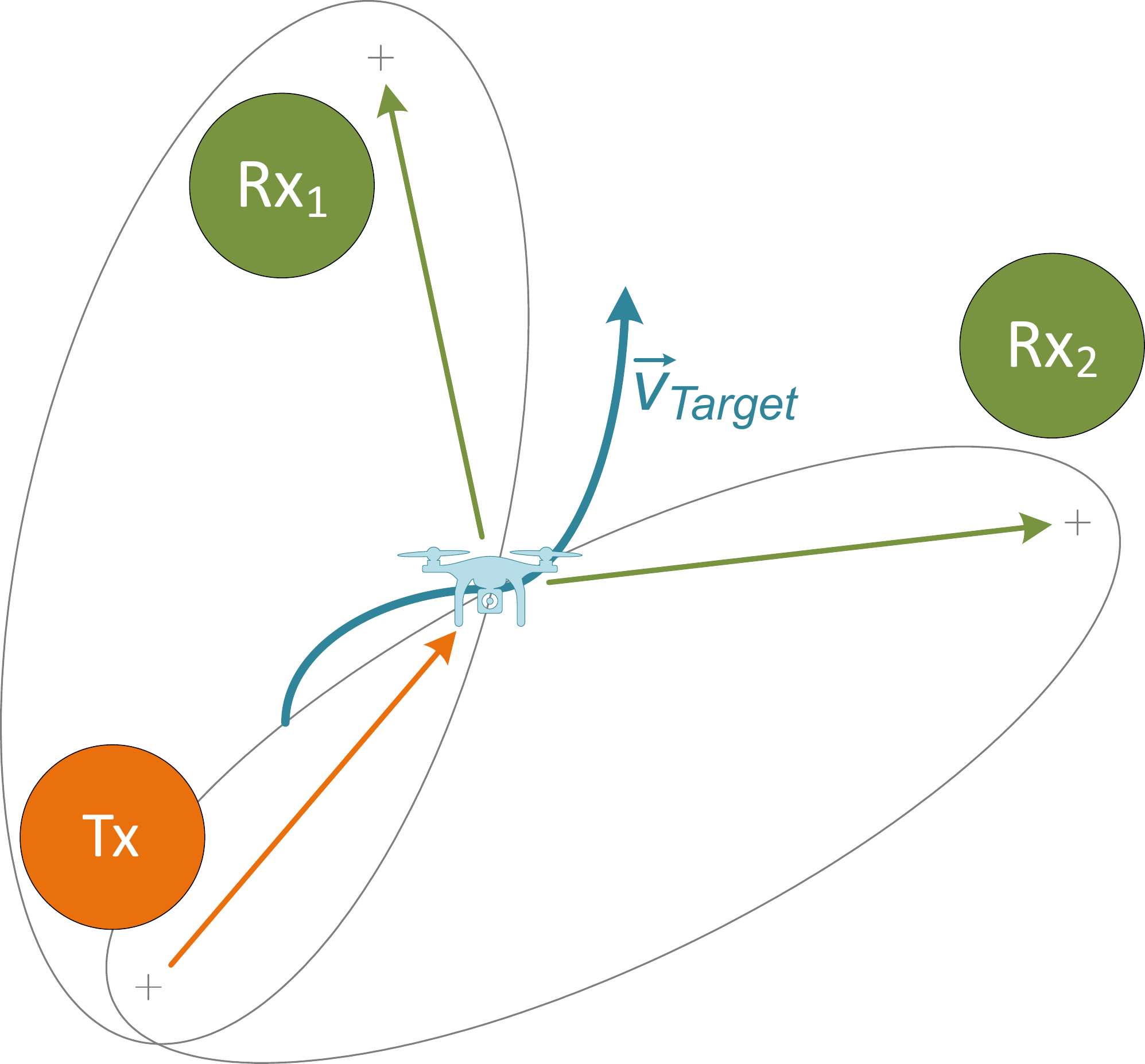}
    \caption{Dual link \gls{isac} in broadcast mode. We obtain two simultaneous bistatic measurements corresponding to the two ellipses.}
    \label{fig:dual_link_broadcast}
\end{figure}

While some basic aspects of radio access have already been discussed in the previous chapter on architectural issues, it becomes clear that we need to differentiate \gls{ms}-\gls{mimo} access into \gls{ms} broadcast and orthogonal \gls{ms} access. \gls{ms} broadcast access (preferably in \gls{dl}), where only one radio node in \Cref{fig:generic_mimo_arch} is transmitting and the others are listening, provides several measurements at the same time, which is very advantageous for dynamic targets although it does not cover all possible sensing links (only $N-1$ out of $N^2$), see \Cref{fig:dual_link_broadcast}.
The orthogonal \gls{ms} mode turns the broadcast \gls{ms} channel into a shared sensing channel.
We have to rely on the nested distribution of \glspl{re}, also referred to as \glspl{rb} in the time-frequency plane (\gls{ofdma} in \gls{dl} or SC-\gls{fdma} in \gls{ul}). 
The orthogonal \gls{ms} mode turns the broadcast \gls{ms} channel into a shared sensing channel.

Another problem is related to the \gls{tdd} scheme in 5G NR.
Dynamic \gls{tdd} \gls{dl}\,/\,\gls{ul}-split is aimed at supporting assignments based on instantaneous traffic demands. 
Consequently, there is a gap in one \gls{tdd} stream when the other is active, and normally the duty cycle of the \gls{dl} is greater than that of the \gls{ul}.
This raises two questions. 
The first is which influence do the \gls{re} distribution and the \gls{dl}\,/\,\gls{ul} balance have on the dynamic target estimation performance, and what is the optimum allocation of \gls{re} and \gls{tdd} gaps? 
The second results in an estimation problem from sparse data. 
Sparse allocation of \glspl{re} and \gls{tdd} gaps renders simple 2D-\gls{fft} target estimators useless since replacing the missing samples in the frequency domain just with zeros would distort the desired shape of the \gls{af} of the correlation reference waveform by disturbing sidelobes.

The flexibility of 5G NR already allows for great waveform adaptability and optimum coordination of \gls{isac} radio access. In addition to the physical \gls{ofdm} parameters defined by the numerology, it is about the time-frequency distribution of the \glspl{re} and \gls{tdd} gaps in the radio frame, which is crucial for resource allocation. This enables the 5G communication system to accommodate a variety of applications with different bandwidth, latency, and reliability needs. We will observe a similar influence of time-frequency resource distribution for radar sensing quality control. \gls{isac} radar sensing quality parameters are related to sensitivity of target detection, target resolution in range and Doppler, robustness against misdetection in clutter, target classification, and tracking performance in dynamic scenarios 

\begin{figure}
    % FIGURE 11
    \centering
    \includegraphics[width=\linewidth,bgcolor=gray!10,rndcorners=5,rndframe={color=gray!50, width=\fboxrule, sep=\fboxsep}{5}]{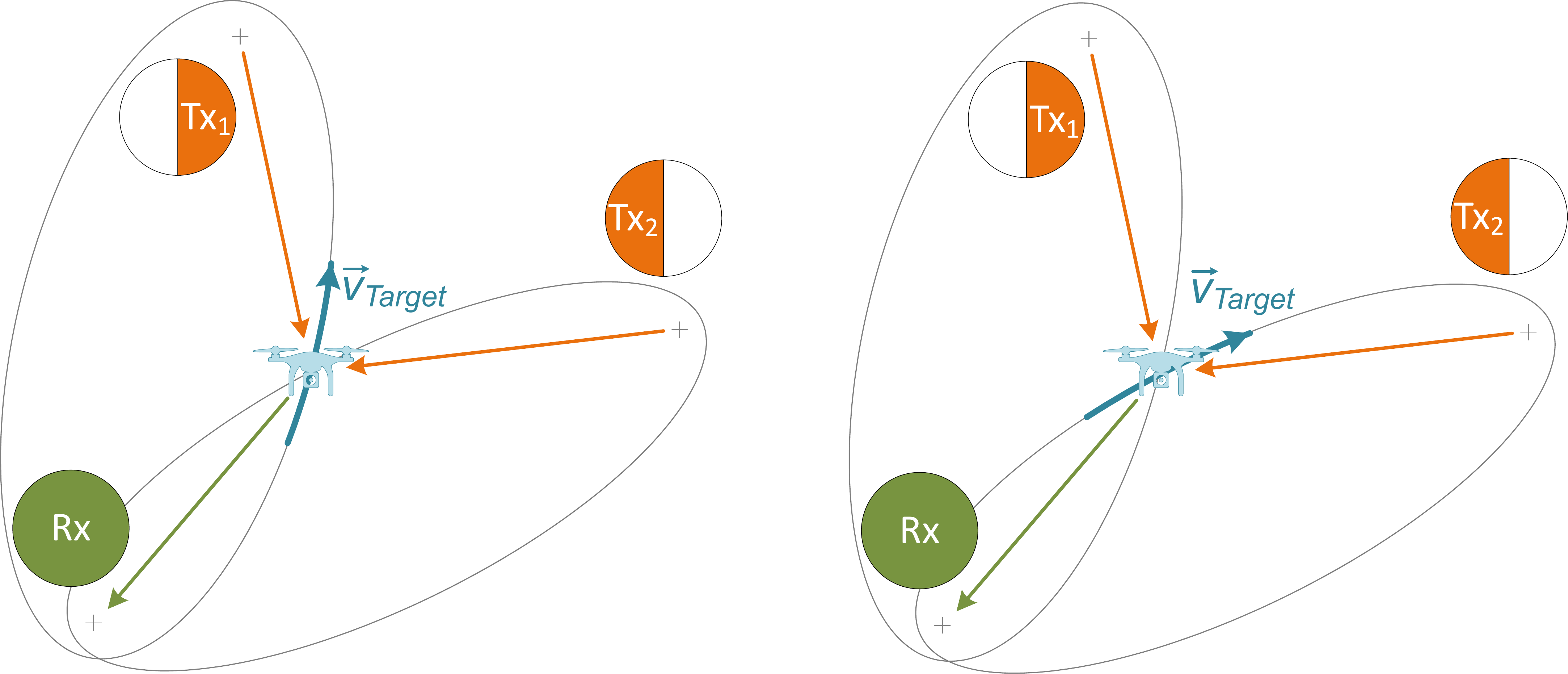}
    \caption{Orthogonal \gls{ms} access with Tx1 and Tx2 sharing the same \gls{ofdma} frame. In both cases, the position of the target is the same, but the direction of its movement is different.}
    \label{fig:orthogonal_access}
\end{figure}

\begin{figure}
    % FIGURE 12
    \includegraphics[width=\linewidth,trim={0 -10 0 -10},bgcolor=gray!10,rndcorners=5,rndframe={color=gray!50, width=\fboxrule, sep=\fboxsep}{5}]{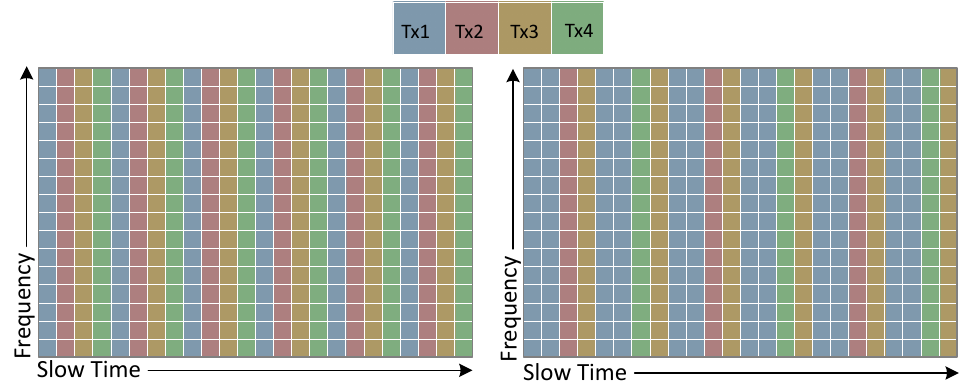}
    \caption{\gls{tdma} allocation of resource blocks. Left: uniform \gls{re} allocation for equal sampling rate in slow time. Right: non-uniform \gls{re} allocation for different slow time sampling regimes.}
    \label{fig:tdma_res_alloc}
\end{figure}

Here we will discuss some basic influential factors of \gls{ms} access schemes and \gls{ms} \gls{re} distribution. \Cref{fig:orthogonal_access} depicts the same three-node scenario as in \Cref{fig:dual_link_broadcast}, however with two transmitters and one receiver. The two transmitters share the same \gls{ofdma} frame. We will recognize that sharing these resources also means a certain division of resolution performance for range and/or Doppler. The optimum sharing of \gls{ofdma} resources depends on the geometry of the sensor cluster relative to the target. In the left case, the direction of the target's velocity vector is tangential to the Tx1\,/\,Rx ellipse, while the right case aligns with the Tx2\,/\,Rx ellipse.  Provided the radio nodes are stationary, the observed Doppler shift would be zero for the Tx1\,/\,Rx link in the left case and zero for the Tx2\,/\,Rx in the right case. At the same time, in the corresponding other cases, the observed Doppler shift would be higher, although not necessarily maximum. The maximum Doppler shift would be observed when the velocity vector is perpendicular to the ellipses. This situation can be exploited at the transmitters for resource block allocation. \Cref{fig:tdma_res_alloc} depicts two examples of slow time sampling by \gls{ofdma} resource blocks with four transmitters. The left case describes uniform, repetitive sampling at the same rate for all four Tx. We see that shared \gls{tdma} access to the medium reduces the sampling rate in slow time and thus the maximum Doppler shift that can be estimated unambiguously. The included timing offset can be compensated for as long as the Nyquist sampling rate is maintained. The right part of \Cref{fig:tdma_res_alloc} explains how the \gls{tdma} scheme can be adjusted to the target dynamics. While Tx3 illuminates the target still with the same sampling rate in slow time, the unambiguously resolved Doppler bandwidth has reduced for Tx2 and Tx4. Tx1 can be used in two ways. If the two neighboring \gls{ofdm} symbols are averaged, the effective overall Doppler sampling rate is kept, the Doppler bandwidth is reduced only a bit, and the sensitivity of target detection is increased. If the consecutive pairs of \gls{ofdm} symbols undergo the \gls{fft}, the \gls{snr} gain by coherent averaging still applies, but we will also have some possibility to increase the resolvable Doppler frequency. The sensitivity of target detection can also be influenced by the radar integration time, i.e., the length of the slow time observation window, which can extend over several coherently processed \gls{ofdm} frames. The limit is given by range migration, if the target moves from one range cell to the next one (in the range-Doppler domain).      

\begin{figure}
    % FIGURE 12
    \includegraphics[width=\linewidth,trim={0 -10 0 -10},bgcolor=gray!10,rndcorners=5,rndframe={color=gray!50, width=\fboxrule, sep=\fboxsep}{5}]{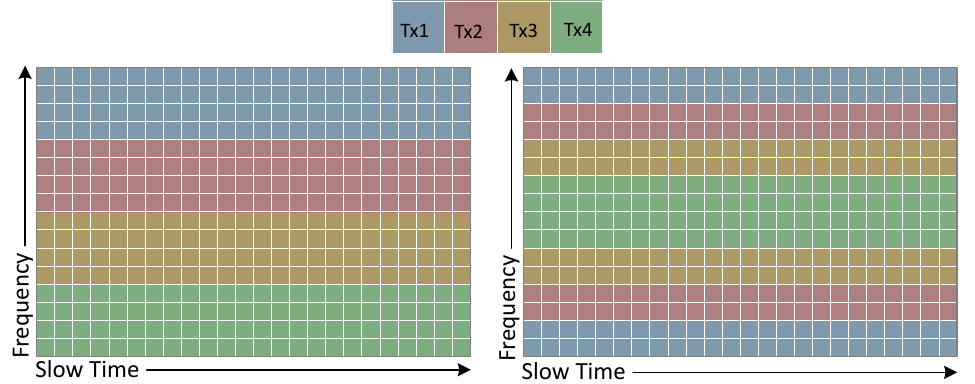}
    \caption{\gls{fdma} allocation of resource blocks. Left: Compact bandwidth allocation, uniform split in frequency. Right: Same bandwidth, but fragmented allocation in frequency.}
    \label{fig:fdma_res_alloc}
\end{figure}

\Cref{fig:fdma_res_alloc} explains the \gls{fdma} split within the \gls{ofdm} frame. The left part shows a uniform bandwidth allocation with compact bands that enable uniform range resolution for all four sensor links. In the right part, the same total bandwidth is allocated per transmitter, but the allocation is not consecutive, except for Tx4. A closer look shows that the range resolution capability for Tx1 is highest. However, because of the fragmented (”sparse”) frequency allocation, straightforward \gls{fft} estimation would not produce viable estimates because of grating lobes in the autocorrelation function. We would need a more sophisticated estimation procedure than just \gls{fft} that can cope with the sparse allocation of frequency resources as described in more detail in the next subsection.    

Some further comments are necessary. The time-frequency allocation described here as an example does not consider the time-frequency selectivity of the sensing channel. Especially the frequency selectivity because of multipath can require a more specific allocation of \glspl{re}. Moreover, adaptive \gls{re} scheduling for joint optimization of sensing and communication quality of service would be necessary. Another interesting observation comes from the fact that for localization, we would need multiple measurements, e.g., at least 3 Tx\,/\,Rx pairs for 3D location and speed estimation. This means that the optimal distribution of resource blocks in the \gls{ofdm} frame depends on the sensor geometry relative to the target and is mutually dependent. Moreover, for multiple targets, it depends on the difference of their dynamic target state vectors in its multiple dimensions. Therefore, the optimal allocation of resources in the \gls{ofdm} frame leads to a joint optimization problem depending on the positions of all sensors used and the dynamic behavior of all relevant targets, given the available resources.   

\begin{figure}
    \centering
    \includegraphics[width=\linewidth,trim={-30 -30 -30 -30},bgcolor=gray!10,rndcorners=5,rndframe={color=gray!50, width=\fboxrule, sep=\fboxsep}{5}]{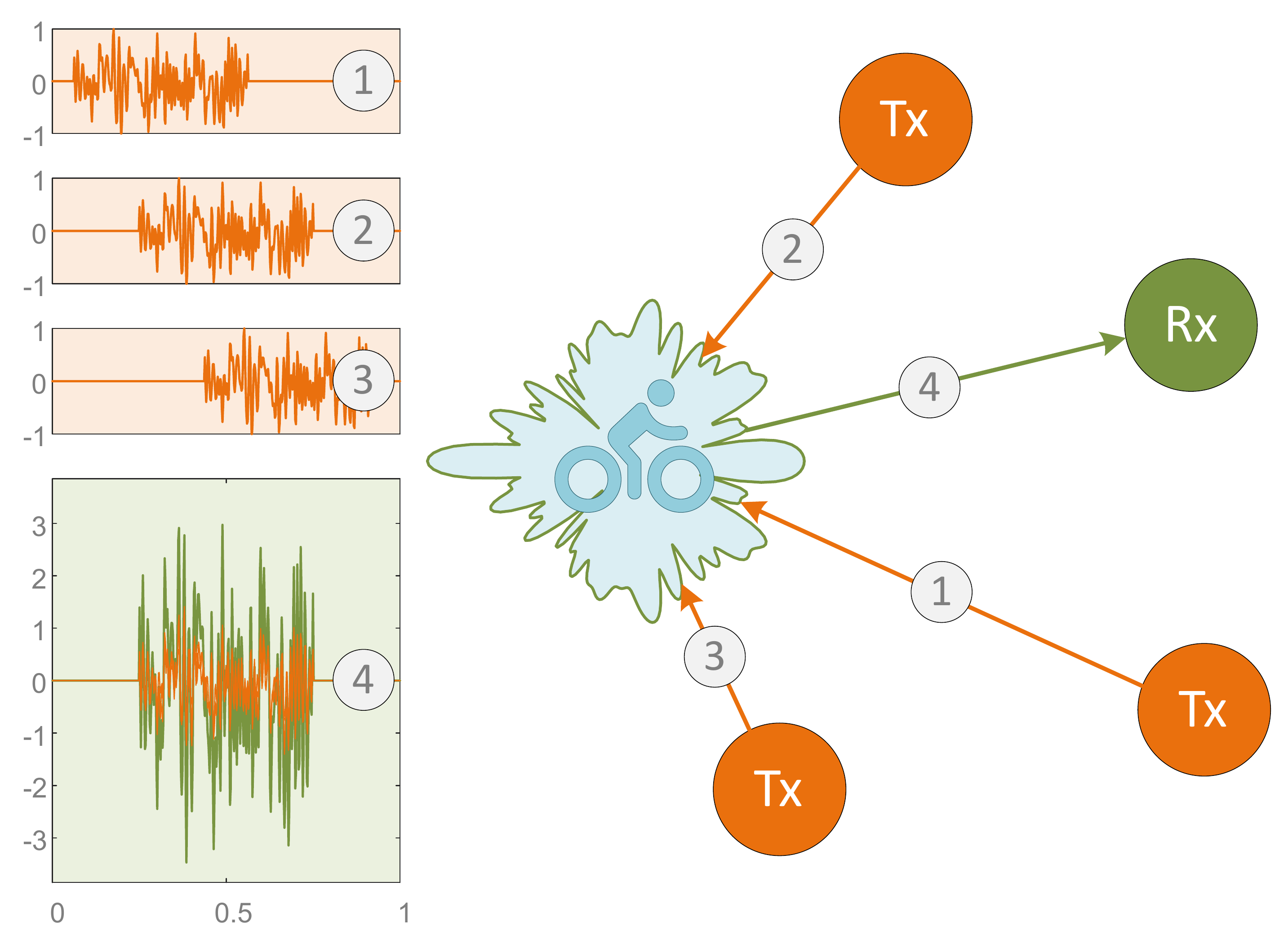}
    \caption{Joint transmission of multiple identical \gls{ofdm} waveforms (orange) towards the target. The delays are predistorted for coincidence in delay at the target. However, due to random backscattering from the surface of the extended target, the received response (green) appears as a non-coherent superposition. }
    \label{fig:jointtransmission}
      
\end{figure}

The discussion of \gls{ms}-\gls{mimo} access in \gls{isac} and any comparison with \gls{mumimo} communication would be incomplete without considering \gls{jt}, where the same illumination signal is transmitted by multiple distributed transmitters. Already throughout research towards 4G, \gls{comp} access, including \gls{jt}, was discussed, especially for enhanced quality of service at cell edges. Although successfully demonstrated~\cite{ref33}, it seems that \gls{comp} has not been widely used up to now. However, the situation may develop differently for \gls{isac} because of its inherent multilink requirements as explained above. \gls{jt} for \gls{ms}-\gls{mimo} \gls{isac} definitely deserves attention in research. There seem to be remarkable differences to \gls{jt} in communications. We have to make sure that focusing only involves the target and not the clutter. This could be achieved by background subtraction, e.g., if the target is moving and the clutter is static. However, due to the statistical reflectivity of typical extended targets, which leads to decorrelated radar returns as described above, coherent focusing to certain reflection points would only be possible for single Tx\,/\,Rx pairs. Therefore, although coherent \gls{jt} promises tremendous spatial resolution that could resolve details from extended targets down to single reflection points, non-coherent focusing seems to be a more reasonable goal. This would lead to increased reflected energy off the target without increasing spatial resolution. At the same time, we get a diversity gain because of the different directions of the waves impinging on the target~\cite{ref34}. Thus, non-coherent \gls{jt} \gls{isac} seems to keep the statistical advantage of distributed \gls{mimo} radar, but the geometrical advantage may get lost since the multistatic Tx\,/\,Rx geometry cannot be traced back if the non-orthogonal waveforms cannot be decomposed at the receiver. However, since the \gls{csi} pilots are orthogonal, simultaneous estimation of the joint target channels is possible, which would allow adjusting the transmit signal delay in a subsequent step for non-coherent superposition as indicated in \Cref{fig:jointtransmission}.

\subsection{Recovery of the Correlation Reference Signal}
Excess time of arrival estimation, also jointly together with Doppler, can be based on specific pilot signals or on using the data payload as correlation reference to compare \gls{los} path and \gls{btp} time of flight. Although pilots and other reference signals are designed for estimation of \gls{csi} in communications or \gls{ue} positioning, they may not be enough for 
%high dynamic range 
target state vector estimation in high dynamic range radar sensing. One reason for this is their sparse structure, which is sufficient for equalization, but interpolation errors in the frequency domain and the related distortion of the \gls{af} may not be tolerated for radar sensing. Moreover, since \gls{csi} pilots use only a fraction of the transmit signal power, the achievable correlation gain is limited. On the other hand, asking for more pilot signals is counterproductive, since this would reduce communication efficiency too much. Even extensive use of \gls{prs} would have the same effect. In contrast to this, \gls{cpcl} as proposed in~\cite{ref1} and introduced already here in \Cref{sec:architectures}, relies on the full data payload for precise target estimation and tracking. Note that radar localization has a systematic disadvantage compared to 3GPP \gls{ue} device positioning since it relies on passive backscattering. This means that the sensing link budget follows the radar equation and not the Friis equation. Furthermore, the target cannot be identified by some user ID. Instead, we need to use much more sophisticated means to detect, recognize, and separate multiple targets. Using the data payload as correlation reference offers not only the possibility to use the total transmitted power (maximizing the \gls{snr}) and system bandwidth (maximizing resolution). It also allows the correlation reference waveform to be adjusted and modified, enabling unprecedented approaches to adaptive radar functions. However, the transmitted payload is not known a priori at the receiver (unlike the pilots) and has to be estimated at the receiver before performing correlation. While conventional passive radar (PCL) uses a dedicated, directional receiver channel to obtain a precise, multipath-free copy of the transmitted signal via \gls{los}, \gls{isac}'s own communication function offers many more options for recovering the transmitted signal without a separate Rx channel. However, we have to be careful in defining what we need as a reference and which recovery approach we apply. We will discuss different approaches and how they build upon the standard \gls{ofdm} transmission approach.

In addition to Tx\,/\,Rx frame and carrier synchronization and cyclic prefix removal, an \gls{ofdm} receiver performs channel equalization in the frequency domain, which provides a symbol-by-symbol estimate of the transmitted \gls{ofdm} signal based on multiplication by the inverse channel frequency response estimated at the \gls{csi} pilot positions and then interpolated to be applied to all carriers. This procedure is followed by decision making out of a finite modulation alphabet and by error correcting channel coding. As a result, the transmitted information can be decoded at the receiver with very low bit error probability. Therefore, we might come up with the idea to use that information for recovering the transmitted waveform for \gls{isac} correlation by remodulation. However, the requirements for reliable and high dynamic range radar target detection and estimation are a bit different from communications. While in communication, we detect coded information per carrier; in radar, we perform a matched filter correlation using the whole \gls{ofdm} symbol to identify the target range. The \gls{ofdm} symbol is treated as an extended spread spectrum radar pulse that is compressed by correlation, which results in a rather high correlation gain. Moreover, we take further advantage of the coherent processing of many subsequent \gls{ofdm} symbols. The result should clearly indicate the target-related propagation parameters delay and Doppler as described in the next subsection. However, the target may be associated only with a very small part of the received energy, perhaps in the neighborhood of strong clutter peaks and hidden by their correlation sidelobes. The dynamic range is therefore generally higher for \gls{isac} than for communication. High detection probability and good resolution of target parameters in a continuous parameter space depend more on precise knowledge of the illumination waveform and its coherence over time than on its information content in terms of bits and bytes. However, RF impairments at the transmitter can affect the waveform and, hence, the shape of its correlation function. For instance, since with \gls{ofdm} the power amplifier is often operated close to saturation, non-linear distortions can occur~\cite{ref37}. Also, linear distortions (complex frequency response, IQ imbalance, etc.) of the modulator, transmit filters, and antennas can prove to be disruptive. For communications, those effects are normally leveled out by the inherent error correction mechanisms. This may not be true for radar, as the detection problem is different. Targets often appear as small peaks in environments with strong clutter. They only become apparent when overlapping system responses can be resolved through iterative subtraction using a model-based estimator, as described in the next subsection. Therefore, precise knowledge of the shape of the transmitted waveform, as well as its correlation function (including the sidelobes), is of the utmost importance. This includes all distortions of the RF chain, including nonlinear ones that appear after digital reconstruction. Thus, it is preferable to rely entirely on the recovery of the correlation reference received over the air for target parameter estimation. 

Inverse filtering is the most important step in correlation reference recovery because it equalizes the frequency response, removes associated multipath, and provides a clean copy of the transmitted signal. However, the usual inverse filter operation in the frequency domain has serious pitfalls. One relates to \glspl{re} (resp. carriers), which are loaded with zero power. First of all, these empty \glspl{re} should not be included in the inverse filter operation. In the fully connected \gls{cpcl} mode, the information about empty \glspl{re} can be directly taken from the \gls{dmrs}. In the more general case of the reduced \gls{cpcl} mode, empty \glspl{re} in the received reference signal are detected at the receiver based on a null hypothesis test~\cite{ref38}. Another related issue when recovering the transmit signal at the receiver side is the non-uniform (instantaneous) carrier magnitude. A simple inverse filter equalizes all received carriers to unit gain of the channel transfer function. This would lead to a non-uniform noise distribution of the carriers used for target parameter estimation. Since the \gls{csi} pilots are usually provided with more power and more robust modulation than the data carriers, these estimates would have a lower variance. On the other hand, multilevel \gls{qam} of the carriers would lead to noise amplification at the carrier positions that are loaded according to the modulation alphabet by a lower magnitude of the constellation diagram. These effects have to be taken into account for optimum target parameter estimation as discussed in the next subsection. 

The quality of the reconstructed transmit signal can be further enhanced by a two-step iterative procedure. In a first step, we perform regular equalization of the received \glspl{re} based on \gls{csi} pilots. The result is used for a first inverse filter step as described above. While these estimated payload data are processed for communication data reception, we use them to perform another channel equalization that should now be more accurate than the regular equalization since it uses the full transmit power and should be less affected by interpolation errors, which can arise because of the sparse structure of \gls{csi} pilots. With the now enhanced reference signal, we can calculate the final channel impulse response, which is used for radar target detection and estimation. The advantage of this approach compared to the alternative based on transmit symbol decision, decoding, and remodulation at the receiver is as follows. All linear and non-linear distortions at the transmitter are now included in the correlation reference, which assures a high dynamic range estimate. Even the time reference is kept, and Tx2Rx synchronization is automatically established since the critical \gls{etof} and eDoppler are effectively calculated as differences between \gls{los} and \gls{btp} from the same measured \gls{cir} (see discussion on synchronization issues later on). We refer to this approach as Turbo \gls{cpcl} because it is a nested iterative approach that takes advantage of the processing gain in the inner estimation loop using the \gls{csi} pilots. In terms of estimation accuracy, we can also understand the outer loop as a refinement of the first estimate. 

Another approach to filter out the remaining multipath is beamforming, which is discussed in the context of spatial precoding later on.  
\subsection{Model-based Bistatic Target Parameter Estimation}
This subsection deals with the estimation of target-related parameters from observed \gls{isac} \gls{ofdm} frames. As described above, the 3D dynamic target state vector may be estimated from a set of concurrent path parameters in delay and Doppler. This is not a restriction on the generality, as we can easily add beamforming and even directional estimation, as will be discussed later on. The aim here is to estimate the statistical delay-Doppler distribution, which is also known as the 2D point spread function of linear time-variant systems. In Bello’s classical terminology~\cite{ref63}, it is called \gls{scf}. This is the magnitude squared of the \gls{sf} $S(\tau,\alpha)$ that results from the inverse 2D Fourier transform of the time-variant transfer function $H(f,t)$. 

The performance of a range-Doppler radar is related to the \gls{af} $\chi$ of the transmit signal $x(t)$ as denoted by
\begin{equation}\label{eq:ambiguity_function}
    \chi(\tau,\alpha) 
        = \int\limits_{-\infty}^{+\infty}
            x(t)x^\ast(t - \tau) 
            \exp(\jmath 2 \pi \alpha t)
        \mathrm{d} t \, .
\end{equation}
The \gls{af} describes the resolution capability of the excitation waveform associated with its impulse compression performance and correlation gain. While the ideal \gls{af} according to \eqref{eq:ambiguity_function} is a 2D Dirac delta function ("thumbtack"), for uniformly occupied \gls{ofdm} frequency-time frames it appears as a 2D $\Sinc$ function because of its limited bandwidth $B$ and frame duration $T$. Although the side lobes of this \gls{af} are often considered as disturbing, the main lobe is the narrowest, and the correlation gain is the highest for a given area of support $BT$,  in the frequency-time domain. The resulting detector may therefore be seen as optimum in terms of maximum likelihood performance in case one path is to be estimated, simple \gls{fft} processing, and constant noise energy distribution in the delay domain. 

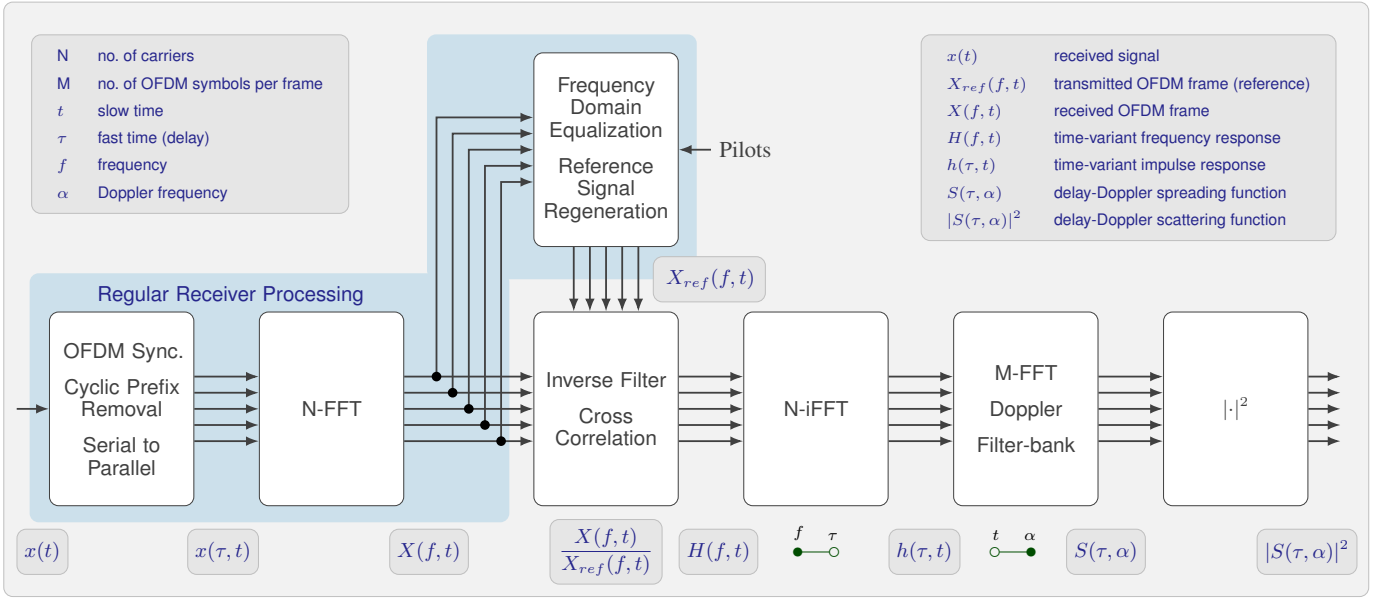
\begin{figure*}[htb]
    \hspace{-3mm}
    \scalebox{0.859}{\tikzset{
  block/.style={
    rectangle,rounded corners,
    fill=white,
    text=darkgray,
    draw=darkgray,
    font=\small\sffamily,
    align=center,
    text width=2cm,
    minimum height=3cm,  % Gesamthöhe des Knotens
    },
  blockF/.style={
    rectangle, rounded corners,
    %fill=blue!10!white,
    draw=gray!50,
    fill=gray!20,
    text=blue!40!darkgray,
    font=\small\sffamily,
    %draw=blue!80!white,
    align=center,%    text width=1.25cm,
    minimum height=0.7cm,  % Gesamthöhe des Knotens
    },
  laplace node/.style 2 args={
    inner sep=0pt,
    align=center,
    text height=1ex,
    text depth=.2ex,
    text=green!30!black,
    % Inhalt:
    label={[xshift=-0.75em, yshift=0.01em]\scriptsize #1},
    label={[xshift=0.75em, yshift=0.01em]\scriptsize #2},
  }
}

\tikzstyle{arrow} = [-{Latex[length=2mm]}, thick,draw=darkgray]

\begin{tikzpicture}[node distance=1cm and 1cm]

% Blocks
\node (ofdm)[block] {OFDM Sync.\\ \vspace{0.2cm} Cyclic Prefix Removal\\ \vspace{0.2cm}Serial to Parallel};
\node (mfft)[block, right= of ofdm] {N-FFT};
\node (inverse) [block, right=2cm of mfft, node distance=10cm] {Inverse Filter\\ \vspace{0.2cm} Cross Correlation};
\node (equalizer) [block, above=of inverse]{Frequency Domain Equalization\\ \vspace{0.2cm} Reference Signal Regeneration};
\node (mifft) [block, right=of inverse] {N-iFFT};
\node (doppler) [block, right=of mifft] {M-FFT\\ \vspace{0.2cm} Doppler\\ \vspace{0.2cm} Filter-bank};
\node (scattering) [block, right=of doppler] {$|\mathbf{\cdot}|^2$};

% Formeln
\node [blockF,anchor=north east] at ($(ofdm.south west)+(0.3,-0.35cm)$) {$x(t)$};
\node [blockF,anchor=north east] at ($(mfft.south west)+(0cm,-0.35cm)$) {$x(\tau,t)$};
\node [blockF,anchor=north east] at ($(inverse.south west)+(-1cm,-0.35cm)$) {$X(f,t)$};
\node [blockF,anchor=north,text width=1.5cm,minimum height=1cm] at ($(inverse.south)+(0cm,-0.2cm)$) {$ \displaystyle\frac{X(f,t)}{X_{ref}(f,t)}$};
\node [blockF,anchor=north west] at ($(inverse.south east)+(0cm,-0.35cm)$) {$H(f,t)$};
\node[laplace node={$f$}{$\tau$},anchor=north] at ($(mifft.south)+(0cm,-0.65cm)$) {$\Laplace$};
\node [blockF,anchor=north west] at ($(mifft.south east)+(0cm,-0.35cm)$) {$h(\tau,t)$};
\node[laplace node={$t$}{$\alpha$},anchor=north] at ($(doppler.south)+(-0.2cm,-0.65cm)$) {$\laplace$};
\node [blockF,anchor=north west] at ($(doppler.south east)+(-0.5cm,-0.35cm)$) {$S(\tau,\alpha)$};
\node [blockF,anchor=north west] at ($(scattering.south east)+(-0.8cm,-0.35cm)$) {$|S(\tau,\alpha)|^2$};

\node [blockF,anchor=north west, text width=1.5cm] at ($(equalizer.east)+(-0.4cm,-1.65cm)$) {$X_{ref}(f,t)$};

%\node[draw=green!30!black, rounded corners,font=\scriptsize\sffamily,text=green!30!black,align=center, anchor=north east] at ($(16,6.3)+(-0.5,-0.5)$) {
\node[draw=gray!50,fill=gray!20, rounded corners,font=\scriptsize\sffamily,text=blue!40!darkgray,align=center, anchor=north west] at ($(0.6,6.3)+(-2,-0.5)$) {
    \renewcommand{\arraystretch}{1.5}
    \begin{tabular}{l l}
      N & no. of carriers \\
      M & no. of OFDM symbols per frame\\
      $t$ & slow time \\
      $\tau$ & fast time (delay) \\
      $f$ & frequency \\
      $\alpha$ & Doppler frequency \\
    \end{tabular}
  };

%\node[draw=green!30!black, rounded corners,font=\scriptsize\sffamily,text=green!30!black,align=center, anchor=north west] at ($(0,6.3)+(-1.5,-0.5)$) {
\node[draw=gray!50,fill=gray!20, rounded corners,font=\scriptsize\sffamily,text=blue!40!darkgray,align=right, anchor=north east] at ($(19.3,6.3)+(-0.5,-0.5)$) {
    \renewcommand{\arraystretch}{1.5}
    \begin{tabular}{l l}
      $x(t)$ & received signal\\
      $X_{ref}(f,t)$ & transmitted OFDM frame (reference)\\
      $X(f,t)$ & received OFDM frame\\
      $H(f,t)$ & time-variant frequency response\\
      $h(\tau,t)$ & time-variant impulse response\\
      $S(\tau,\alpha)$ & delay-Doppler spreading function\\
      $|S(\tau,\alpha)|^2$ & delay-Doppler scattering function\\
    \end{tabular}
  };

% Arrows
\begin{pgfonlayer}{background} 
%    \fill[blue!20!green!20,rounded corners] (ofdm.south west)+(-0.7cm,-1.4cm) rectangle (19.3,6.3);
    \draw[draw=gray!50,fill=gray!10,line width=\fboxrule,rounded corners] (ofdm.south west)+(-0.7cm,-1.4cm) rectangle (19.3,6.3);
    
    \fill[blue!60!green!20,rounded corners] (ofdm.south west)+(-0.3cm,-0.25cm) rectangle (6,2.1);
    \fill[blue!60!green!20,rounded corners] (ofdm.north west)+(5.85cm,+0.5cm) rectangle (8.9,5.8);

    \draw[arrow] (ofdm.west)+(-0.5cm,0) -- (ofdm.west);
    \draw[arrow] node (pilots) [text=darkgray,right=0.5cm of equalizer.east] {Pilots} (pilots.west)-- (equalizer.east);
    \foreach \i in {-2,...,2}{ 
        \draw[arrow] ([yshift=\i * 0.25 cm]ofdm.east) -- ([yshift=\i * 0.25 cm]mfft.west);
        \draw[arrow] ([yshift=\i * 0.25 cm]mfft.east) -- ([yshift=\i * 0.25 cm]inverse.west) coordinate[midway] (myarrowH);
        \draw[arrow] ([yshift=\i * 0.25 cm]inverse.east) -- ([yshift=\i * 0.25 cm]mifft.west);
        \draw[arrow] ([yshift=\i * 0.25 cm]mifft.east) -- ([yshift=\i * 0.25 cm]doppler.west);
        \draw[arrow] ([yshift=\i * 0.25 cm]doppler.east) -- ([yshift=\i * 0.25 cm]scattering.west);
        \draw[arrow] ([yshift=\i * 0.25 cm]scattering.east) -- ++(0.5cm,0);
     
        \draw[arrow] ([xshift=-\i * 0.25 cm]myarrowH) node[fill=black, circle, inner sep=1.5pt] {} |- ([yshift=\i * 0.25 cm]equalizer.west) coordinate[midway](myarrowV);
    
        \draw[arrow] ([xshift=\i * 0.25 cm]equalizer.south) -- ([xshift=\i * 0.25 cm]inverse.north);
        }
    \end{pgfonlayer}

\begin{pgfonlayer}{background} 
\end{pgfonlayer}
\node[font=\small\sffamily, text=blue!40!darkgray,text width=5cm, anchor=west] at (-0.5,1.75) {Regular Receiver Processing};

\end{tikzpicture}}
    %\captionsetup{justification=centering}
    \caption{Basic \gls{isac}-\gls{ofdm} receiver signal processing in frequency-time domain and estimation of the scattering function. The blue marked area indicates standard \gls{ofdm} processing.     
      }
    \label{fig:ofdmproc}
\end{figure*}

The calculation of the \gls{sf}, respectively $H(f,t)$, is most effective from standard \gls{ofdm} frames since it can be carried out in the frequency domain by a factorized 2D \gls{fft}, which is not a full 2D \gls{fft} but rather a significant simplification. First, the instantaneous frequency response is calculated symbol by symbol from the received signal in the frequency domain  $Y(f)$  relative to the recovered reference signal $X_{ref}(f)$ via
\begin{equation}\label{eq:inverse_filter}
H(f) 
    = \frac{
        R_{Y,X_{ref}}(f)
    }{
        R_{X_{ref}}
    }
    = \frac{
        Y(f)(X_{ref})^\ast(f)
    }{
        \left\vert X_{ref}(f)\right\vert^2
    }
    = \frac{Y(f)}{X_{ref}(f)}.
\end{equation}
The first equality is the classical Wiener-Hopf equation. It is also known as Wiener-deconvolution. In the general framework of frequency domain system identification~\cite{ref30}, estimation of the auto\,/\,cross power spectral densities  $R_{X}(f)$ and $R_{Y,X}(f)$ of stationary random signals would need a separate statistical averaging procedure. However, it follows directly from the \gls{ofdm} paradigm that the orthogonality of the received \gls{ofdm} carriers is kept, and any leakage noise of the estimated samples in the frequency domain is avoided. This results from the \gls{cpx}, which is added before transmission and removed at the receiver before \glspl{fft} processing. The absence of further statistical averaging allows for rigorous simplification, which finally leads to the well-known inverse filter approach, which consists of dividing by the recovered correlation reference signal as summarized in equation \eqref{eq:inverse_filter}. 
Actually, \gls{pcl} applies a similar processing scheme~\cite{ref38}. This also corresponds to standard zero-forcing equalization. Another reason that allows for the factorization of the frequency\,/\,delay and time\,/\,Doppler transform is related to the “underspread” characteristics of the mobile radio channel~\cite{ref95}. This means it is a slowly time-variant system, which can be handled as block-wise time-invariant as long as the \gls{ofdm} symbols are chosen to be short enough to avoid time-selective fading (the block fading assumption). This finally allows the factorization, which avoids full 2D \glspl{fft} and matches perfectly to the \gls{ofdm} receiver processing as depicted in \Cref{fig:ofdmproc}. The adjacent Doppler analysis is carried out by a subsequent block of \glspl{fft} along slow time over $M$ stacked \glspl{cir}. These \glspl{fft} can also be interpreted as a Doppler filter bank. The number of transformed consecutive \gls{ofdm} symbols (the \gls{ofdm} frame) determines the coherent radar integration time $T$. Further noncoherent integration can be added. 
Although the simple 2D \gls{fft} based frequency domain system identification scheme has several advantages, it also has some drawbacks and pitfalls. Among those are: (i) Even if the \gls{ofdm} frames were completely filled with carriers of equal strength, the strong sinc sidelobes in the 2D delay/Doppler plane may mask weak reflections. The resolution of components of similar strength is limited by the \gls{af} mainlobe width, which is given by the $1/B \times 1/T$ aperture size, where $B$ is the bandwidth and $T$ is the coherent processing interval in slow time. (ii) The estimated \gls{sf} is calculated at the predefined \gls{fft} grid in the $(\tau,\alpha)$ plane. However, the maxima of  $\left\vert S(\tau,\alpha)\right\vert^2$, which mark the $(\tau,\alpha)$ target parameters, are not bound to this grid. Therefore, we would need efficient local interpolation schemes if we apply the \gls{fft}. (iii) Dynamic targets that are revealed by a non-zero Doppler shift will also change their observed delay (resp. range). Therefore, the estimated \gls{sf} may be blurred by range-Doppler migration if the delay resolution (i.e., bandwidth) is high and the chosen coherent integration time is too long, since the Doppler filter bank assumes constant delays. (iv) The most fundamental drawback, however, comes into play when we are faced with a sparse, non-uniform distribution of the \glspl{re} in the frequency\,/\,time frame that is so important for \gls{isac} as discussed above. In this case, the \gls{af} is distorted in an uncontrolled manner, which may cause detection ambiguity and make correct estimation impossible. We should be aware that application of non-rectangular window functions (tapering) in frequency and\,/\,or slow time aperture space is always suboptimal since that reduces resolution and increases noise induced variance. They also do not help at all in case of a sparse \gls{re} distribution.  

In the sequel, we describe an example that explains the problems of \gls{ft} based estimation of the \gls{scf} that arise with aperture-limited and sparse observations in the frequency\,/\,slow time domain. At the same time, we introduce a model-based estimation approach, which can solve both problems. We consider a propagation situation which is determined by $P$ specular paths with delay\,/\,Doppler parameters $\tau_p$, $\alpha_p$ and path weight $\gamma_p$, respectively
%
\begin{comment}
\begin{align}
H(f,t) &= \sum\limits_{p=1}^{P} 
    \gamma_p 
    {\rm e}^{-\jmath 2 \pi f \tau_p}
    {\rm e}^{-\jmath 2 \pi t \alpha_p}
    \Rect\left(\tfrac fB\right)
    \Rect\left(\tfrac tT\right)
    \label{eq:datamodel_t_f} \\
    \hspace{-10mm}\overset{\parbox{0mm}{\rotatebox{+90}{$\quad\overset{\tau \quad f}{\Laplace}$}}}{S}
    (\tau,\alpha) &= \sum\limits_{p=1}^{P} 
    \gamma_p 
    (
        \delta(\tau-\tau_p) \ast \Sinc(\tau B)
    ) \notag \\
    & \hspace{17mm} \cdot (
        \delta(\alpha-\alpha_p) \ast \Sinc(\alpha T)
    ). \label{eq:datamodel_tau_alpha}
\end{align}
\end{comment}

\hspace{-7mm}\resizebox{1.05\linewidth}{!}{
\parbox{1.15\linewidth}{
\begin{align}
H(f,t) &= \sum\limits_{p=1}^{P} 
    \gamma_p 
    {\rm e}^{-\jmath 2 \pi f \tau_p}
    {\rm e}^{-\jmath 2 \pi t \alpha_p}
    \Rect\left(\frac fB\right)
    \Rect\left(\frac tT\right)
    \label{eq:datamodel_t_f} \\
    \tikz[remember picture,overlay,baseline] 
        {
        \node[rotate=-90] at (6.5mm,7mm) {$\Laplace$};
        }
    {S}(\tau,\alpha) &= \sum\limits_{p=1}^{P}
    \gamma_p 
    \Bigl(
        \delta(\tau\mathord{-}\tau_p)\mathord{\ast}\Sinc(\tau B)
    \Bigr)
    \Bigl(
        \delta(\alpha\mathord{-}\alpha_p)\mathord{\ast}\Sinc(\alpha T)
    \Bigr)
    \label{eq:datamodel_tau_alpha}
\end{align}
}
}

% 1d/2d aufdroeseln, faltung benamsen
where $H(f,t)$ denotes the slowly time-variant radio channel's system response, limited to finite observation apertures, both in frequency and (slow) time. After a 2D \gls{ft}, we obtain the complex \gls{sf} $S(\tau,\alpha)$ which is affected by a $\Sinc$ function in both dimensions. The symbol $\ast$ is a shortcut for a convolution operation in the dimensions $\tau$ and $\alpha$, respectively.  

\begin{figure}[t]
    % FIGURE 14
%    \includegraphics[width=\linewidth,bgcolor=gray!10,rndcorners=5,rndframe={color=gray!50, width=\fboxrule, sep=\fboxsep}{5}]{figures/Sketches/DelayDopplerDistribution.pdf}
    \includegraphics[width=\linewidth,trim={0 -10 0 -10},bgcolor=white,rndcorners=5,rndframe={color=gray!50, width=\fboxrule, sep=\fboxsep}{5}]{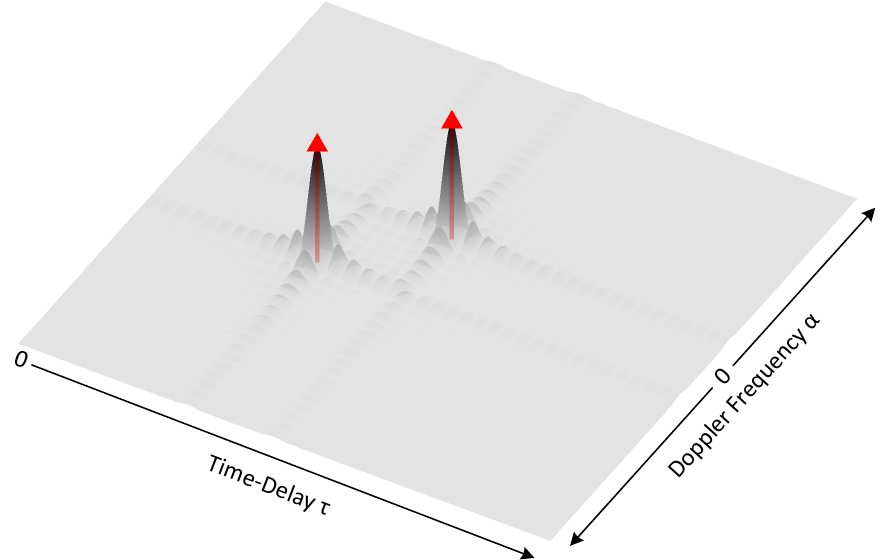}
    \caption{Observed delay-Doppler distribution. The frequency-time resource grid is limited in \gls{bt} and fully occupied. Red arrows represent Dirac deltas standing for an ideal delay-Doppler distribution (no \gls{bt} limitation).      
    Calculated from 48x48 carriers in frequency and time, padded with zeros for 512x512 FFT and interpolation in the delay-Doppler domain.
   }
    \label{fig:delay_doppler_distribution}
\end{figure}

Here, $\left\vert S(\tau,\alpha)\right\vert^2$ is the so-called \gls{scf}, where the detection takes place. It is also referred to as the delay-Doppler distribution. For the example $P=2$, the red arrows in \Cref{fig:delay_doppler_distribution}
indicate the ideal case of two Diracs if no \gls{bt} limitation in the $(f,t)$ aperture plane takes place. The rippled surface is the magnitude of $H(f,t)$ according to \eqref{eq:datamodel_t_f} that results from the superposition of the two equal power propagation paths. The complex valued $H(f,t)$ is sampled and used for the \gls{fft} transform. The 2D rectangular limited support in the $(f,t)$ aperture domain describes the observed data taken by the \gls{ofdm} frame. The effect of \gls{bt} limitation is also included in \eqref{eq:datamodel_t_f} and \eqref{eq:datamodel_tau_alpha}. If the \gls{ofdm} carriers are uniformly power loaded, this appears to be the optimal processing scheme. Some interpolation in delay-Doppler for better detection of maxima can be achieved by zero filling in the $(f,t)$ aperture space outside the observed \gls{ofdm} frame. Although the two paths appear to be smeared in the delay-Doppler domain by the 2D convolution with a 2D $\Sinc$ function (see \Cref{fig:delay_doppler_distribution} ), they can be easily identified since they are well separated. 

\begin{figure}[t]
    % FIGURE 16
    \includegraphics[width=\linewidth,trim={0 -10 0 -10},bgcolor=white,rndcorners=5,rndframe={color=gray!50, width=\fboxrule, sep=\fboxsep}{5}]{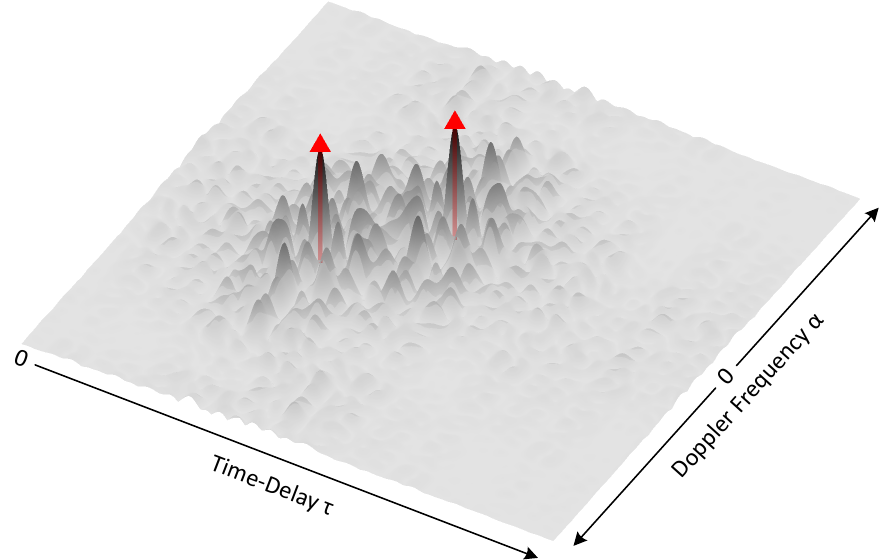}
    \caption{Estimated delay-Doppler distribution, frequency-time aperture limited in \gls{bt}, sparsely occupied resource grid. Calculated with the same \gls{fft} parameters as in \Cref{fig:delay_doppler_distribution}.}
    \label{fig:delay_doppler_distribution_limited}
\end{figure}

\begin{figure}
    % FIGURE 17
    \includegraphics[width=\linewidth,trim={10 20 10 10},bgcolor=white,rndcorners=5,rndframe={color=gray!50, width=\fboxrule, sep=\fboxsep}{5}]{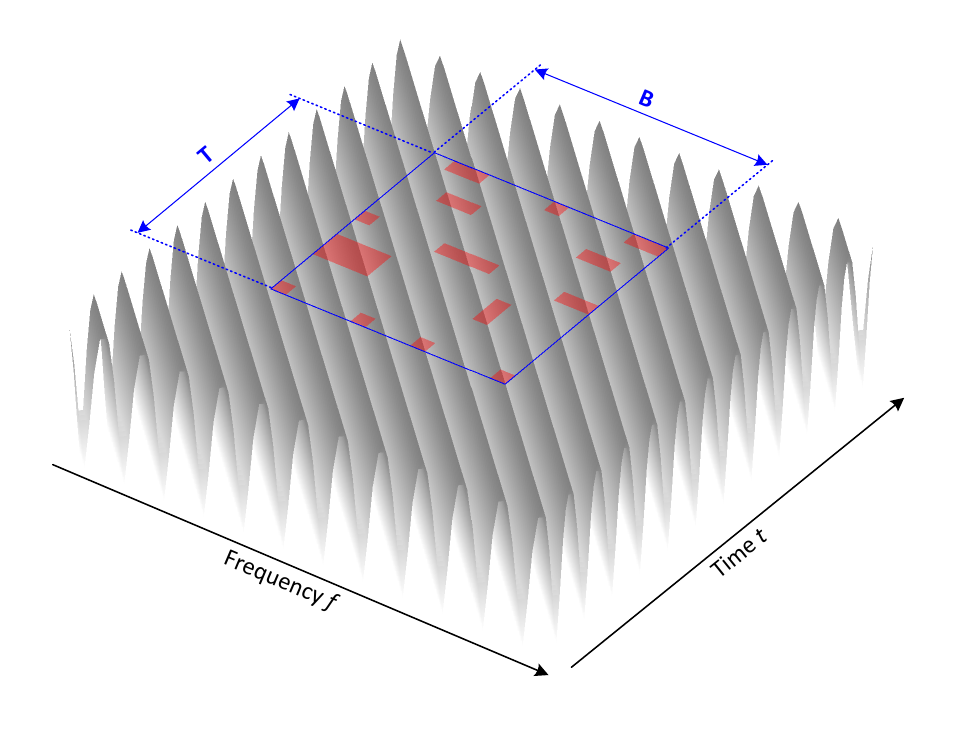}
    \caption{Sparse sampling in the \gls{bt} aperture-constrained $(f,t)$ \gls{ofdm} resource grid. The rippled surface is $|H(f,t)|$ for the two-path propagation model in \eqref{eq:datamodel_t_f} and \eqref{eq:datamodel_tau_alpha}. The red rectangles indicate the active resource blocks where $H(f,t)$ is sampled. } 
    \label{fig:TBRes}   
\end{figure}

Now we consider a \gls{ofdma} resource grid with sparsely allocated \glspl{re} as indicated by the red dots in the $(f,t)$-frame in \Cref{fig:TBRes}.  $H(f,t)$ is now effectively sampled by the sparsely allocated \glspl{re}. The missing samples due to empty carriers are set to zero for \gls{fft}-processing. This is common practice since the \gls{fft} requires a uniform sample grid. However, simply replacing samples that are not measured with zeros is a bad choice, as zeros are definitely wrong. The resulting estimated delay-Doppler distribution is therefore subject to corresponding distortions, which are characterized by uncontrolled side lobes and can make detection impossible as depicted by the example in \Cref{fig:delay_doppler_distribution_limited}. 
 
Fortunately, the solution is obvious. A quick look at \eqref{eq:datamodel_t_f} explains that we can simply consider the equation given there for the analytical description of the problem as a parameterized model of the complex transfer function. So we just need to estimate the parameters of this model. If we have enough well-allocated \glspl{re} in the $(f,t)$ resource frame, we can estimate the model parameters by fitting the $H(f,t)$  model at the measured RE positions. This is very well supported by the orthogonality of the system response in the frequency domain with respect to the \gls{ofdm} carriers. 
Parameter estimation is achieved by minimizing a properly defined least squares cost function. 
In some general sense, this estimation procedure constitutes a multidimensional maximum likelihood parameter estimation scheme~\cite{ref64,ref67}. A deep learning based approach for joint delay-Doppler model parameter optimization from frequency and time samples can be found here~\cite{schieler2025prop_est_cnn}. Efficient real-time implementation is another problem and remains an active area of research. While batch processing may not be feasible, recursive estimation seems to be more appropriate. It is also worth mentioning that the model order $P$ is, in most cases, not known a priori and has to be determined. Finally, the complex path weights $\gamma_p$ are estimated. Moreover, as $\gamma_p$ is polarization sensitive, it can be given by a $2 \times 2$ Jones matrix that accounts for HH, VV co-polar and HV, VH cross-polar I/O responses.

Model-based estimation allows us to obtain the set of discrete delay-Doppler parameters ($\tau$, $\alpha$) without having to apply an \gls{fft}. This avoids any aperture truncation and sparsity gap errors, which would reduce resolution and lead to spurious sidelobes, as discussed above. \Cref{fig:TBRes} explains that the propagation data model effectively interpolates between the observed sparse samples in $(f, t)$, thus avoiding the artifacts that would result from a simple \gls{fft} application. 
At the same time, the effect of high resolution is explained, since the model function implicitly extrapolates the observed data beyond the $BT$ limits, where $H(f,t)$ is measured, thus evading the Fourier resolution limit (beyond Rayleigh resolution). A key aspect of this estimation procedure is that it decomposes the relevant propagation components in the complex linear signal domain. This way, their mutual interaction in the delay-Doppler domain, because of limited support and sparsity in the observed frequency-time aperture domain, is removed. The cancellation of overlapping non-resolved contributions and sidelobes in the complex signal domain seems to be more powerful than the well-known CLEAN procedure~\cite{hoegbom1974clean}, which works in the real-valued point spread or correlation domain.

\begin{figure}
    % FIGURE 18
    \centering
    \includegraphics[width=\linewidth,trim={-10 -30 -10 -30},bgcolor=gray!10,rndcorners=5,rndframe={color=gray!50, width=\fboxrule, sep=\fboxsep}{5}]{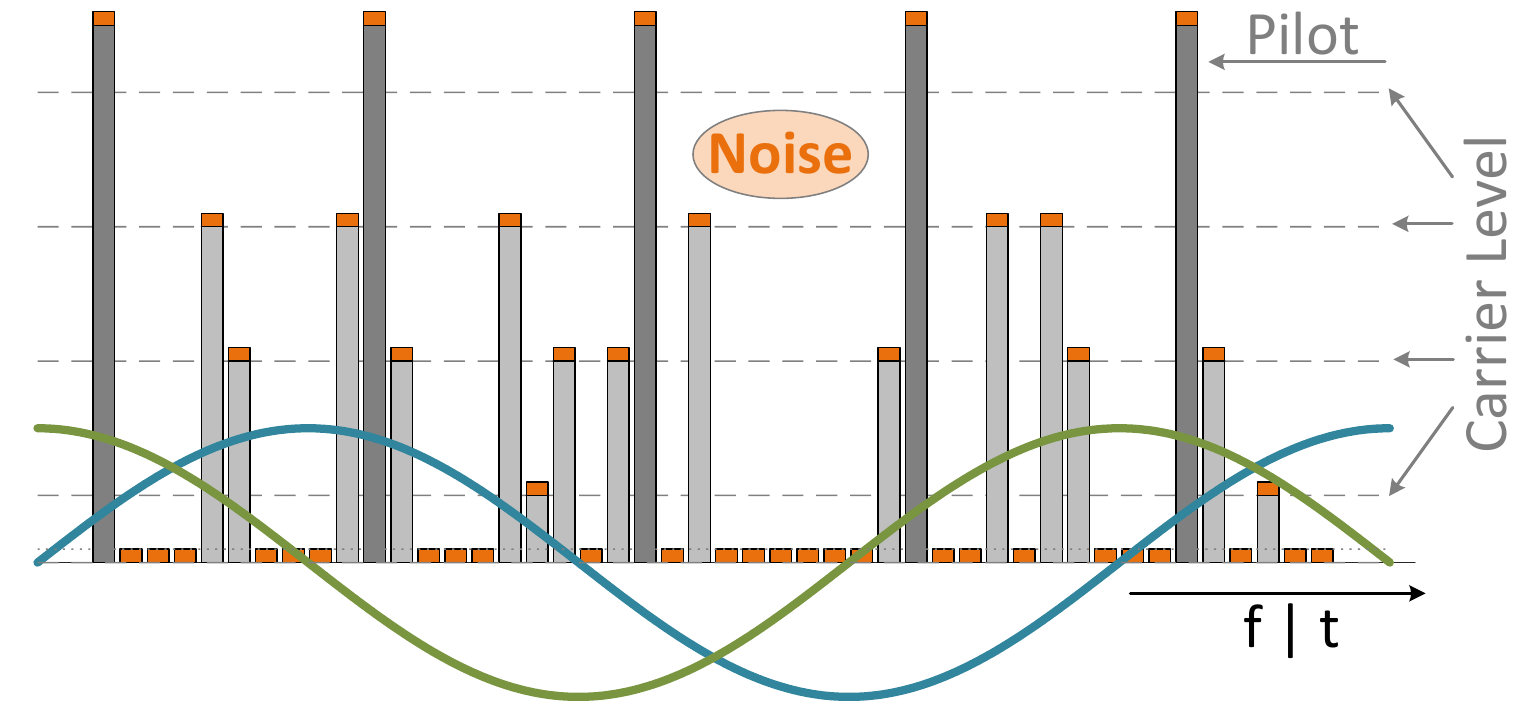}
    \caption{Sparse sampling of $H(f,t)$. The sampling pattern shows empty carriers (noise only) and carriers with different magnitudes because of multilevel modulation.  The complex sinusoid (blue and green) depicts the response of a single path in the  $(f,t)$ domain whose parameters frequency, phase, and magnitude have to be estimated.}
    \label{fig:Sparse-Sampling}
\end{figure}

However, when considering the waveform of the data payload as an excitation signal for system identification, further considerations are necessary. 
\Cref{fig:Sparse-Sampling} shows samples of $H(f,t)$ at some cross section in $f$ or $t$. The problems are as follows: Empty \glspl{re} have to be identified and excluded from the inverse filter operation. A null hypothesis test allows empty but noisy resource elements to be detected and removed from the estimation. The corresponding gap positions are effectively filled by interpolation. Another problem relates to the non-uniform power level of data carriers due to stronger \gls{csi} pilots and because of adaptive power loading and multi-level \gls{qam} coding. While the simple inverse filter would be affected by uneven noise levels, the \gls{snr} of the carriers can be estimated at the receiver and considered in parameter estimation as weighted least squares fitting, which gives the samples with lower \gls{snr} a smaller weight. 

The formulated signal model \eqref{eq:datamodel_t_f} also provides a theoretical basis for \glspl{kpi} that quantify the accuracy and resolution of the parameter estimates. 
This allows, for example, to predict the estimation performance based on available \glspl{re}, but also to judge which \glspl{re} are most beneficial to solve the estimation problem. This is important for allocation of resource blocks in the \gls{ofdma} frame.  
The simplest way to assess the performance of an estimator is to quantify its bias, i.e., the systematic estimation error, and its variance, i.e., the statistical distribution of the estimates. 
Luckily, the often-employed maximum likelihood estimators for propagation parameter estimation~\cite{ref64, ref67} possess the beneficial statistical features of unbiasedness and efficiency, such that the assessment of estimation performance reduces to analyzing theoretical lower bounds for the estimation variance. 
That is why the \gls{crb} is frequently employed as a design criterion for waveform optimization, judging the achievable estimation variance. 
The analytical expressions for this bound aid, for example, in the design of waveforms explicitly tailored to the \gls{ofdm}-related estimation problem given here~\cite{gedschold2025waveformDesignVerification, ref36, Eijnde1991_SignalDesign}. 
The \gls{crb} results from the inverse of the so-called Fisher information matrix
\begin{equation}
    \mathrm{var}(\boldsymbol\theta_i) \geq \mathrm{CRB}(\boldsymbol\theta_i) = (\boldsymbol{F}^{-1}(\boldsymbol\theta))_{ii} \, .
\end{equation}
For conciseness, all model parameters are summarized in $\boldsymbol{\theta} = [\boldsymbol{\tau}_p, \boldsymbol{\alpha}_p, \boldsymbol{\gamma}_p]$.
Conveniently, the Slepian-Bangs formula~\cite{stoica2005slepian} allows us to directly calculate the Fisher information matrix from the partial derivatives of the signal model in the time-frequency aperture domain with respect to the target parameters in delay-Doppler  
\begin{equation}
    \boldsymbol{F}_{i,j}(\boldsymbol{\theta}) = \frac{2}{\sigma^2} \mathcal{R}\left\{ \left\langle \frac{\partial}{\partial \boldsymbol\theta_i} \bm H, \frac{\partial}{\partial \boldsymbol\theta_j} \bm H \right\rangle \right\} \, .
\end{equation}
Here, we consider $\bm H$ as a sampled, i.e., finite-dimensional, version of $H(f,t)$ (see \Cref{fig:TBRes}), and it can be understood as discrete in the carrier and symbol dimensions, sampled by the sparse resource blocks as indicated in \Cref{fig:TBRes}. $\langle \cdot, \cdot \rangle$ is the inner product, $\mathcal{R}$ the real part, and $\sigma^2$ the variance of an additive white Gaussian noise process.

There are different possibilities for getting knowledge about $\bm H$. In the simplest case of one target (one propagation path, no superposition of multiple paths), $H(f,t)$ is not frequency selective (constant magnitude). In this case, the \gls{crb} can be calculated only by assuming a certain bandwidth and \gls{snr}. This can be used for allocation of resource blocks in the \gls{ofdma} frame, as will be discussed below. For this end, we can directly assess the contribution of each carrier or resource block to the final parameter estimate. Since the Fisher information is additive over the samples of the signal model~\cite{javorszky1996} (resulting from the inner product), we can also make the same comparison for several superimposed paths. However, now  $H(f,t)$ becomes frequency selective, and the Fisher information will be subjected to some variance. In a two-cycle estimation procedure, we can get an initial estimate of $H(f,t)$ from pilot-based \gls{csi} estimates, which enhances the estimation in the next step, or we make predictive use of former estimates, which continuously updates the predistortion along the observed track.      

The simple single target case leads to the result that delay parameter estimation benefits from concentrating transmit power on the edges of the available spectrum~\cite{ref36, Mura2025waveformDesignOFDM_ISAC} as already intuitively stated in the section on \gls{ms} Radio Access and Scheduling. The same is also known from radar target estimation, where the lower bound for the estimation variance is proportional to the inverse of the \gls{rms}-bandwidth of the signal~\cite[Ch.~7.2.1]{richards2014RadarSignalProcessing}. In contrast to the plain bandwidth definition, the \gls{rms}-bandwidth is the second-order moment of the power spectral density of the signal. Hence, the \gls{rms}-bandwidth increases when concentrating more transmit power on the edges of the signal's spectrum.

The key takeaways from this subsection are as follows. (i) Range-Doppler resolution of target parameters resembles a 2D harmonic retrieval problem, which extends to higher dimensions when spatial dimensions are added. (ii) Multidimensional estimation should be solved as a joint estimator rather than a sequential 1D estimator, as this avoids any parameter association problems. In addition, resolution in a dimension with a small aperture can benefit from the resolution of another aperture dimension. Finally, joint multi-dimensional resolution counts. (iii) Estimation must be able to handle the sparse distribution of resource elements in  $(f,t)$ due to \gls{ofdma} and \gls{tdd}. (iv) Model-based estimation is the preferred approach for high-resolution estimation of radar parameters from finite and sparsely observed apertures~\cite{ref64,ref67,miranda2025joint_est_ofdma,miranda2025mbpe_sparse_ofdma}. (v) Performance figures based on the \gls{crb} can be used to allocate and control the resource elements of multiple sensing links in the \gls{ofdm} frame according to the aspired detection, resolution, and tracking performance (see also subsection below).     

The fundamental advantage of model-based estimation is its physical relevance. The model consists of two parts. The propagation data model in \eqref{eq:datamodel_tau_alpha} reflects the structural information of multipath in terms of delay and Doppler. This can be seen as a priori information about the physical process to be identified. The unknown information is contained in the set of parameters to be estimated. The propagation data model can be extended to include propagation directions and refined, e.g., to include coupling between parameters such as delay and Doppler or \gls{doa} and delay, etc.~\cite{ref67}. The second part is the device data model that describes the measurement device and the process of data acquisition. In addition to the basic aperture parameters, bandwidth, and observation time (which are considered in delay-Doppler estimation), it can include various linear and nonlinear static and time-variant \gls{isac} sensor response functions (at \gls{tx} and \gls{rx}). This opens the possibility to consider any hardware impairments, far beyond just bandwidth and observation time constraints~\cite{ref35}. Also, specific waveform features resulting from the primary communication function of \gls{isac} can be taken into account. This topic will be continued in subsection Spatial Precoding, Beamforming, and Bidirectional Estimation below when directional filtering and estimation (\gls{doa}, \gls{dod}) is discussed.  
\subsection{ Sensing Cycles – OFDMA Resource Allocation, Time-frequency Precoding and Link Adaptation}
It seems beneficial to adjust \gls{isac} radio access parameters and resources to maximize location and tracking parameters according to the current sensing mode. However, at the beginning of a reconnaissance mission, we generally have no a priori knowledge of the existence, position, and speed of the target. We would need dedicated radar sensing cycles following gradual levels of perception. In a first surveillance and acquisition cycle, the environment is scanned, and when a target is detected, quick, rough, and sometimes even incorrect estimates of the target parameters are made (e.g., many false alarms). Then, in a second step, the \gls{isac} radio parameters will be adjusted to provide more accurate and less ambiguous estimates. This can be achieved, e.g., by a higher sensing symbol repetition rate, a longer coherent processing interval (c.f. radar integration time), which goes along with higher Doppler resolution, and by beamforming, which adapts to the directivity of propagation. However, since the dynamics and contributions of the multiple radio links depend on their relative positions with respect to the target, which change over time, \gls{isac}-specific \gls{csi} estimation and feedback are needed for dynamic transmitter resource allocation, precoding, and link adaptation to support dynamic target tracking. Since this is a two-stage process based on previous \gls{csi} measurements, the effectiveness depends on the \gls{csi} pilot rate and the speed at which the channel changes.

\textbf{Link adaption in radar.} Although the next generation mobile radio may include dedicated tools for \gls{isac} quality of service control, it seems promising to explore how existing mechanisms and tools for \gls{csi} estimation and signaling can be used for \gls{isac} performance control. Traditionally, intersymbol interference in mobile communications caused by multipath is compensated at the receiver by an equalizer. Later, it was realized that precoding at the transmitter side may be more effective, especially if multiple users are included. Over the decades of its maturation, mobile communications has developed sophisticated methods for precoding and resource allocation. Link adaptation has evolved into an unprecedentedly flexible multi-domain technique that includes scalable numerology, flexible resource element allocation in frequency and time, waveform selection, carrier power loading, modulation and coding, and multiple antenna precoding techniques including beamforming~\cite{ref47}. Together with advanced techniques for \gls{csi} estimation and signaling, the control channel allows reconfiguration and adaptation of the transmitter to the instantaneous link conditions for optimum quality of communication service. The latest stage has been set with 5G NR, which is designed to support enhanced mobile broadband, ultra-reliable low latency communications, and massive machine type communications. Since coordination of multiple links and multipath exploitation will be even more important for performance optimization in \gls{isac} (due to the geometric relevance of the multiple distributed links), we expect that precoding and link adaptation will be of similar or even greater importance. 

Let's take another quick look at the link adaptation in radar. The situation is slightly different here. Although multistatic and networked radars have a long-lasting history~\cite{ref49}, single-station monostatic radars are still the default design~\cite{ref48}. The reasons for this seem to be manifold. Monostatic radars are independent and can act autonomously (even mobile). The underlying geometry, a circle or a sphere, seems to be simple. Most of all, signal processing is local, which makes Tx\,/\,Rx synchronization and resource adaptation (bandwidth, Tx power, pulse rate, dwell time, beam management, etc.) a local task. There are very advanced resource management principles for monostatic radars applied that can cope with dynamic scenarios of multiple targets~\cite{ref46} and are able to adapt their operation based on cognitive procedures~\cite{ref44}. Attempts to overcome the geometric limitations of coverage and target visibility by a network of radar sensors often end up with non-coherent cooperation of multiple monostatic radars~\cite{ref45} (fusion of multiple directions from dislocated radars is a non-coherent operation). References to radar networks with multi-static cooperation are rare~\cite{ref48}. Even if the publicly available literature may not completely reflect the entire state of knowledge, the radar-centric approach to \gls{isac}, which requires a separate (probably proprietary) communication network or additional communication functions integrated into the radar's air interface, does not appear to be a very promising role model for large scale application of \gls{isac} in public networks. At the very least, \gls{isac}'s communications-centric approach seems more promising, as it reuses the already available communication network for data transport and fusion. In addition, the well-designed and constantly evolving sensing, signaling, and control schemes of mobile communications reveal a kind of hitchhiking support for radar functionality. This is most obvious for the \gls{cpcl} approach, which seems to be the closest coexistence of communication and sensing.

\textbf{Radio access coordination of multiple sensing links.} As we have explained above, \gls{isac} radar detection and tracking of 3D dynamic target state vectors in general requires multiple sensing links that have to be coordinated and adapted to the dynamic target scenario to meet the required sensing performance. It is obvious that resource management and link adaptation must take place in a distributed \gls{msisac} system at the radio and network level. While the primary goal of communication is efficient transfer of data between two dedicated radio nodes acting as the source and the destination, the goal of \gls{isac} is the localization and recognition of target objects. Multiuser communication links, in essence, can be seen as independent single links just sharing common radio resources according to rules of fairness or priority. Sensing rather requires the proactive cooperation of multiple radio nodes (illuminators and sensors) that share the same resources and work together to fulfill a common sensing task. Already when selecting the nodes involved, their spatial availability (both current and predicted), expected influence on localization accuracy due to \gls{gdop} and \gls{snr}, and detection functionality due to heterogeneous configuration and connectivity have to be taken into account. Furthermore, the allocation of communication and computation resources is based on the decision on the data collection task and the type of data fusion. It seems important to recognize that not a single radio node (e.g., a \gls{ue}) is a beneficiary of an \gls{isac} service, but the fused information from several sensing links that is made available to authorized users represents the added value. Therefore, \gls{msisac} resource allocation is always a joint, respectively cooperative activity of a sensing network.  

Once we have defined which radio nodes contribute to the execution of a sensing mission, we need to decide how to allocate radio access resources. This is especially interesting when multiple sensing links share the same radio channel. Which signals and access schemes do we have available for target scene illumination in \gls{isac}? While dedicated pilot signals are required to establish communication links (Tx\,/\,Rx synchronization, demodulation reference signaling, and \gls{csi} estimation), and therefore are essential also for \gls{isac} operation, the use of \gls{csi} pilots alone will in general not be enough for \gls{isac} sensing as already explained in the subsection on Recovery of the Correlation Reference Signal. The use of the Positioning Reference Signal would be another option. The \gls{prs} is a dedicated signal that enables self-localization of mobile \glspl{ue} in the \gls{dl}. It can carry network-specific data to support localization services and allows \gls{ofdma} frame mapping to support multiple \glspl{ue}~\cite{ref74}. Although it is not intended for radar sensing, it can be used in the \gls{dl} of \gls{msisac} to temporarily enhance target illumination. However, excessive use of \gls{prs} would only block resources that may not be available for communication services. This means that radio resource efficient \gls{isac} should aim to reuse the communication payload for sensing, as already proposed by the \gls{cpcl} principle. It is also possible to use the communication payload-based sensing results for \gls{csi} feedback and sensing link adaptation. The use of \gls{csi} pilots has the advantage that orthogonal multi-sensor access is easy to achieve.

\textbf{Broadcast and multi-sensor access.} According to the subsection on Multi-Sensor Radio Access and Scheduling, it is reasonable to distinguish between the broadcast mode and multi-sensor \gls{isac} access. In the first case, each transmitter serves several sensing receivers, which provides simultaneous measurements and, thus, instantaneously multiplies the information about the target. This applies, e.g., if the sensing nodes are receive-only sniffers deployed as part of a distributed infrastructure or acting as deployable sniffers in the reduced \gls{cpcl}-mode as described in \cref{sec:architectures}. In case of \gls{msisac}, the orthogonal multiplexed sensing links must share the same channel by proper allocation of the \glspl{rb} in the \gls{ofdm} frequency-time frame as described in the subsection on Multi-Sensor Radio Access and Scheduling. Obviously, resource sharing should be made adaptive according to the number and dynamics of targets, as well as the multipath propagation situation. 

For the purpose of \gls{msisac} radio access in \gls{dl} or \gls{ul}, we must adjust the distribution of \glspl{rb} in the frequency-time frame by \gls{ofdma}. From the instantaneous \gls{isac} \gls{csi} and \gls{kpi} estimates transmitted via the \gls{ul} to the \gls{cu} of a distributed base station, a resource scheduler can decide how many resources per link have to be spent to follow the dynamically changing parameters of the respective target. Later on, when the moving target is tracked, and the dynamic situation changes according to the target's position and velocity vector relative to the baseline of the sensor links, the \gls{rb} distribution will be updated. 

\textbf{CRB driven allocation of OFDMA resource blocks.} Since the target resolution performance depends on the \gls{crb} as discussed in the subsection on Model-based Bistatic Target Parameter Estimation, it is recommended to use it as a cost function for \gls{msisac} resource scheduling. For the \gls{ms} access, the bottom line is about optimum arrangement and average power weighting of carriers in the \gls{ofdma} frame for any sensing link. The total number of carriers to be allocated per link depends on both the required processing gain of the matched filter (according to the \gls{snr} of the respective link) and the resolution performance. The \gls{crb} criterion tends to shift the carriers in blocks to the edges of the available frequency-time frame. Since we may have to host multiple sensing links, we would need to interleave them in two dimensions, frequency and time. A simple rule of thumb for \gls{rb} scheduling of multiple cooperating sensing links may be to allocate the \glspl{rb} of links with temporarily lower resolution requirements to the inner space of the \gls{ofdma} frame. The advantage of the 2D nesting is that virtual simultaneous measurements from multiple transmitters become possible, with the restrictions that a shared medium entails. 

The weighting applied due to the \gls{crb} \gls{kpi} becomes more complex when multiple targets or target-related bounced propagation paths need to be resolved. Moreover, besides binary power loading within one \gls{ofdma} symbol, we can apply continuous power loading~\cite{ref85}. 
This will be the case if only \gls{tdma} (hence no \gls{fdma}) multiplexing within the \gls{ofdma} frame is applied. While continuous power loading can be seen as equivalent to maximizing data transfer capacity, respectively mutual information, in multiuser \gls{ofdm} (by the water filling algorithm), binary loading is equivalent to carrier avoidance power loading. The latter strategy prioritizes the use of high-performing subcarriers and disables low-performing subcarriers. The spared subcarriers can then be used for other links that can benefit more from the respective empty frequencies. For example, in a joint multilink measurement, the links with the highest resolution requirements in terms of delay and\,/\,or Doppler would be assigned \glspl{re} in the outer regions of the frequency/time resource frame, while the links with lower resolution requirements would get \glspl{re} more in the central part.

\textbf{AI enabled resource allocation.} The allocation of \gls{msisac} resources from the perspective of sensing alone already appears to be a complex optimization task. However, since the same radio resources are used for sensing and communication, the balance between communication and radar sensing performance must also be taken into account when scheduling resources. While in communication, it is all about the data rate and latency; in radar sensing, the issue is how many dynamic targets can be reliably detected and tracked in a given scenario. Due to the dual use of radio resources inherent in the \gls{cpcl} principle, the choice between communication and sensor technology regarding resource consumption is not a simple either\,/\,or since it is not even per se competitive. Moreover, cooperative sensing will entail an independent (additional) requirement for the transport of intermediate sensing data for fusion with special requirements in terms of latency and capacity. In addition, if there are not enough loadable carriers to provide the required sensing performance, it would be appropriate to include dummy carriers that serve only for sensing. Eventually, \gls{msisac} resource scheduling would depend on the required sensing task and respective sensing cycle, making it a challenging joint multilink optimization task. The multi-criteria nature of the optimization problem yields a non-convex cost function that cannot be minimized by simple means. This suggests the need for an AI-supported solution. Another challenge results from the sparsity in frequency and time because of the multiple interleaved measurements. Therefore, estimation methods that can deal with sparse frames are required. A simple inverse filter and \gls{fft} application would fail. This is also due to the non-uniform carrier magnitudes due to multilevel modulation and power loading. The solution is model-based target parameter estimation using weighted least squares as described above. 

\textbf{Multipath exploiting time reversal precoding.} As already described above, multipath propagation can be considered either a 'foe', reducing target visibility and location accuracy, or a 'friend', exploited to improve \gls{isac} performance. Besides explicit model-based methods requiring a priori information from the geometric structure of the propagation environment and implicit in situ training-based methods for multipath exploitation, there are frequency-time precoding schemes that use observed channel response functions and \gls{csi} feedback. These methods are known as \gls{trf} of waves. Originally, \gls{trf} was proposed by M. Fink in the context of ultrasonic waves~\cite{ref87}. Later on, it was also used to focus electromagnetic waves, with applications in communication and radar~\cite{ref88,ref89,ref34}.
In short, the basic idea is as follows: When a spatial wave field is emitted from a source in a complex ('inhomogeneous') medium and recorded by a set of sensors around the edge of the medium, then mirrored along the time axis ('time reversal') and retransmitted, all wave components will refocus at the source, despite the time delay to the sensors and regardless of dispersive propagation effects (multipath reflections, scattering, diffraction, etc.). In communications, a basic application scenario for \gls{trf} assumes a \gls{ue} that estimates the channel $H(f,t)$ as $\hat{H}(f,t)$ in the \gls{dl} and transmits the \gls{csi} to the \gls{gnb}, where it is mirrored in time and convolved with the waveform to be transmitted. 
This process can be formulated in the frequency and time domains as
\begin{align}\label{eq:timereversal}
    Y(f,t) & = X(f) H(f,t) \hat{H}^\ast(f,t) = X(f) \hat{R_H}(f, t) ,\notag \\[3mm]
    \tikz[remember picture,overlay,baseline] 
        {
        \node[rotate=-90] at (4.5mm,7mm) {$\Laplace$};
        }
    {y}(\tau, t) & = \int\limits_{-\infty}^{+\infty} x(\tau - \sigma) \hat{r_h}(\sigma, t) \mathrm{d}\sigma
\end{align}

%\overset{\parbox{0mm}{\rotatebox{+90}{$\quad\overset{{\tau \quad f}}{\Laplace}$}}}

where $\hat{R_H}(f,t)$ denotes an estimate of the \gls{ft} of the channel's \gls{acf} at time $t$ and $\hat{r_h}(\tau, t)$ its (fast) time domain counterpart.

\gls{tr} is always a two-step predictive procedure that relies on the \gls{csi} estimation in \gls{dl} and signaling in \gls{ul}, assuming that the channel does not change too fast. \gls{tr} precoding changes the resulting channel impulse response to the \gls{acf} of the unmodified response \eqref{eq:timereversal}.

It appears that the focusing effect becomes better the stronger the delay spread is. Therefore, reverberant multipath rich channels transform to low delay spread channels by precoding (as the channel \gls{acf} becomes short). In this sense, \gls{trf} can be interpreted as a “channel-matched filter”, which can help to simplify the receiver complexity by effectively shifting the equalization operation from the receiver to the transmitter. Obviously, focusing in delay goes along with true spatial focusing since the channel-matched filter condition depends solely on the local superposition of multiple reflected paths impinging from different directions with different delays. This way, a multipath-rich propagation environment acts like a huge reflect array (a \gls{trf} mirror) that causes a sharp spatial focus. If there is not enough multipath for focusing available, multiple cooperating \gls{msisac} transceivers or \gls{ris} can be deployed to achieve a similar effect. In addition to the shift in implementation effort from the \gls{ue} to the infrastructure, there are further advantages, such as increased energy efficiency, since more reflected energy is collected and coherently combined at the Rx antenna. Moreover, in communications, spatial focusing reduces the probability of non-desirable interception.

\begin{figure}
    \centering
    \includegraphics[width=1\linewidth,bgcolor=gray!10,rndcorners=5,rndframe={color=gray!50, width=\fboxrule, sep=\fboxsep}{5}]{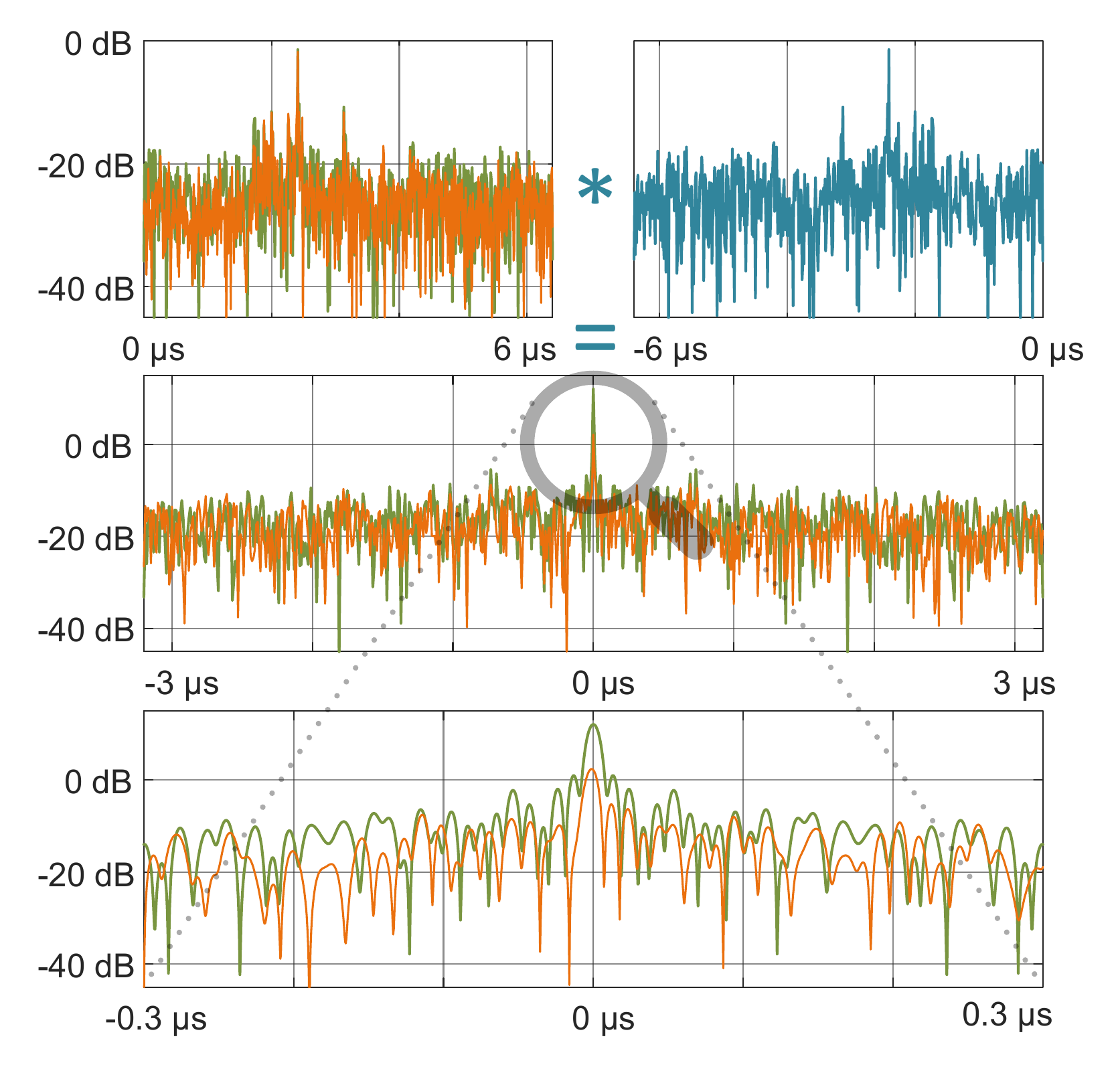}
    \caption{Time reversal focusing from measured \glspl{cir}. Upper row: Green/orange: Two measured \glspl{cir}, ca. \SI{32}{\centi\meter} apart; Blue: Time reversed \gls{cir} from measured green \gls{cir}. Middle and lower row: \gls{tr} channel-matched response: Green matched to the correct \gls{cir}, orange matched to the displaced \gls{cir}. 
    }
    \label{fig:timereversal}
\end{figure}

\Cref{fig:timereversal} shows an example of \gls{tr} focusing calculated from measured channel impulse response functions. The sounding measurement took place in a \gls{nlos} street scenario in Cologne, Germany. The center frequency of the sounder was \SI{3.75}{\giga\hertz} and the bandwidth was \SI{100}{\mega\hertz}. The sectorial BS-Antenna was above the rooftop, and the omnidirectional \gls{ue} antenna was mounted on a slowly moving car on street level. The instantaneous power delay profiles (magnitude squared \gls{cir}) depicted in the upper left figure for two consecutive positions (green and orange) about 32 cm apart show that \gls{los} is obstructed. We can identify about 6 to 7 more or less relevant multipath components, with one strong reflection dominating. The green \gls{cir} was then taken as a \gls{tr} reference depicted mirrored in delay in the upper right figure. The resulting \gls{tr} channel matched response is shown below (green). Note that in practice, the response will appear on the right-hand side because of causality. Compared to the original \gls{cir}, we see a clear and strong focus in the delay domain, as was to be expected, since the original \gls{cir} has a noise-like structure. The orange response results from the correlation of the same reference with the \gls{cir} at the displaced position. The degradation of about 10 dB in correlation gain because of the reference mismatch is obvious. This describes the expected spatial focusing. The \gls{cir} mismatch of the \gls{cir} is mainly because of the phase difference that is not visible in the power delay profile shown here. The distance of about 4 wavelengths leads to strong decorrelation, even if we only have this small number of paths with a somewhat limited angular spread.

\begin{figure}
    \centering
    \includegraphics[width=\linewidth,trim={-30 -30 -30 -30},bgcolor=gray!10,rndcorners=5,rndframe={color=gray!50, width=\fboxrule, sep=\fboxsep}{5}]{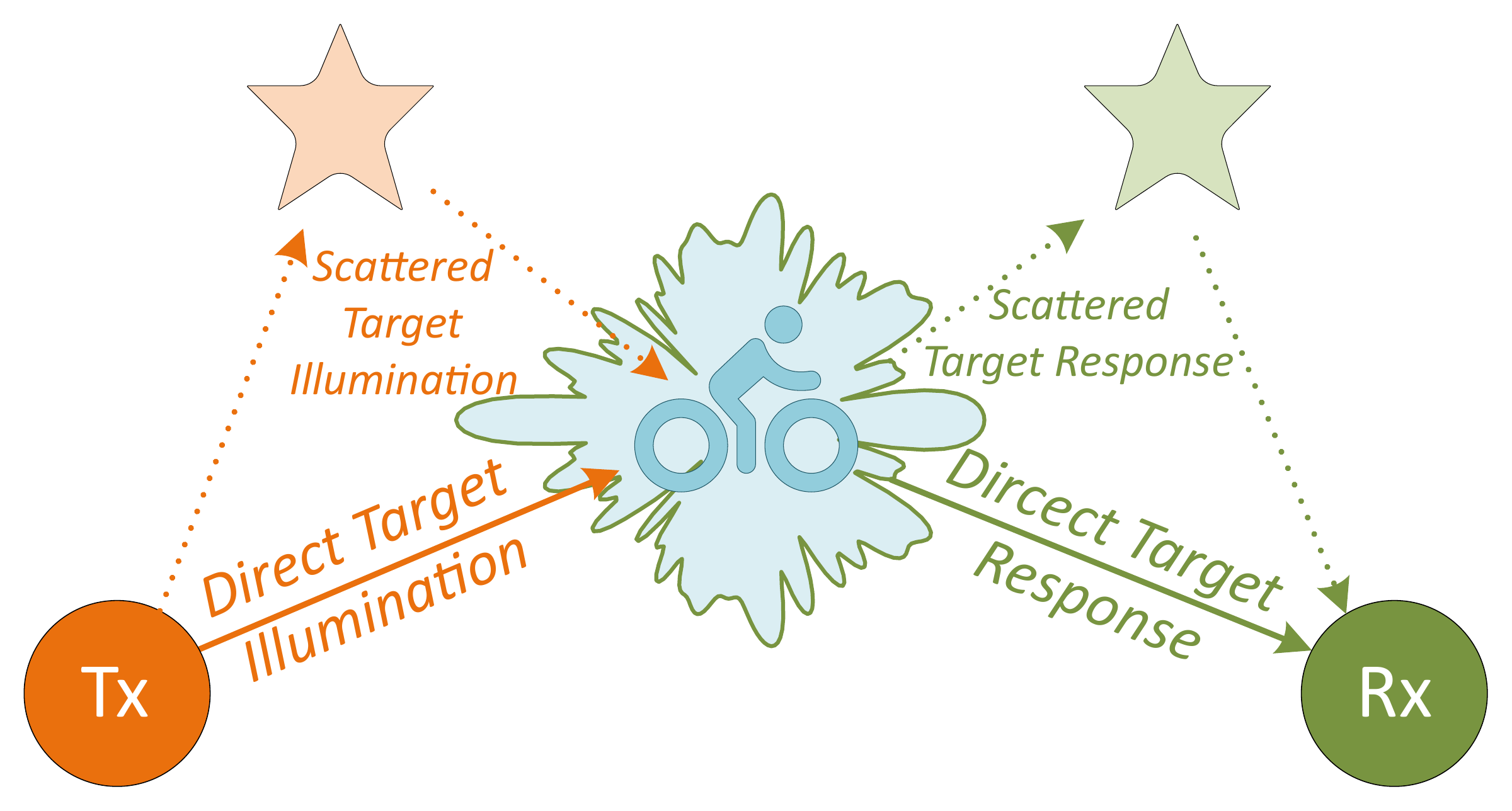}
    \caption{Remaining multipath propagation after background subtraction (c.f. \Cref{fig:multipath_geom}).}
    \label{fig:backgroundsubtraction}
\end{figure}

The situation for radar is somewhat different. The reason is that now we should focus on the passive target object in order to enhance its illumination and visibility against clutter. However, in most applications, we will not have a transceiver for \gls{csi} sensing and reporting in place of the target. If we were now to perform end-to-end focusing of the transmit and receive antennas in \Cref{fig:multipath_geom}, we would obtain a focus at the receive antenna, which, however, includes all multipath interference, that has nothing to do with the target. One promising solution to get rid of the not-target-related clutter influence is background subtraction that removes or at least reduces the target-independent clutter, including direct \gls{los}. Background subtraction is easiest to use when the sensor network is static (e.g., in the case of infrastructure-only sensing) and the target is moving. It will become more complicated if the sensor nodes are moving, and perhaps can be supported by beamforming. In any case, only propagation paths routed via the target must be retained. If the end-to-end response is now precoded by time reversal, all and only the propagation paths that are routed via the target are used for the focusing at the sensor, see \Cref{fig:backgroundsubtraction}.

\textbf{Doppler precoding.} This finally leads us to the question of whether a similar focusing technique is possible in the Doppler domain. Bringing the observed Doppler shift to zero for a moving target would support a longer coherent integration time for correlation processing. This can be important for target detection and reliable tracking, and again follows the idea of shifting the processing effort from the sensor as “precoding” to the infrastructure. The result can be seen as a 2D channel matched filter that focuses the received signal in the joint 2D delay-Doppler domain. Obviously, focusing on the time delay as described above will also impact the Doppler domain, since the delay corresponds to range and the Doppler shift corresponds to speed, i.e., the range derivative. Therefore, if mirroring in delay works perfectly for range focusing, the Doppler contribution by the different paths would also collapse to zero. However, in a practical situation, this might not be so easy. This is because the delay resolution and variance are limited by the bandwidth and noise constraint of each \gls{ofdm} symbol. Doppler, however, relates to the carrier phase, which is much more sensitive to range variance. In 2D delay-Doppler estimation, Doppler is obtained by \gls{ft} along the slow time \Cref{fig:ofdmproc}, which may include several hundred or even more consecutive \gls{ofdm} symbols. Phase stability along the coherent processing interval is of highest importance to get a stable Doppler shift that describes a linear movement. While \gls{fft} filtering along the slow time reduces the variance, the inherent variance of the channel estimates used for the \gls{tr} operation can actually worsen the situation. Therefore, model-based joint delay-Doppler estimation should work better as the inherent model would cause some smoothing along slow time. Moreover, since the model order can be controlled, this would pave the way towards using only the dominating and persisting paths for delay-Doppler precoding, e.g., those paths that are stable and reflected by objects at known positions, which can be traced back for explicit multipath exploiting localization.

\textbf{Target related precoding.} Regarding the influence on the target object, we can deduce from  \Cref{fig:backgroundsubtraction} that the diversity advantage for increased target detection probability applies here as well, as was discussed for distributed \gls{msisac}. However, the geometric advantage for localization only applies if we can use the true positions of scatterers for explicit, model-based localization. Here, the circle closes, and it becomes clear that \gls{trf} and ray optical modeling of propagation can be combined. However, recalling the special role of the target as a solitary interacting object in the propagation environment (see Section 4, 'Multipath Propagation'), it deserves even more attention for waveform adaptation. While delay-Doppler precoding can be understood as a 2D channel-matched filter, a similar and equivalent matching operation can be assigned to the target. Without going into every detail, we will summarize some aspects. For extended targets (where the radar bandwidth allows resolving several delay bins that are related to the same target), it may make sense to adapt the illuminating signal to the target response by time reversal. The result can be considered a target-matched filter that maximizes coherently reflected energy while avoiding its spread to several delay bins. This seems to be a better metric (because it is ultimate) for calculating the radar link budget than the radar cross-section, which presumes non-extended targets. 

It seems that a wider bandwidth can resolve more structural details about the target, which would also favor higher frequencies. But this could be a serious pitfall since, besides some material-related frequency dependency of reflectivity, there may also be a strong structural dependency of reflectivity. This is very often related to cavities being a part of the target structure. Then the target can respond with strong Eigenmodes if properly excited. Since Eigenmodes are related to resonance effects, the proper frequency of the illumination signal is crucial. So we have to make sure that the resonance frequency is covered. Depending on the size of interesting targets, these Eigenfrequencies can be quite low (e.g., FR1). Therefore, multiband target illumination as described in the subsection on Heterogeneous Radio Nodes can be useful.

Another domain for target response estimation and adaptation is polarization. According to the Jones matrix calculus, we can estimate the $2 \times 2$ polarimetric target response matrix by illumination with two orthogonal polarizations (dual linear or circular) and sensing of the scattered waves in orthogonal polarization orientations~\cite{ref68}. The input/output Jones vector is defined by the dual polarimetric antennas at Tx and Rx. In general, orthogonal access at the two Tx antenna ports (e.g., by \gls{tdma}) would be appropriate. Depending on its shape, the reflectivity of the target may strongly depend on the polarization orientation of the illuminating waves, and somehow, the target structure acts as a more or less polarization-sensitive filter. This may require adaptation of polarization to maximize target reflectivity. The polarization orientation of the illuminating wave can be controlled by proper weighting and simultaneous transmission of the waveform from two polarimetric Tx antenna ports. With phase control included, circular or elliptic polarization can be generated from two orthogonal linear polarized antenna ports. The same adaptation principle also applies to the Rx chain. Note that usual definitions of linear polarization as Vertical (V) and Horizontal (H) apply to the local coordinate system of Tx and Rx antennas, which may not be aligned in H and V between each other. The same applies to the local coordinate system of the target, and also does not apply to the Earth-centered coordinate system. This means that just simplifying to vertical (VV) or horizontal (HH) polarization at the transmitter (Tx) and receiver (Rx) is not enough in most cases if the Tx and Rx antennas are randomly oriented in space, as it may be the case if agile \gls{isac} sensor platforms are included. 
\subsection{Spatial Precoding, Beamforming, and Bidirectional Estimation}
Spatial precoding is a well-established method in mobile radio that came into play with the introduction of multiple antennas. It helps increase link performance, reduce interference, host and address multiple users, and increase data rate and bandwidth efficiency. Spatial precoding will also be very important for \gls{isac}, especially for \gls{msisac}~\cite{BF1}. Since 5G NR uses a flexible architecture based on codebook and non-codebook (full adaptive) schemes, spatial radio access can be controlled for omnidirectional or directive sensing. Without going too much into the technical details, we should emphasize that spatial precoding, especially beamforming, becomes more important at higher frequencies (FR2), where the need for higher beamforming gain increases because of link budget issues. However, hybrid (mixed analog and digital) antenna interfaces seem to be the dominating solution, where digitization takes place at the output of the beamformer, not at the antenna ports. Multiple simultaneous beams for the same user are also possible. 5G NR allows transmit precoding and receive combining, and also supports beamforming in the \gls{ul}, which means that the \gls{gnb} can assign a spatial precoding matrix to the \gls{ue}, which can be a dedicated device acting as a sniffer sensor, as described above. In the sequel, we will discuss some striking applications of beamforming and spatial access for \gls{msisac}, though we do not claim to be exhaustive.

\textbf{Beamforming.} The initial purpose of beamforming is to enhance the \gls{snr} (and consequently the range), as it is associated with an antenna gain. In the context of distributed cooperative \gls{msisac} networks, one application is to balance the \gls{snr} to achieve smooth coverage and performance distribution irrespective of the density of the spatial distribution of the sensor nodes. \gls{msisac} network densification will also go along with a proper choice of the frequency. Higher frequencies will be more appropriate for sensing hotspots since the wider bandwidth at FR2 offers more resolution capabilities in delay. But, because of the Friis and radar equations, higher frequencies are more subjected for transmission loss since the effective antenna area is becoming smaller for the same gain. This can be well compensated for by antenna arrays. However, this is accompanied by directivity. Hence, we would need time-consuming angular scanning in the surveillance cycle. Moreover, it will be appropriate to allocate different beams for the target and for the UE direction. Multifrequency sensing can help to get a fast (quick and dirty) overview picture of the scene and some first indication of possible targets using wide-open beam antennas at lower frequencies that do not require expensive angular scanning. The result is then used to instruct the more selective FR2 beamformer in a second step~\cite{ref83}.

\begin{figure}
    \centering
    \includegraphics[width=\linewidth,trim={0 0 0 0},bgcolor=gray!10,rndcorners=5,rndframe={color=gray!50, width=\fboxrule, sep=\fboxsep}{5}]{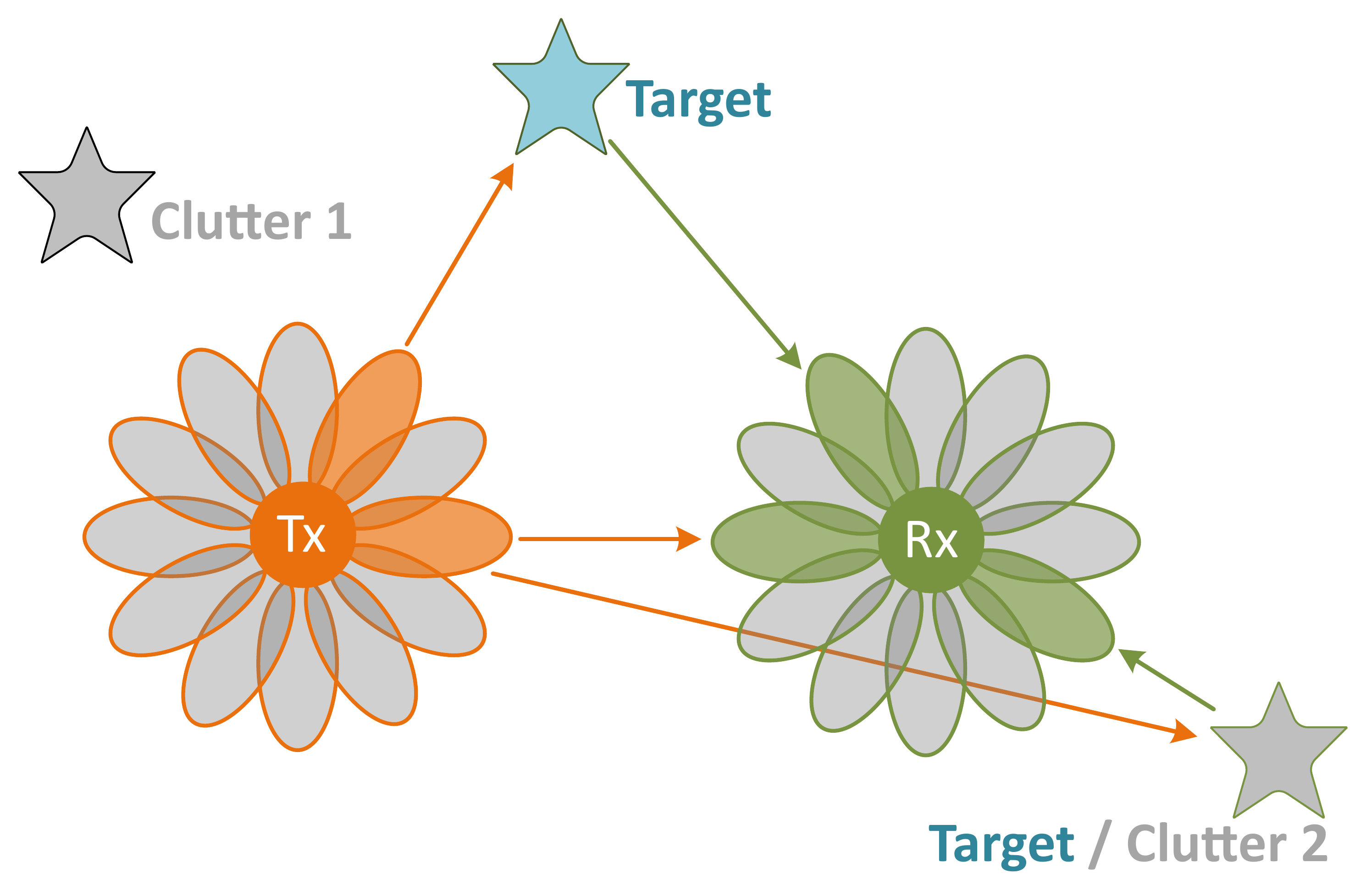}
    \caption{Bidirectional beamforming supporting clean \gls{los} filtering and clutter suppression.}
    \label{fig:doubledirbeamforming}
\end{figure}

Beamforming directivity can be applied deliberately to reduce interference and multipath clutter through spatial filtering, especially when executed in a bidirectional manner at both the Tx and Rx. The generic antenna architecture depicted in \Cref{fig:doubledirbeamforming} is circular on both sides, making it ideal for symmetric D2D scenarios, such as V2V, with arbitrary spatial orientation of the devices. An equivalent antenna setup can be found for V2I communication, where planar arrays at the infrastructure side may be more appropriate. The direct \gls{los} link is important for data transfer and for obtaining the correlation reference. \gls{los} beamforming suppresses multipath, which supports transmit signal recovery. A second surveillance channel points to the target, while clutter 1 is not detected by both beams and is therefore not visible. The case of target\/\,clutter 2 is particularly interesting. Reflections from this object could not be filtered out by the Tx beam, but can be well suppressed by the Rx beam pointing towards Tx. Therefore, the Rx-\gls{los} beam is important for receiving a clean, multipath-free copy of the transmitted signal. When the same object is considered as the target, the corresponding Rx beam pointing to the target makes it visible and also reduces \gls{los} interference. This example shows how the two bidirectional beams support each other and may increase directive filtering a lot. However, it also increases beam search effort. It is worth mentioning that, in a 3D geometry, matching the two opposite beams must also include polarization orientation, especially when azimuth and elevation matter and the orientation of the antennas may be more or less arbitrary~\cite{ref68}. Since the two beams are served by the same radio interface, mutual coherence is maintained, which is important for eToF and eDoppler estimation in \gls{cpcl}. 

\textbf{High resolution DoD/DoA estimation.} Beamforming can be very important to support ISAC radio access and also provide some directional information about impinging waves, especially if the beam is narrow. However, it does not directly allow high-resolution directional estimation beyond the Rayleigh limit. The decisive factor for high-resolution \gls{doa} estimation is phase difference at displaced antenna outputs. This would enable a much more accurate \gls{doa} estimation than simply selecting the radiation pattern or relying on magnitude differences for \gls{doa} interpolation within the main beam lobe. The problem is that the analog\,/\,hybrid 5G beamforming interface hides the antennae in\,/\,outputs for individual access and processing, which prevents the estimation phase differences. However, high-resolution estimation schemes such as MUSIC or \gls{esprit} would require mutual antenna output correlation, which is a quadratic (hence nonlinear) operation and not linear beamforming for building the array correlation matrix. 

In the case of hybrid beamforming, the outputs of the analog multibeam network map a huge number of hundreds or even thousands of physical antennas by a linear weight-and-sum network to a small number of virtual antenna ports where digitization takes place. \cite{ref97}, for instance, describes the beamformer design based on limited-resolution phase shifters. In the following, we introduce an approach for high-resolution \gls{doa} estimation that makes intelligent use of the precoding beamforming vectors. We assume that the beam patterns overlap and the beam outputs explicitly differ in phase, which can then be used for \gls{doa} estimation. The required number of virtual antenna ports is relatively small (much smaller than the number of antennas). This is because we only need to resolve those paths by angle that are not resolved by other parameters (delay, Doppler, or the transmit beamformer).

\begin{figure}
    \centering
    \includegraphics[width=\linewidth,trim={60 -20 60 60},bgcolor=gray!10,rndcorners=5,rndframe={color=gray!50, width=\fboxrule, sep=\fboxsep}{5}]{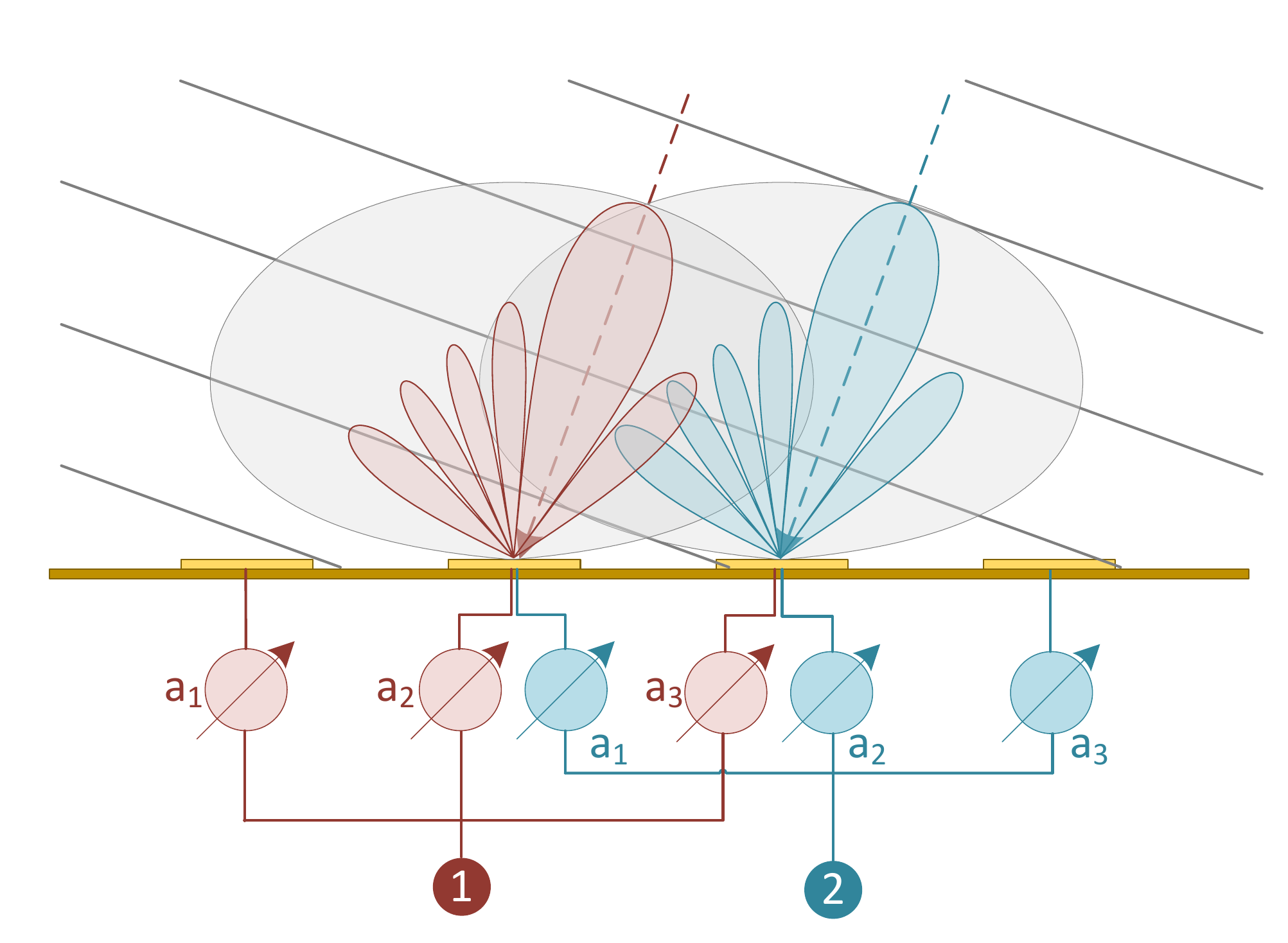}
    \caption{Emulating two shift-invariant radiation patterns for high resolution direction finding. The wider beams allow covering a wide angular sector, while the narrower beams reduce the angular field of view by spatial filtering.}
    \label{fig:shift_invariance}
\end{figure}

A simple example is given in \Cref{fig:shift_invariance}. There are two beams formed, both of which have the same gain pattern. The beamforming vector is effectively shifted by one antenna element, which emulates two identical, overlapping subarrays effectively shifted by one antenna element. This results in so-called translation-invariant radiation patterns, meaning that any impinging wave is weighted by the same gain pattern. However, as the two beams have different phase centers, they are very well suited for high-resolution \gls{doa} estimation. The well-known monopulse technique for high-resolution \gls{doa} estimation in radar is reminiscent of a simple replica of the same principle in the analog world. On a more recent time scale, the two arrays directly correspond to the shift-invariant pair of arrays that forms the basic principle of the \gls{esprit} algorithm. Obviously, the proposed approach can be further modified by increasing the number of subarrays and varying the phase center distances. Multiple overlapping shift-invariant subarrays would allow resolving multiple wave directions by subarray smoothing and not only increasing the accuracy of \gls{doa} estimation for one impinging path. This method can be considered as a hybrid implementation of the \gls{esprit} algorithm. It is limited to linear and planar arrays, which is not a serious drawback since integrated arrays primarily come as planar structures. Another advantage is that it does not require a specific array design with very sharp pencil beams. Since \gls{doa} estimation accuracy is gained by phase comparison, it is more important that the phase shifters are identical. 
Resolution of multiple paths within the same common beam is possible by exploiting the phase differences. On the other hand, beamforming can suppress waves from outside of the uniquely resolvable sector. This is indicated in \Cref{fig:shift_invariance} by the narrower beam. This would allow the distance between antenna elements to be extended beyond the half-lambda limit, which would increase estimation accuracy.

Even if the goal is to estimate the \gls{dod}/\gls{doa}   based on the maximum likelihood parameter estimation framework~\cite{ref64, ref67}, the radiation pattern design can be relaxed since the specific pattern is not as important as long as the pattern is known. Therefore, we need to know how the complex pattern transforms from the specified coefficients to the radiated field to enable the iterative cancellation of overlapping beam patterns. Thus, the focus shifts from optimizing quantized coefficients as in \cite{ref97} to calibrating the radiation pattern. Lastly, estimation of \gls{dod} requires identifying the waveforms transmitted via the different antenna ports at the receive (Rx) side. Therefore, some kind of orthogonalization in the frequency-time frame is necessary.

\textbf{Multidimensional estimation.} The multidimensionality aspect of parameter estimation refers to the bistatic propagation scenario, which includes the bidirectional geometry (in azimuth and elevation at Tx and Rx), \gls{tof} and Doppler for each propagation path, weighted with a $2 \times 2$ complex-valued polarimetric Jones matrix. Thus, the bidirectional estimation extends the \gls{tof}-Doppler estimation described in the subsection on Model-based Bistatic Target Parameter Estimation to a joint 6D estimation procedure. When performing \gls{doa}\/\,\gls{dod} estimations, impairments of antennas and antenna arrays can be of highest importance as they are most susceptible to RF effects. In this case, the device data model will be determined by calibration measurements of the installed radiation patterns. The traditional approach of model-based estimation decomposes into a local gradient-based and a global parameter optimization procedure with constraints. Actual research challenges address efficient real-time implementation in networks with distributed computational resources and AI support to multi-domain convex optimization with detection of relevant propagation paths. A related issue is in situ training, which effectively updates calibration throughout operation~\cite{ref67,schieler2025prop_est_cnn}. Finally, advanced AI-based methods can be used to adjust the model parameters, as this can be a challenging problem of non-convex optimization, which can become even more complex when device data from calibration is not perfect or not fully available from calibration measurements in reference setups~\cite{ref35}, but has to be learned implicitly by \gls{ml} in operational situations.

\section{Synchronization, Data Fusion, and Tracking in Distributed MS ICAS Networks}
\label{sec:synchronization}
Advanced \gls{msisac} is highly cooperative.
This begins with the bistatic sensing architecture, which is quite close to the basic radio access scheme of mobile communication.
It already has all the advantages of bistatic radar with respect to arget detection and identification.
We refer to this first level as (i) \gls{slc}, which refers to processes that can be performed by each sensor pair independently of each other. Although it does not involve the exchange of processed data with other sensor nodes, it is already cooperative, e.g., when the sensor receives measurement signals from several illuminating radio nodes, e.g., in the \gls{msisac} up-link mode.
The next level of cooperation is achieved when multiple sensing links share processed data to estimate and track the 3D dynamic target state vector of one or more targets.
This level already provides important benefits of \gls{msisac}, which include spatially smooth performance distribution by multisensor data fusion, multilink access coordination, and local radio resource allocation.
According to our initial discussion on \gls{msisac} architectures, the \gls{isac} nodes may belong to a distributed infrastructure as an \gls{ms} sensing cluster site, include mobile \glspl{ue} (in \gls{dl} or \gls{ul}), or belong to an independent, self-contained cloud of meshed radio nodes.
We refer to this level as (ii) Intra Site (or Intra Cloud) Cooperation (IntraSC).
The third level of cooperation is called (iii) Intersite (or Inter Cloud) Cooperation (InterSC).
It fuses the estimated target state vectors from different sensing sites or sensing clouds and can combine or fuse \gls{isac} data with other (including non \gls{isac}) information to produce a comprehensive situation report.
In the sequel, we will at first discuss some synchronization issues and then shortly refer to distributed data fusion and tracking in \gls{msisac}.

\subsection{Differential Coherent Operation in \gls{msisac} at Sensor Level}

At the sensor level, parameters are estimated that are closely related to the dynamic target state vector, such as time delay (rsp. range), Doppler frequency shift (rsp. relative speed), or directions (\gls{dod}, \gls{doa}).
While the latter in general requires more complicated RF interfaces (because of the antenna arrays), the former two require synchronization with respect to sampling and \gls{lo} frequencies offsets of the bistatic radio nodes~\cite{ref90}.
Array processing is easier in this respect as only local (or platform-related) coherent processing is necessary.
Therefore, distributed \gls{ms}-\gls{mimo} \gls{isac} is often claimed to have disadvantages due to a lack of synchronization between sensor nodes.
On closer inspection, however, many solutions can be found if we avoid excessive and unnecessary requirements.
As already explained above, \gls{ms}-\gls{mimo} \gls{isac} requires special efforts to establish sensor access coordination and synchronization.
However, since the Tx and Rx are connected as a communication transceiver pair in a multiuser or mesh communication network, the receiving radio nodes are already basically synchronized to the transmitter by using the Primary and Secondary Synchronization Signals (PSS, SSS)~\cite{ref91,ref92}.
Therefore, multisensor operation, \gls{ofdm} frame coordination, and carrier frequency adjustment are already accomplished according to the requirements for communication.
This ensures orthogonal, mutual interference-free medium access and synchronous \gls{fft} processing, which is a fundamental advantage for low variance system identification~\cite{ref30}.
This is further supported by the design of the \gls{ofdm} symbol, which is not chosen too long to avoid time selectivity and supplemented by a cyclic prefix.
For further synchronization update, one to four Demodulation and Phase Tracking Reference Signals (PT-RS and DM-RS) can be allocated per slot within the Physical Downlink\,/\,Uplink Shared Channel (PDSCH\,/\,PUSCH)~\cite{ref91,ref92}.

Although these timing and clock synchronization schemes are very flexible and powerful for communication even in situations with high Doppler shift, they seem not to be enough for \gls{isac} on the sensor level cooperation.
\gls{tof} and Doppler estimation call for sub-ns delay and precise carrier phase synchronization (the latter becoming more demanding the higher the frequency).
Fortunately, these problems can be mitigated by carrying out differential measurements.
A striking historical example is given by the well-known passive radar concept, which uses a ``transmitter of opportunity'' as an illuminator and a ``passive'' (receiving only) sensor.
\gls{pcl} does not presume any explicit signaling scheme for synchronization between Tx and Rx.
The bistatic \gls{etof} is estimated from the received communication signal, which is correlated against the \gls{los}-reference.
Classical passive radar receivers use two synchronous channels, an omnidirectional surveillance channel and a directive channel receiving the clean \gls{los} reference, which must be free of multipath.
In case of the more advanced \gls{cpcl}, the \gls{etof} and eDoppler can be actually calculated from the same \gls{cir} by the delay and frequency difference between \gls{los} and \gls{btp}, provided that the correlation reference is correctly reconstructed and \gls{los} and target peaks are clearly identified.
This eliminates the remaining Tx and Rx clock and \gls{lo} offset and even avoids drift and deviations between parallel receiver channels.
Another possibility for differential operation would be the calculation of \gls{tof} differences at a non-synchronized Rx if we have two precisely synchronized Tx (time difference of arrival estimation).
Finally, we can calculate differences between two synchronized receivers relative to a non-synchronized Tx.
In both cases, we have to make sure that at least parts of the distributed Tx or Rx \gls{isac} nodes are absolutely synchronized, e.g., as a part of the infrastructure. 

Another coherent receiver operation relates to accurate and high-resolution Doppler estimation.
This can again be referred to as a differential operation, as it is based on the relative phase between Tx and Rx.
Therefore, we need a stable \gls{los} between Tx and Rx, but coherency needs to be maintained only over a limited time interval (also called radar integration time), which may be as long as one \gls{ofdm} subframe or extend over multiple frames.
If the \gls{isac} radio nodes are stationary, the estimated target Doppler shift represents the bistatic target speed relative to the pair of nodes.
The same requirements apply to background subtraction, \gls{sar} imaging, and coherent averaging of the received signal.
The whole thing gets a little more complicated when the radio nodes are moving, since the frequency offset between Tx and Rx \gls{lo} and the Doppler shift between Tx and Rx cannot be distinguished.
However, if the total frequency shift (or phase drift) for the composite sum of all received paths is compensated, the estimated target-related Doppler corresponds directly to the eDoppler as described above in subsection Multistatic Dynamic Target State Vector Estimation.

The compensation procedure for \gls{los} phase drift requires some comments.
The correction should relate exclusively to the \gls{los} path and not be influenced by other multipaths.
If we have a sequence of transfer function estimates within the coherent processing interval, we can derive the \gls{los} phase slope along slow time from consecutively estimated \glspl{cir}.
Within the \gls{dft}\,/\,\gls{fft} framework, the same is achieved from the integral sum of samples of the frequency response function (FRF).
However, if there are multiple unresolved dominant reflections, the \gls{los} phase slope will be impaired.
In this case, a better estimate can be achieved by first applying a high-resolution, model-based delay-Doppler estimation approach to the \gls{los} path.
Since this implicitly assumes constant speed (a linear phase slope model), some smoothing operation is also included.
This may be desirable if the traveling radio nodes perform some wiggling movements that are not relevant for the global trajectory, particularly if it is a drone.
Similar is also important for hovering drones, which are assumed to be stationary on average. 

\subsection{ Absolute Synchronization in \gls{msisac} Networks }

At the higher levels of cooperation (may it be IntraSC or InterSC), a common time base and unique time stamps for absolute synchronization are necessary since the exchange of processed data takes place.
At these levels, position-related measurements such as delay (\gls{etof} or bistatic range), Doppler (eDoppler or bistatic velocity), and directional information (\gls{doa}, \gls{dod}), which are already generated at the lower \gls{slc} level, are fused.
Therefore, we no longer have coherency requirements on the delay and carrier phase level.
Instead, the remaining requirements arise from multilink access coordination and fusion, target dynamics, and possibly sensor node dynamics which is on a relaxed scale. 

Precise timing synchronization of distributed clocks is a well-known problem in the widespread field of communication, sensing, and control networks.
While GNSS (such as GPS) offers a medium accuracy service (some tens of ns) with global availability, other solutions for more precise synchronization have been proposed based on the IEEE1588 Precision Time Protocol.
The White Rabbit (WR) project has demonstrated even sub-ns accuracy synchronization in wired local networks using an Ethernet-based protocol~\cite{ref39} based on forward-backward transmission. 

The infrastructure of 5G networks is already very well synchronized.
According to 3GPP/TR 38.816, we can assume a time synchronization of $\pm$~\SI{130}{\ns} and frequency synchronization of \SI{0.05}{ppm}, achieved based on GNSS/GPS Primary Reference Time Clocks (PRTC) or Precision Time Protocol, IEEE 1588v2 (PTP).
This seems more than enough for data fusion in infrastructure-based sensing, where the different sensing sites are stationary (infrastructure only or \gls{dl}\,/\,\gls{ul} sensing relative to the infrastructure).
The situation changes a little when the synchronization master clock of a sensing site is moving, e.g., in the case of a meshed \gls{isac} established by a moving cloud of drones.
The stability of \gls{gpsdo} may degrade for several reasons. Since direct (differential) synchronization at the sensor level is still maintained for every sensing link through the correct use of the reference signal as described above, the time reference for IntraSC data fusion continues to pose no serious problem.
While there are a variety of robust \glspl{gpsdo} available on the market for use in dynamic situations, WR has also proven to be compatible with wireless transmission~\cite{ref40}.
This would, e.g., allow a sidelink-based meshed sensing network to be synchronized relative to the supervising network instance (a remote \gls{gnb}).
Another ad-hoc method for mutual synchronization of distributed radar nodes that does not require a centralized distribution of reference time is described in~\cite{ref43}.
This all clearly shows that besides GNSS, we have other approaches at hand that are capable of absolute time synchronization on the fusion level and partly on the PHY-Level.
These are also applicable when GNSS is not available or not reliable. 

As discussed in the subsection on Multi-Sensor Radio Access and Scheduling, we may have broadcast or orthogonal multisensor access.
While the broadcast mode allows simultaneous measurement of a subset of distributed \gls{ms}-\gls{mimo} \gls{isac} links (see \Cref{fig:dual_link_broadcast}), the orthogonal \gls{ms} access always deteriorates due to the limitations of the shared channel.
\gls{fdma} will reduce range resolution, while \gls{tdma} influences the availability of measurements in slow time.
While the influence on coherent processing (such as \gls{tof} or Doppler estimation) has already been discussed, non-isochronous measurements influence the correct assignment of multiple measurement data to the target trajectory.
Correct target location requires fusion of isochronous position-related data.
This may have several issues. Time alignment to achieve coincidence can be achieved by retroactive interpolation or, better, forward-looking by prediction on the parameter level.
Further reasons for data outages here and there can originate from failing target reflections or from target \gls{los} shadowing.
In any case, data fusion has to consider timing issues for correct alignment and for labeling the age of data, since the validity of the prediction decreases with time.  

\subsection{ Distributed Data Fusion, Functional Split, Tracking, and Sensor Mission Control}

Although data fusion and tracking are of paramount importance to the \gls{msisac} network, we can only provide very limited, general insights here.
Data fusion aims to combine related measurements to infer about the dynamic target state vector.
Tracking takes into account the temporal evolution of measurements and finishes estimation by smoothing and error correction.
The underlying network architecture suggests a multilevel approach for both fusion and tracking.
Although one of the main advantages of \gls{isac} is that it provides a network for transport and computing resources for data fusion, this raises several questions.
The transport links may have different capacity and latency performance depending on the available medium, which can range from a mobile or fixed wireless link to an optical RF over Fiber (RFoF) link with the highest performance.
There can also be significant differences in terms of computing resources, ranging from low performance in small drones to medium performance in cars up to high performance in base stations.
The Open RAN paradigm seems promising for \gls{isac} data transport and control flow routing within an infrastructure site due to its transparent and multi-vendor approach, which allows interoperability of components at different levels, also including specialized sensing hardware such as dedicated SDR-based sniffers or high-performance computing instances.
Sensing functional split should consider availability and allocation of resources for both communication and computation, see~\cite{ref41,ref66}.

The obvious, albeit rather conventional approach, is to detect the targets echos on the sensor level and transmit the preprocessed data set to the fusion center, where further fusion takes place.
This effectively establishes a multilayer approach for data fusion.
However, the hierarchy of fusion needs to be defined.
For example, we may have incomplete measurements at the lower level that nevertheless can contribute to the final result.
Another problem is data association, which means that we may not know (especially at the lower level) which measurements belong to the same target.
Eventually, the capacity and latency of the data links between the sensors and sensor clusters and the fusion may vary, which would have a significant impact on the functional split of data fusion across the different levels of the \gls{isac} network.
In distributed \gls{msisac} there are specific reasons to decide about the functional split of fusion on the sensor level (\gls{slc}) or on the network level (IntraSC or InterSC).
Local fusion on the sensor level enables early evaluation of the sensor performance, thereby supporting decisions regarding the relevance of the respective node and its data for the final fusion.
An overdetermined set of measurements (excessive number of links) can reduce variance, increase robustness to missing information, and help to sort out unreliable and probably wrong measurements.
In this way, the network can be relieved of the transmission and processing of irrelevant data to the fusion center.
Data reduction can have a decisive influence in cases of wireless transport links with their varying capacity and latency.
In addition to the impact on location accuracy (\gls{gdop}), the selection of sensors for fusion would also depend on the sensor configuration, such as available bandwidth, frequency, power, antenna, and array configuration, etc.
The immediate evaluation of local data can help in deciding on the allocation of local radio resources and on the sensor mission.
The latter seems to be most important in cases of movable, mobile, or otherwise reconfigurable \gls{isac} radio nodes to decide about the future sensor mission.
Moreover, control over local information to be shared can also support keeping privacy and the technical integrity of sensing entities. 

It is interesting to see that AI-based data fusion approaches also make a general distinction between early and late fusion, see~\cite{ref57}.
While early fusion, which is primarily sought after in deep learning, attempts to combine data at a raw data input level, late fusion begins combining data after independent processing on the sensor level.
Early fusion is often considered to be more powerful.
On the other hand, late fusion preserves more modularity and structural flexibility.
Deeper research into communication and computation resource constraints for early and late AI-based data fusion would be interesting.

Tracking accounts for the movement of targets relative to sensor nodes.
It establishes the temporal and spatial connection of hypothesized parameters.
Tracking takes advantage of target inertia for smoothing and prediction of noisy and otherwise erroneous estimates.
The Kalman paradigm~\cite{bar2004estimation} recursively updates the target state against disturbed observations.
The dynamic state vector can include a more or less constrained model in terms of 3D position, orientation, speed, acceleration, etc.
Target tracking helps to sort out wrong position hypotheses and infer about correct data association in multi-target tracking by \gls{mht} ~\cite{reid1979MultiHypothesisTracker} or \gls{jpda} filtering ~\cite{barshalom2009JPDAMultiTargetTrackin}.
Since tracking can already take place on the lower level with respect to position-related parameters like \gls{tof} of Doppler shift, it can help to pre-process data on the sensor level, contribute to data reduction, and help to predictively allocate resources on the sensor level.   

Fusion and tracking on the intersite level aims more at a global picture in a spacious area which is covered by multiple independent (perhaps not overlapping) sensor sites.
Cooperation at this level no longer involves direct sensor data fusion.
Instead, it involves the fusion of multiple local scenarios of dynamic target state vectors and trajectories.
Global perception and situation recognition are generated and made available to users, or even used by the \gls{isac} network for cognitive conclusions about actions to be taken.
The latter also opens doors for cognitive \gls{isac}.
This means that sensing targets are identified at the global scenario level, translated into local sensing missions, and then sent as requests to the lower levels of the multi-layer \gls{isac} network, where decisions about the allocation of resources are made.
Therefore, cross-layer signaling and cooperation between the sensing layers (rsp. levels) are important.   

The hierarchical concept of data fusion also supports division of work and interoperability between different operators if infrastructure related sensing is concerned (infrastructure-only or UL/DL sensing). Since a sensor cluster usually belongs to a base station, it is operated by a mobile network operator. This includes sensor configuration, maintenance, performance management, and data compression. Sensing results with well-defined quality can be further delivered via a customized interface to a higher level fusion center, e.g. to aggregate nation-wide situation maps or to monitor extended locations of critical infrastructure for safety critical application.

\section{Measured Example for Bistatic Joint Range-Doppler Estimation }
\label{sec:measurements}
\newcommand{\ovalcaptionlabel}[1]{%
  \tikz[baseline]{\node[draw, ellipse, thick, inner sep=1pt, anchor=base] {#1};}%
}

\begin{figure}
    \centering
    \includegraphics[width=\linewidth,bgcolor=gray!10,rndcorners=5,rndframe={color=gray!50, width=\fboxrule, sep=\fboxsep}{5}]{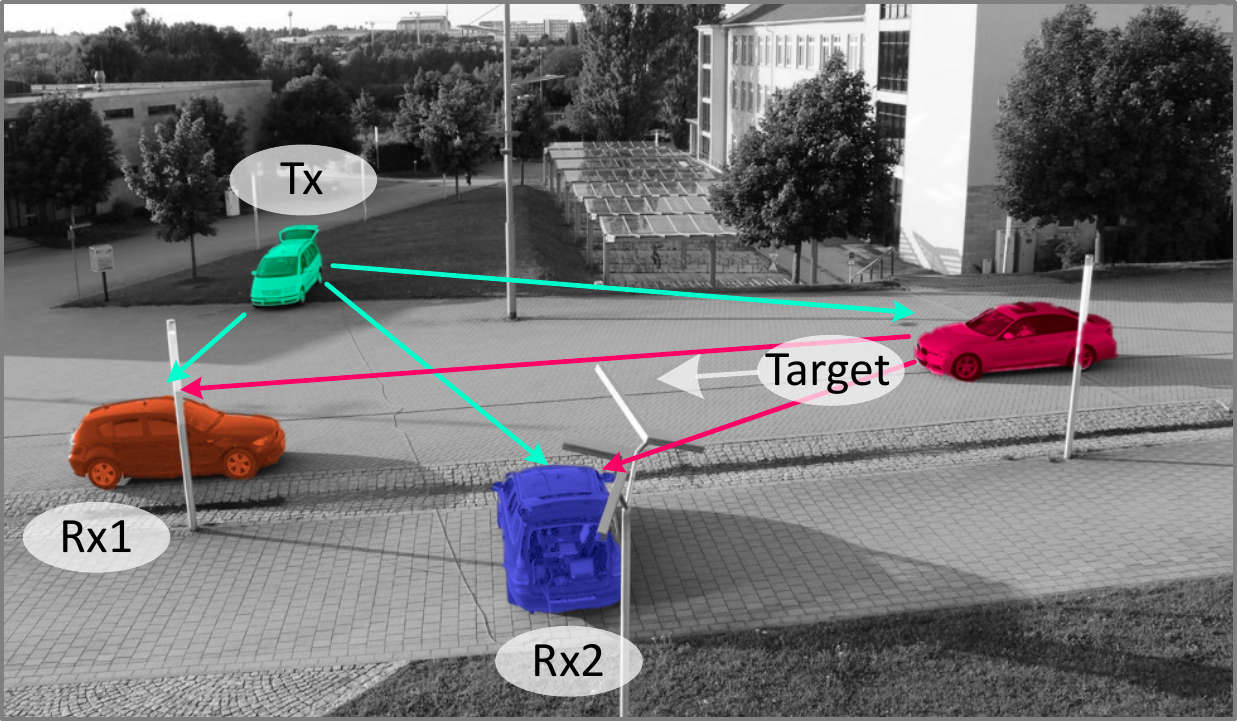}
    \caption{Dual link \gls{isac} demonstration setup. The radio nodes Rx1, Rx2, and Tx are stationary; the target car was moving.}
    \label{fig:meas-setup}
\end{figure}

\begin{figure}[t]
    \centering
    \includegraphics[width=\linewidth,trim={-10 -10 -10 0},bgcolor=gray!10,rndcorners=5,rndframe={color=gray!50, width=\fboxrule, sep=\fboxsep}{5}]{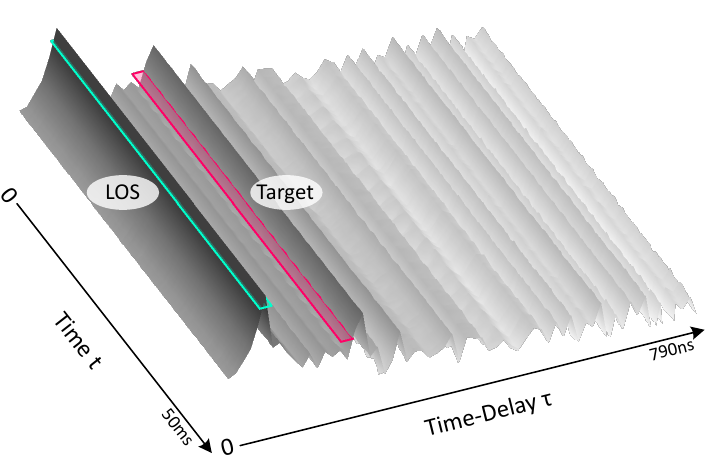}
    \caption{Received \gls{pdp} sequence at Rx1.}
    \label{fig:measRx1-1}
\end{figure}

Here we will shortly review a dual link \gls{isac} demonstration by a channel sounding experiment on the campus of TU Ilmenau.
The setup is shown in \Cref{fig:meas-setup}.
The \gls{isac} network consisted of three stationary radio nodes, two Rx and one Tx, mounted on three cars.
The target car was moving.
The \gls{isac} radio nodes were emulated by \gls{usrp} X310.
The \SI{33}{\dBm} output power transmit signal was a periodically repeated multisine at \SI{5.2}{\GHz}, bandwidth \SI{80}{\MHz}, period \SI{3.2}{\micro\second}, uniform carrier magnitude, phase optimized for minimum peak to average power ratio.
Since the multisine was periodically repeated, a cyclic prefix is not necessary to ensure leakage-free \gls{fft} processing.
The transmitted waveform was a priori known at the receivers for matched filter reception, respectively correlation, and the three radio nodes were synchronized by a \gls{gpsdo}.
The recorded \gls{pdp} (which is the magnitude squared \gls{cir}) at Rx1 is shown in \Cref{fig:measRx1-1} recorded along a slow time interval of $\approx$~\SI{50}{\ms}.
The \gls{pdp} sequence shows a stable, strong \gls{los} signal at \SI{63}{\ns}.
Another strong reflection is visible at $\approx$~\SI{200}{\ns}.
This is clutter and seems to result from scattering on one wall of the buildings on either side.
The target echo is almost invisible since it is masked by the clutter.

\begin{figure}[b]
    \centering
    \includegraphics[width=\linewidth,trim={-10 -10 -10 0},bgcolor=gray!10,rndcorners=5,rndframe={color=gray!50, width=\fboxrule, sep=\fboxsep}{5}]{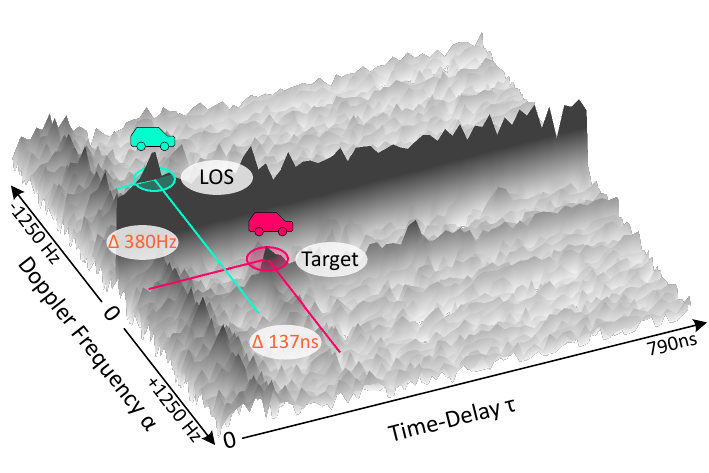}
    \caption{ Delay-Doppler spectrum with \gls{etof} and eDoppler parameter pair at Rx1.}
    \label{fig:measRx1-3}
\end{figure}

\begin{figure}[t]
    \centering
    \includegraphics[width=\linewidth,trim={-10 -10 -10 0},bgcolor=gray!10,rndcorners=5,rndframe={color=gray!50, width=\fboxrule, sep=\fboxsep}{5}]{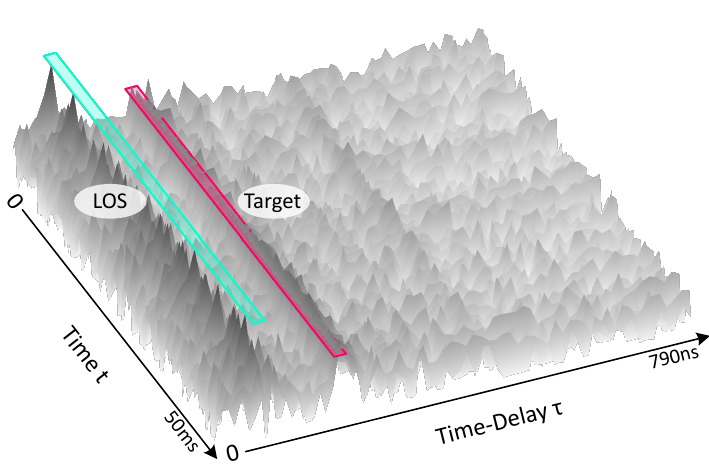}
    \caption  {Same \gls{pdp} sequence as in \Cref{fig:measRx1-1}, but with background subtraction.}
    \label{fig:measRx1-2}
\end{figure}

It only appears as a small ripple.
Even background subtraction cannot clearly reveal the target echo in the \gls{pdp}, as seen in \Cref{fig:measRx1-2}.
The \glspl{pdp} in \Cref{fig:measRx1-1} and \Cref{fig:measRx1-2} have the same scaling in \si{\dBm\per\ns}.
The location of both \gls{los} and the target echo is indicated by the two colored markers.
The two markers (turquoise-green and magenta) are at the same \si{\dBm\per\ns} level in both pictures.
The difference between them is about \SI{18}{\dB}.
The visibility of the target is enhanced, but still bad.

\begin{figure}[b]
    \centering
    \includegraphics[width=\linewidth,trim={-10 -10 -10 0},bgcolor=gray!10,rndcorners=5,rndframe={color=gray!50, width=\fboxrule, sep=\fboxsep}{5}]{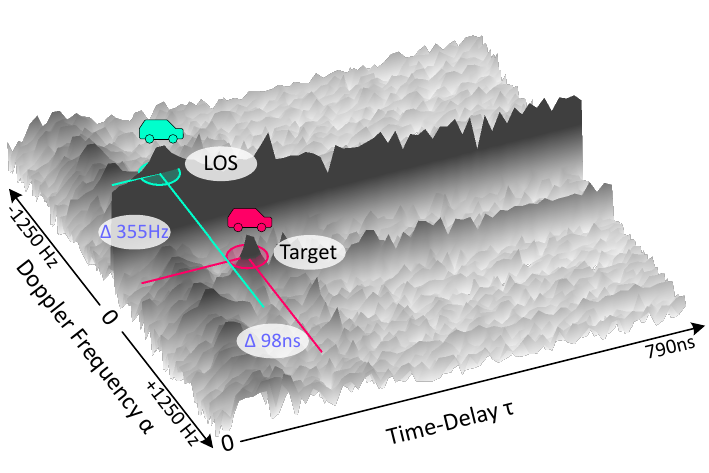}
    \caption{Delay-Doppler spectrum with eToF and eDoppler parameter pair at Rx2.}
    \label{fig:measRx2-3}
\end{figure}

\begin{figure}
    \centering
    \includegraphics[width=0.6\linewidth,bgcolor=gray!10,rndcorners=5,rndframe={color=gray!50, width=\fboxrule, sep=\fboxsep}{5}]{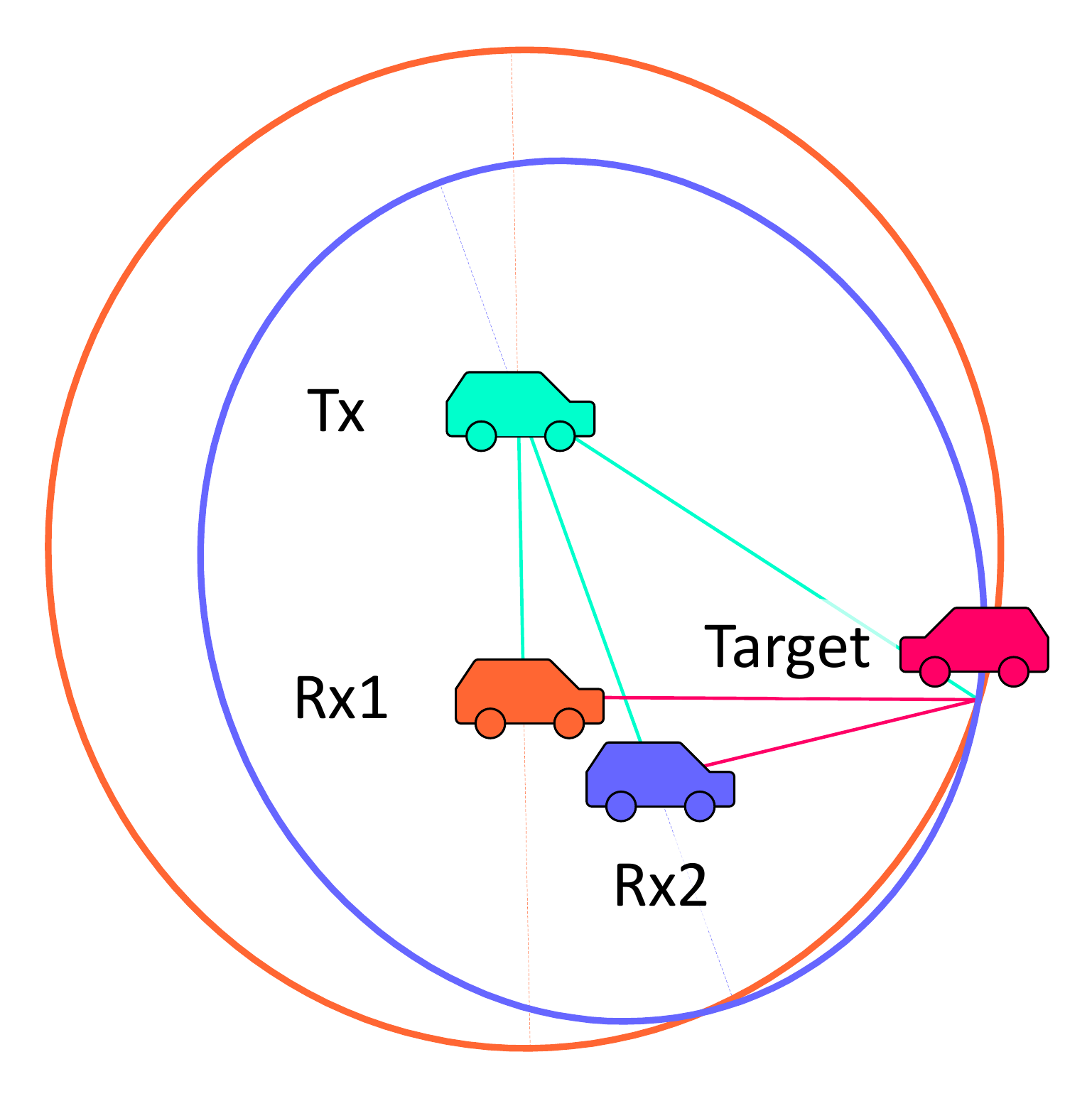}
    \caption{ Resulting pair of eToF ellipses for target car location estimation.}
    \label{fig:meas-ell}
\end{figure}

However, the target reveals very well in the delay-Doppler distribution (also known as the delay-Doppler spectrum or the scattering function), which is calculated by the magnitude squared result of another set of \glspl{fft} at every delay bin over a \SI{50}{\ms} interval of slow time according to the algorithm in \Cref{fig:ofdmproc}. 
The target echo is now very well separated by its Doppler shift, while the static contribution (\gls{los} and non-target related clutter) collapses at the zero Doppler intersection in the delay-Doppler map.
The target is clearly identified at \SI{137}{\ns} eToF and \SI{380}{\Hz} eDoppler (Rx1) and at \SI{98}{\ns} eToF and \SI{355}{\Hz} eDoppler (Rx2) in \Cref{fig:measRx1-3} and \Cref{fig:measRx2-3} respectively.
From these two measurements, two eToF ellipses are drawn to identify the position of the car in 2D.
The result is depicted in \Cref{fig:meas-ell}.
Unfortunately, the intersection is not unique, and the \gls{gdop} would be high.
This tells us that the positions of the sensor nodes in our example was not very appropriate.
In an operational situation, we would have to look for a better combination of nodes or observe and track the position to enhance the result.

%\setkeys{Gin}{transparent,transcolor=white}

\section{Outlook and Conclusion}
\label{sec:conclusion}

\gls{msisac} has been shown to be the most promising and powerful approach to integrate radar sensing functionality into next generation mobile radio networks. While multiple radio nodes are readily available in cellular \gls{mu} \gls{mimo} communication systems, \gls{msisac} also covers sidelink based mesh network approaches. However, use cases and sensing architectures (in terms of multi-sensor access and coordination, precoding, link adaptation, target parameter estimation and location tracking) will be very different if infrastructure only sensing, UL/DL sensing (with UE included as illuminator or sensor), or meshed sidelink sensing is pursued. Throughout this paper, we put a strong focus on the \gls{cpcl} principle, first proposed in~\cite{ref1}. Not only does \gls{cpcl} resemble and extend the classical passive radar paradigm, but it also appears to be pivotal in resolving bistatic synchronisation issues and enabling the proactive utilization of all mobile radio resources and functions for radar sensing and data processing.

With regard to the sensing function, \gls{msisac} constitutes a distributed \gls{mimo} radar. Multiple sensors bring the diversity advantage for more reliable target detection and better target recognition. We also gain the geometrical advantage for estimation and tracking of the 3D target state vector. \gls{msisac} has the additional benefit of reusing the ubiquitary existing, and constantly evolving resources of the mobile network. This gives us the performance gain in terms of coverage and resolution that can be achieved with the network densification and with other new 6G paradigms anyway (higher frequencies, wider bands, flexible radio access, precoding schemes, multiple antennas, including massive \gls{mimo} and \gls{ris}). In addition, heterogeneous radio interfaces and link adaptation tools, inherent data fusion and computation capabilities, and mobile sensor nodes that can be assigned a dedicated sensing mission, pave the way for a cognitive multifunction radar sensor network. We tried to unfold a comprehensive picture of \gls{msisac}. Because of the deep relation and the virtual equivalence of \gls{ms} \gls{mimo} \gls{isac} to \gls{mu} \gls{mimo} communications we firmly belief that the first will play a similar role to \gls{isac} as \gls{mu} \gls{mimo} did to mobile communication. 

However, we were unable to appreciate all the possible principles and advantages of \gls{msisac} on the same level.
For example, the discussion on the large-scale use of AI to control resources and enable powerful estimation methods that can cope with interference and proactively exploit multipath propagation would deserve deeper discussion.
In addition, many questions relating to multipath propagation were only addressed very briefly.
For instance, the topics of targets' bistatic reflectivity function measurements and modeling would need deeper discssion.
This includes micro Doppler properties of targets that may be important for target detection, tracking and recognition~\cite{ref3,ref69,ref70,ref77,ref98}. %ref71

From the perspective of resource efficiency, \gls{isac} has the advantage of reusing network resources and frequencies.
Bandwidth is a very scarce and valuable resource.
From a commercial perspective, communication-centric \gls{isac} (which is integrated into an existing communication network) naturally provides access to the many frequency bands that are already designated for mobile communications.
In contrast, it would be very difficult to obtain dedicated spectrum for ubiquitous radar sensor services, especially in the valuable and scarce lower frequency bands (FR1 and perhaps FR3).
However, the reuse of these communication bands has several consequences.
Sensing signal processing and target state vector estimation has to cope with the fragmented frequency bands.
Moreover, bandwidth efficiency and economic frequency management become serious concerns.
Obviously, excessive allocation of additional reference signals solely for sensing would be contradictory.
The reuse of the communication payload for radar sensing, closely linked to the \gls{cpcl} principle, seems to be the most efficient and economical solution. 

Cosidering future developments, we realize that the introduction of \gls{isac} will be a longer process accompanied by technical and economic decisions.
We will see these reflected in standardization processes.
Since there are reservations about technical and regulatory issues, suspected impairments of communication performance, missing business models and even privacy concerns, it would be an exaggeration to expect that the road towards a perfect and comprehensive \gls{isac} service would be a fast track.
However, there are also other examples (apart from pure voice telephony) in the long history of mobile communication that have finally caught on, albeit not in a straight line.  

We think that infrastructure-only sensing, which does not directly include a user terminal under contract to the \gls{mno} in the sensing service cycle, could be a pilot example promising the fastest route to economic success.
Reason is that \gls{isac} is promising and reasonable solutions for urgent needs in society, such as road and lower air traffic regulation, public security and safety, critical infrastructure protection, etc.
Especially proprietary campus networks with dedicated applications and specific economic conditions could be a great place for \gls{isac} use cases to grow. For public networks, operator models of ISAC based surveillance services to support public safety issues could be developed similar to other Mission Critical Services (MCX) in 5G. 

In the long term, the technical and economic advantages of the synergies between communication and sensor technology, the inherent resource efficiency, and the ubiquitous availability of the mobile network, which enables a comprehensive perception of a spatially extended scenario, will make \gls{isac} a success, enabling mobile network operators to offer radio sensor technology as a service at reasonable investment and operating costs.
%
%\section*{Declarations}
%
%\subsection*{Funding Information}
\section{Acknowledgments}
The authors acknowledge the financial support by the Federal Ministry of Research, Technology and Space of Germany in the projects ``Open6GHub'' (grant number: 16KISK015), ``6G-ICAS4Mobility'' (grant number: 16KISK241),  ``KOMSENS-6G'' (grant number: 16KISK125), and the Deutsche Forschungsgesellschaft (DFG) under the research projects JCRS CoMP with grant number. TH 494/35-1.
%
%\subsection*{Acknowledgments}
%n/a
%
%\subsection*{Competing Interests}
%n/a
%
%\subsection*{Author Contributions}
%
%\input{content/contributions}
%
\printbibliography

@INPROCEEDINGS{miranda2025joint_est_ofdma,
  author={Miranda, Marc and Semper, Sebastian and Schneider, Christian and Thomä, Reiner and Del Galdo, Giovanni},
  booktitle={2025 Joint European Conference on Networks and Communications \& 6G Summit (EuCNC/6G Summit)}, 
  title={Joint Delay-Doppler Estimation Using OFDMA Payloads for Integrated Communications and Sensing}, 
  year={2025},
  volume={},
  number={},
  pages={577-582},
  doi={10.1109/EuCNC/6GSummit63408.2025.11037009},
  ISSN={2575-4912},
  month={6},}

@INPROCEEDINGS{miranda2025mbpe_sparse_ofdma,
  author={Miranda, Marc and Semper, Sebastian and Schneider, Christian and Thomä, Reiner},
  booktitle={2025 19th European Conference on Antennas and Propagation (EuCAP)}, 
  title={Model-Based Joint Delay-Doppler Estimation for ICAS with Sparse OFDM Resources}, 
  year={2025},
  volume={},
  number={},
  pages={1-5},
  doi={10.23919/EuCAP63536.2025.10999889}
}

@article{schieler2025prop_est_cnn,
  title = {Wireless Propagation Parameter Estimation with Convolutional Neural Networks},
  ISSN = {1759-0795},
  doi = {10.1017/s1759078725000431},
  journal = {International Journal of Microwave and Wireless Technologies},
  publisher = {Cambridge University Press (CUP)},
  author = {Schieler,  Steffen and Semper,  Sebastian and Thomä,  Reiner},
  year = {2025},
  month = {4},
  pages = {1–8},
}

@book{richards2014RadarSignalProcessing, edition={2}, title={Fundamentals of Radar Signal Processing}, publisher={Mc Graw Hill Eductation}, author={Richards, Mark A.}, year={2014}}

@ARTICLE{Mura2025waveformDesignOFDM_ISAC,
  author={Mura, Silvia and Tagliaferri, Dario and Mizmizi, Marouan and Spagnolini, Umberto and Petropulu, Athina},
  journal={IEEE Transactions on Wireless Communications}, 
  title={Optimized Waveform Design for OFDM-based ISAC Systems Under Limited Resource Occupancy}, 
  year={2025},
  volume={},
  number={},
  pages={1-1},
  doi={10.1109/TWC.2025.3546475}}

@inproceedings{gedschold2025waveformDesignVerification,
  title = {Experimental Performance Validation of Fisher Information-Optimized Multicarrier Waveforms for Sub-THz Channel Sounding},
  url = {},
  DOI = {},
  booktitle = {2025 19th European Conference on Antennas and Propagation (EuCAP)},
  publisher = {IEEE},
  author = {Gedschold,  Jonas et al.},
  year = {2025},
  month = mar,
  pages = {1–5}
}

@article{Eijnde1991_SignalDesign,
    title = {On the Design of Optimal Excitation Signals},
    journal = {IFAC Proceedings Volumes},
    volume = {24},
    number = {3},
    pages = {1139-1144},
    year = {1991},
    note = {9th IFAC/IFORS Symposium on Identification and System Parameter Estimation 1991, Budapest, Hungary, 8-12 July 1991},
    issn = {1474-6670},
    doi = {https://doi.org/10.1016/S1474-6670(17)52503-3},
    author = {E. {Van den Eijnde} and J. Schoukens}
}

@book{stoica2005slepian,
  author    = {Stoica, Petre and Moses, Randolph L. and others},
  title     = {Spectral analysis of signals},
  date      = {2005},
  publisher = {Pearson Prentice Hall Upper Saddle River, NJ}
}

@inproceedings{javorszky1996,
author = {Jávorszky, GB and Boyd, Saxon and Kollár, István and Vandenberghe, L and Wu, SP},
booktitle = {8th IMEKO TC4 Symposium on Recent Advances in Electrical Measurements},
year = {1996},
month = {09},
pages = {192-197},
title = {Optimal Excitation Signal Design for Frequency Domain System Identification Using Semidefinite Programming}
}

@book{bar2004estimation,
  title={Estimation with Applications to Tracking and Navigation: Theory Algorithms and Software},
  author={Bar-Shalom, Y. and Li, X.R. and Kirubarajan, T.},
  isbn={9780471465218},
  url={https://books.google.de/books?id=xz9nQ4wdXG4C},
  year={2004},
  publisher={Wiley}
}

@ARTICLE{reid1979MultiHypothesisTracker,
  author={Reid, D.},
  journal={IEEE Transactions on Automatic Control}, 
  title={An algorithm for tracking multiple targets}, 
  year={1979},
  volume={24},
  number={6},
  pages={843-854},
  doi={10.1109/TAC.1979.1102177}}

@ARTICLE{barshalom2009JPDAMultiTargetTrackin,
  author={Bar-Shalom, Yaakov and Daum, Fred and Huang, Jim},
  journal={IEEE Control Systems Magazine}, 
  title={The probabilistic data association filter}, 
  year={2009},
  volume={29},
  number={6},
  pages={82-100},
  doi={10.1109/MCS.2009.934469}}

@ARTICLE{hoegbom1974clean,
    author = {{H{\"o}gbom}, J.~A.},
    title = "{Aperture Synthesis with a Non-Regular Distribution of Interferometer Baselines}",
    journal = {Astronomy and Astrophysics Supplement Series},
    year = 1974,
    month = jun,
    volume = {15},
    pages = {417},
    adsurl = {https://ui.adsabs.harvard.edu/abs/1974A&AS...15..417H}
}

@ARTICLE{BF1,
  author={Nguyen, Nhan Thanh and Nguyen, Ly V. and Shlezinger, Nir and Eldar, Yonina C. and Swindlehurst, A. Lee and Juntti, Markku},
  journal={IEEE Journal of Selected Topics in Signal Processing}, 
  title={Joint Communications and Sensing Hybrid Beamforming Design via Deep Unfolding}, 
  year={2024},
  volume={18},
  number={5},
  pages={901-916},
  doi={10.1109/JSTSP.2024.3463403}}

@inproceedings{EUMW,
  title = {Distributed ISAC Systems – Multisensor Radio Access and Coordination},
  url = {http://dx.doi.org/10.23919/EuRAD58043.2023.10289611},
  DOI = {10.23919/eurad58043.2023.10289611},
  booktitle = {2023 20th European Radar Conference (EuRAD)},
  publisher = {IEEE},
  author = {Thom\"{a},  Reiner and Dallmann,  Thomas},
  year = {2023},
  month = sep,
  pages = {351–354}
}

@article{A1,
  title = {Cooperative ISAC: An End-to-End Perspective},
  volume = {1},
  ISSN = {3066-2494},
  url = {http://dx.doi.org/10.1109/JSTEAP.2025.3603535},
  DOI = {10.1109/jsteap.2025.3603535},
  number = {1},
  journal = {IEEE Journal of Selected Topics in Electromagnetics,  Antennas and Propagation},
  publisher = {Institute of Electrical and Electronics Engineers (IEEE)},
  author = {I,  Chih-Lin and Han,  Zixiang and Xi,  Rongyan and Wang,  Tianxiong and Zhang,  Xiaozhou and Jiang,  Tao and Wang,  Sen and Zhao,  Hanting and Gui,  Xin and Jin,  Jing and Wang,  Qixing},
  year = {2025},
  month = sep,
  pages = {179–192}
}

@article{A2,
  title = {Network-Level ISAC Design: State-of-the-Art,  Challenges,  and Opportunities},
  volume = {1},
  ISSN = {3066-2494},
  url = {http://dx.doi.org/10.1109/JSTEAP.2025.3603139},
  DOI = {10.1109/jsteap.2025.3603139},
  number = {1},
  journal = {IEEE Journal of Selected Topics in Electromagnetics,  Antennas and Propagation},
  publisher = {Institute of Electrical and Electronics Engineers (IEEE)},
  author = {Han,  Kawon and Meng,  Kaitao and Wang,  Xiao-Yang and Masouros,  Christos},
  year = {2025},
  month = sep,
  pages = {65–83}
}

@inproceedings{A3,
  title = {Distributed Intelligent Integrated Sensing and Communications: The 6G-DISAC Approach},
  url = {http://dx.doi.org/10.1109/EuCNC/6GSummit60053.2024.10597016},
  DOI = {10.1109/eucnc/6gsummit60053.2024.10597016},
  booktitle = {2024 Joint European Conference on Networks and Communications; 6G Summit (EuCNC/6G Summit)},
  publisher = {IEEE},
  author = {Strinati,  Emilio Calvanese and Alexandropoulos,  George C. and Giyyarpuram,  Madhusudan and Sehier,  Philippe and Mekki,  Sami and Sciancalepore,  Vincenzo and Stark,  Maximilian and Sana,  Mohamed and Denis,  Benoit and Crozzoli,  Maurizio and Amani,  Navid and Mursia,  Placido and D’Errico,  Raffaele and Boldi,  Mauro and Costanzo,  Francesca and Rivet,  Francois and Wymeersch,  Henk},
  year = {2024},
  month = jun,
  pages = {392–397}
}

@article{ref0,
  title = {Joint Communication and Sensing Toward 6G: Models and Potential of Using MIMO},
  volume = {10},
  ISSN = {2372-2541},
  url = {http://dx.doi.org/10.1109/JIOT.2022.3227215},
  DOI = {10.1109/jiot.2022.3227215},
  number = {5},
  journal = {IEEE Internet of Things Journal},
  publisher = {Institute of Electrical and Electronics Engineers (IEEE)},
  author = {Fang,  Xinran and Feng,  Wei and Chen,  Yunfei and Ge,  Ning and Zhang,  Yan},
  year = {2023},
  month = mar,
  pages = {4093–4116}
}

@article{ref1,
  title = {Cooperative Passive Coherent Location: A Promising 5G Service to Support Road Safety},
  volume = {57},
  ISSN = {1558-1896},
  url = {http://dx.doi.org/10.1109/MCOM.001.1800242},
  DOI = {10.1109/mcom.001.1800242},
  number = {9},
  journal = {IEEE Communications Magazine},
  publisher = {Institute of Electrical and Electronics Engineers (IEEE)},
  author = {Thoma,  Reiner S. and Andrich,  Carsten and Galdo,  Giovanni Del and Dobereiner,  Michael and Hein,  Matthias A. and Kaske,  Martin and Schafer,  Gunter and Schieler,  Steffen and Schneider,  Christian and Schwind,  Andreas and Wendland,  Philip},
  year = {2019},
  month = sep,
  pages = {86–92}
}

@misc{ref3,
      title={Characterization of Multi-Link Propagation and Bistatic Target Reflectivity for Distributed Multi-Sensor ISAC}, 
      author={Reiner S. Thomä and Carsten Andrich and Julia Beuster and Heraldo Cesar Alves Costa and Sebastian Giehl and Saw James Myint and Christian Schneider and Gerd Sommerkorn},
      year={2023},
      eprint={2210.11840},
      archivePrefix={arXiv},
      primaryClass={eess.SP},
      url={https://arxiv.org/abs/2210.11840}, 
}

@inproceedings{ref9,
  title = {Joint Communication and Radar Sensing: An Overview},
  url = {http://dx.doi.org/10.23919/EuCAP51087.2021.9411178},
  DOI = {10.23919/eucap51087.2021.9411178},
  booktitle = {2021 15th European Conference on Antennas and Propagation (EuCAP)},
  publisher = {IEEE},
  author = {Thoma,  Reiner and Dallmann,  Thomas and Jovanoska,  Snezhana and Knott,  Peter and Schmeink,  Anke},
  year = {2021},
  month = mar,
  pages = {1–5}
}

@article{ref10,
  title = {MIMO Radar with Widely Separated Antennas},
  volume = {25},
  ISSN = {1053-5888},
  url = {http://dx.doi.org/10.1109/MSP.2008.4408448},
  DOI = {10.1109/msp.2008.4408448},
  number = {1},
  journal = {IEEE Signal Processing Magazine},
  publisher = {Institute of Electrical and Electronics Engineers (IEEE)},
  author = {Haimovich,  Alexander and Blum,  Rick and Cimini,  Leonard},
  year = {2008},
  pages = {116–129}
}

@article{ref10a,
  title = {MIMO Radar with Colocated Antennas},
  volume = {24},
  ISSN = {1053-5888},
  url = {http://dx.doi.org/10.1109/MSP.2007.904812},
  DOI = {10.1109/msp.2007.904812},
  number = {5},
  journal = {IEEE Signal Processing Magazine},
  publisher = {Institute of Electrical and Electronics Engineers (IEEE)},
  author = {Li,  Jian and Stoica,  Petre},
  year = {2007},
  month = sep,
  pages = {106–114}
}

@article{ref10b,
  title = {20 Years of MIMO Radar},
  volume = {39},
  ISSN = {1557-959X},
  url = {http://dx.doi.org/10.1109/MAES.2023.3349228},
  DOI = {10.1109/maes.2023.3349228},
  number = {3},
  journal = {IEEE Aerospace and Electronic Systems Magazine},
  publisher = {Institute of Electrical and Electronics Engineers (IEEE)},
  author = {Kalkan,  Yılmaz},
  year = {2024},
  month = mar,
  pages = {28–35}
}

@article{ref10c,
  title = {Spatial Diversity in Radars—Models and Detection Performance},
  volume = {54},
  ISSN = {1053-587X},
  url = {http://dx.doi.org/10.1109/TSP.2005.862813},
  DOI = {10.1109/tsp.2005.862813},
  number = {3},
  journal = {IEEE Transactions on Signal Processing},
  publisher = {Institute of Electrical and Electronics Engineers (IEEE)},
  author = {Fishler,  E. and Haimovich,  A. and Blum,  R.S. and Cimini,  L.J. and Chizhik,  D. and Valenzuela,  R.A.},
  year = {2006},
  month = mar,
  pages = {823–838}
}

@article{ref12,
  author       = {Thomä, R. and Dallmann, T. and Semper, S.},
  title        = {Distributed Multisensor ISAC},
  journal      = {ISAC-Focus Newsletter},
  year         = {2025},
  number       = {7},
  url          = {https://isac.committees.comsoc.org/publications/newletters/}
}

@inproceedings{ref19,
  title = {6G Integrated Sensing and Communication: From Vision to Realization},
  url = {http://dx.doi.org/10.23919/EuRAD58043.2023.10289474},
  DOI = {10.23919/eurad58043.2023.10289474},
  booktitle = {2023 20th European Radar Conference (EuRAD)},
  publisher = {IEEE},
  author = {Wild,  Thorsten and Grudnitsky,  Artjom and Mandelli,  Silvio and Henninger,  Marcus and Guan,  Junqing and Schaich,  Frank},
  year = {2023},
  month = sep 
}

@article{ref20,
  title = {Enhancing Direct Position Determination in Distributed Base Station Systems Through Time-Varying Quantization Design},
  volume = {24},
  ISSN = {2379-9153},
  url = {http://dx.doi.org/10.1109/JSEN.2024.3353804},
  DOI = {10.1109/jsen.2024.3353804},
  number = {5},
  journal = {IEEE Sensors Journal},
  publisher = {Institute of Electrical and Electronics Engineers (IEEE)},
  author = {Ni,  Lihua and Nyantakyi,  Isaac Osei and Liu,  Ning and Wan,  Qun},
  year = {2024},
  month = mar,
  pages = {6953–6963}
}

@article{ref21,
  title = {Distributed Base Station: A Concept System for Long-Range Broadband Wireless Access},
  volume = {10},
  ISSN = {2079-9292},
  url = {http://dx.doi.org/10.3390/electronics10192396},
  DOI = {10.3390/electronics10192396},
  number = {19},
  journal = {Electronics},
  publisher = {MDPI AG},
  author = {Gencel,  Muhammed Faruk and Eslami Rasekh,  Maryam and Madhow,  Upamanyu},
  year = {2021},
  month = sep,
  pages = {2396}
}

@article{ref22,
  title = {Coverage and capacity improvement of millimetre wave 5G network using distributed base station architecture},
  volume = {8},
  ISSN = {2047-4962},
  url = {http://dx.doi.org/10.1049/iet-net.2018.5059},
  DOI = {10.1049/iet-net.2018.5059},
  number = {4},
  journal = {IET Networks},
  publisher = {Institution of Engineering and Technology (IET)},
  author = {Al‐Falahy,  Naser and Alani,  Omar},
  year = {2019},
  month = jul,
  pages = {246–255}
}

@article{ref23,
  title = {Toward Software-Based,  MIMO,  Open-RAN PHY Architectures with Both Linear and Non-Linear Processing},
  volume = {62},
  ISSN = {1558-1896},
  url = {http://dx.doi.org/10.1109/MCOM.001.2300572},
  DOI = {10.1109/mcom.001.2300572},
  number = {8},
  journal = {IEEE Communications Magazine},
  publisher = {Institute of Electrical and Electronics Engineers (IEEE)},
  author = {Nikitopoulos,  Konstantinos and Katsaros,  George N. and Filo,  Marcin and Jayawardena,  Chathura and Tafazolli,  Rahim},
  year = {2024},
  month = aug,
  pages = {133–139}
}

@book{ref24,
    author = {Franchi, Norman and {Dressler et al.}, Falko},
    title = {German Perspective on 6G -- Use Cases, Technical Building Blocks and Requirements. Insights by the 6G Platform Germany},
    publisher = {FAU Studien aus der Elektrotechnik Band 28},
    year = 2024,
    doi = {10.25593/978-3-96147-811-8},
}

@techreport{ref24a,
  author = {{European Telecommunications Standards Institute}},
  title = {Group Report on Integrated Sensing And Communications (ISAC); Use Cases and Deployment Scenarios},
  institution = {ETSI},
  year = {2025},
  month = {03},
  number = {ETSI GR ISC 001 V1.1.1}
}

@article{ref25,
  title = {Integration of Sensing and Localization in V2X Sidelink Communications},
  volume = {62},
  ISSN = {1558-1896},
  url = {http://dx.doi.org/10.1109/MCOM.001.2300748},
  DOI = {10.1109/mcom.001.2300748},
  number = {8},
  journal = {IEEE Communications Magazine},
  publisher = {Institute of Electrical and Electronics Engineers (IEEE)},
  author = {Bartoletti,  Stefania and Decarli,  Nicolò and Masini,  Barbara M. and Giovannetti,  Caterina and Zanella,  Alberto and Bazzi,  Alessandro and Stirling-Gallacher,  Richard A.},
  year = {2024},
  month = aug,
  pages = {185–191}
}

@article{ref26,
  title = {Performance Analysis of Sidelink 5G-V2X Mode 2 Through an Open-Source Simulator},
  volume = {9},
  ISSN = {2169-3536},
  url = {http://dx.doi.org/10.1109/ACCESS.2021.3121151},
  DOI = {10.1109/access.2021.3121151},
  journal = {IEEE Access},
  publisher = {Institute of Electrical and Electronics Engineers (IEEE)},
  author = {Todisco,  Vittorio and Bartoletti,  Stefania and Campolo,  Claudia and Molinaro,  Antonella and Berthet,  Antoine O. and Bazzi,  Alessandro},
  year = {2021},
  pages = {145648–145661}
}

@article{ref27,
  title = {NR Sidelink Performance Evaluation for Enhanced 5G-V2X Services},
  volume = {5},
  ISSN = {2624-8921},
  url = {http://dx.doi.org/10.3390/vehicles5040092},
  DOI = {10.3390/vehicles5040092},
  number = {4},
  journal = {Vehicles},
  publisher = {MDPI AG},
  author = {Tabassum,  Mehnaz and Bastos,  Felipe Henrique and Oliveira,  Aurenice and Klautau,  Aldebaro},
  year = {2023},
  month = nov,
  pages = {1692–1706}
}

@book{ref28,
  editor = {Fan Liu, Christos Masouros, Yonina C. Eldar},
  title = {Integrated Sensing and Communications},
  ISBN = {9789819925018},
  url = {http://dx.doi.org/10.1007/978-981-99-2501-8},
  DOI = {10.1007/978-981-99-2501-8},
  publisher = {Springer Nature Singapore},
  year = {2023}
}

@book{ref29,
  title={An Introduction to Passive Radar, Second Edition},
  author={Griffiths, H.D. and Baker, C.J.},
  isbn={9781630818418},
  series={Artech House radar Series},
  url={https://books.google.de/books?id=6uF8EAAAQBAJ},
  year={2022},
  publisher={Artech House}
}

@BOOK{ref30,
  title     = "System identification: A Frequency Domain Approach",
  author    = "Pintelon, Rik and Schoukens, Johan",
  isbn      = {9780470640371},
  publisher = "Wiley-Blackwell",
  edition   =  2,
  month     =  mar,
  year      =  2012,
  address   = "Hoboken, NJ",
}

@article{ref31,
  title = {Shifting the MIMO Paradigm},
  volume = {24},
  ISSN = {1053-5888},
  url = {http://dx.doi.org/10.1109/MSP.2007.904815},
  DOI = {10.1109/msp.2007.904815},
  number = {5},
  journal = {IEEE Signal Processing Magazine},
  publisher = {Institute of Electrical and Electronics Engineers (IEEE)},
  author = {Gesbert,  David and Kountouris,  Marios and Heath,  Robert W. and Chae,  Chan-byoung and Salzer,  Thomas},
  year = {2007},
  month = sep,
  pages = {36–46}
}

@article{ref32,
  title = {Performance of MIMO Radar With Angular Diversity Under Swerling Scattering Models},
  volume = {4},
  ISSN = {1941-0484},
  url = {http://dx.doi.org/10.1109/JSTSP.2009.2038971},
  DOI = {10.1109/jstsp.2009.2038971},
  number = {1},
  journal = {IEEE Journal of Selected Topics in Signal Processing},
  publisher = {Institute of Electrical and Electronics Engineers (IEEE)},
  author = {Aittomaki,  T. and Koivunen,  V.},
  year = {2010},
  month = feb,
  pages = {101–114}
}

@article{ref33,
  title = {Coordinated multipoint: Concepts,  performance,  and field trial results},
  volume = {49},
  ISSN = {1558-1896},
  url = {http://dx.doi.org/10.1109/MCOM.2011.5706317},
  DOI = {10.1109/mcom.2011.5706317},
  number = {2},
  journal = {IEEE Communications Magazine},
  publisher = {Institute of Electrical and Electronics Engineers (IEEE)},
  author = {Irmer,  Ralf and Droste,  Heinz and Marsch,  Patrick and Grieger,  Michael and Fettweis,  Gerhard and Brueck,  Stefan and Mayer,  Hans-Peter and Thiele,  Lars and Jungnickel,  Volker},
  year = {2011},
  month = feb,
  pages = {102–111}
}

@article{ref34,
  title = {Time Reversal in Multiple-Input Multiple-Output Radar},
  volume = {4},
  ISSN = {1941-0484},
  url = {http://dx.doi.org/10.1109/JSTSP.2009.2038983},
  DOI = {10.1109/jstsp.2009.2038983},
  number = {1},
  journal = {IEEE Journal of Selected Topics in Signal Processing},
  publisher = {Institute of Electrical and Electronics Engineers (IEEE)},
  author = {Jin,  Yuanwei and Moura,  JosÉ M. F. and O’Donoughue,  Nicholas},
  year = {2010},
  month = feb,
  pages = {210–225}
}

@article{ref35,
  title = {Impact of Incomplete and Inaccurate Data Models on High Resolution Parameter Estimation in Multidimensional Channel Sounding},
  volume = {60},
  ISSN = {1558-2221},
  url = {http://dx.doi.org/10.1109/TAP.2011.2173446},
  DOI = {10.1109/tap.2011.2173446},
  number = {2},
  journal = {IEEE Transactions on Antennas and Propagation},
  publisher = {Institute of Electrical and Electronics Engineers (IEEE)},
  author = {Landmann,  Markus and Kaske,  Martin and Thoma,  Reiner S.},
  year = {2012},
  month = feb,
  pages = {557–573}
}

@inproceedings{ref36,
  title = {Excitation Signal Design for THz Channel Sounding and Propagation Parameter Estimation},
  url = {http://dx.doi.org/10.23919/EuCAP60739.2024.10501423},
  DOI = {10.23919/eucap60739.2024.10501423},
  booktitle = {2024 18th European Conference on Antennas and Propagation (EuCAP)},
  publisher = {IEEE},
  author = {Gedschold,  Jonas and Semper,  Sebastian and D\"{o}bereiner,  Michael and Thom\"{a},  Reiner S.},
  year = {2024},
  month = mar,
  pages = {1–5}
}

@article{ref37,
  title = {Channel Estimation and Equalization in Multiuser Uplink OFDMA and SC-FDMA Systems Under Transmitter RF Impairments},
  volume = {65},
  ISSN = {1939-9359},
  url = {http://dx.doi.org/10.1109/TVT.2015.2397277},
  DOI = {10.1109/tvt.2015.2397277},
  number = {1},
  journal = {IEEE Transactions on Vehicular Technology},
  publisher = {Institute of Electrical and Electronics Engineers (IEEE)},
  author = {Kiayani,  Adnan and Anttila,  Lauri and Zou,  Yaning and Valkama,  Mikko},
  year = {2016},
  month = jan,
  pages = {82–99}
}

@article{ref38,
  title = {Reference Signal Fractionation for LTE-Based Passive Radar Facing MIMO and OFDMA Challenges},
  volume = {24},
  ISSN = {2379-9153},
  url = {http://dx.doi.org/10.1109/JSEN.2024.3371529},
  DOI = {10.1109/jsen.2024.3371529},
  number = {8},
  journal = {IEEE Sensors Journal},
  publisher = {Institute of Electrical and Electronics Engineers (IEEE)},
  author = {Dan,  Yangpeng and Xu,  Kun and Yi,  Jianxin and Wan,  Xianrong},
  year = {2024},
  month = apr,
  pages = {12904–12916}
}

@inproceedings{ref39,
  title = {White rabbit: Sub-nanosecond timing distribution over ethernet},
  url = {http://dx.doi.org/10.1109/ISPCS.2009.5340196},
  DOI = {10.1109/ispcs.2009.5340196},
  booktitle = {2009 International Symposium on Precision Clock Synchronization for Measurement,  Control and Communication},
  publisher = {IEEE},
  author = {Moreira,  Pedro and Serrano,  Javier and Wlostowski,  Tomasz and Loschmidt,  Patrick and Gaderer,  Georg},
  year = {2009},
  month = oct 
}

@article{ref40,
  title = {White Rabbit Time and Frequency Transfer Over Wireless Millimeter-Wave Carriers},
  volume = {67},
  ISSN = {1525-8955},
  url = {http://dx.doi.org/10.1109/TUFFC.2020.2989667},
  DOI = {10.1109/tuffc.2020.2989667},
  number = {9},
  journal = {IEEE Transactions on Ultrasonics,  Ferroelectrics,  and Frequency Control},
  publisher = {Institute of Electrical and Electronics Engineers (IEEE)},
  author = {Gilligan,  Jane E. and Konitzer,  Eric M. and Siman-Tov,  Elad and Zobel,  Justin W. and Adles,  Eric J.},
  year = {2020},
  month = sep,
  pages = {1946–1952}
}

@inproceedings{ref41,
  title = {Redefining Next Generation Fronthaul for the Interplay Between Communication and Sensing Data},
  url = {http://dx.doi.org/10.1109/ICTON62926.2024.10647261},
  DOI = {10.1109/icton62926.2024.10647261},
  booktitle = {2024 24th International Conference on Transparent Optical Networks (ICTON)},
  publisher = {IEEE},
  author = {Gelabert,  Xavier and Klaiqi,  Bleron and Busquets,  Noè Bernadas i},
  year = {2024},
  month = jul,
  pages = {1–1}
}

@article{ref43,
  title = {Accuracy study for over-the-air frequency synchronization of continuous wave signals},
  ISSN = {1759-0795},
  url = {http://dx.doi.org/10.1017/S1759078724001090},
  DOI = {10.1017/s1759078724001090},
  journal = {International Journal of Microwave and Wireless Technologies},
  publisher = {Cambridge University Press (CUP)},
  author = {Dallmann,  Thomas and Thom\"{a},  Reiner},
  year = {2024},
  month = nov,
  pages = {1–7}
}

@article{ref44,
  title = {The Development From Adaptive to Cognitive Radar Resource Management},
  volume = {35},
  ISSN = {1557-959X},
  url = {http://dx.doi.org/10.1109/MAES.2019.2957847},
  DOI = {10.1109/maes.2019.2957847},
  number = {6},
  journal = {IEEE Aerospace and Electronic Systems Magazine},
  publisher = {Institute of Electrical and Electronics Engineers (IEEE)},
  author = {Charlish,  Alexander and Hoffmann,  Folker and Degen,  Christoph and Schlangen,  Isabel},
  year = {2020},
  month = jun,
  pages = {8–19}
}

@inproceedings{ref45,
  title = {Fully Adaptive Resource Management in Radar Networks},
  url = {http://dx.doi.org/10.1109/RadarConf2043947.2020.9266440},
  DOI = {10.1109/radarconf2043947.2020.9266440},
  booktitle = {2020 IEEE Radar Conference (RadarConf20)},
  publisher = {IEEE},
  author = {Oechslin,  Roland and Wieland,  Sebastian and Zutter,  Andreas and Aulenbacher,  Uwe and Wellig,  Peter},
  year = {2020},
  month = sep,
  pages = {1–6}
}

@inproceedings{ref46,
  title = {Adaptive Radar Resource Management for All-Digital Multi-Function Phased Array Radar Using Proximal Policy Optimization},
  url = {http://dx.doi.org/10.1109/RadarConf2458775.2024.10548047},
  DOI = {10.1109/radarconf2458775.2024.10548047},
  booktitle = {2024 IEEE Radar Conference (RadarConf24)},
  publisher = {IEEE},
  author = {Witherell,  Brianna and Yu,  Tian-You and Schvartzman,  David and Goodman,  Nathan and Stringer,  Alexander and Dolinger,  Geoffrey},
  year = {2024},
  month = may,
  pages = {1–6}
}

@article{ref47,
  title = {Emerging Tools for Link Adaptation on 5G NR and Beyond: Challenges and Opportunities},
  volume = {9},
  ISSN = {2169-3536},
  url = {http://dx.doi.org/10.1109/ACCESS.2021.3111783},
  DOI = {10.1109/access.2021.3111783},
  journal = {IEEE Access},
  publisher = {Institute of Electrical and Electronics Engineers (IEEE)},
  author = {Martin-Vega,  Francisco J. and Ruiz-Sicilia,  Juan Carlos and Aguayo,  Mari Carmen and Gomez,  Gerardo},
  year = {2021},
  pages = {126976–126987}
}

@inproceedings{ref48,
  title = {Advanced cognitive networked radar surveillance},
  url = {http://dx.doi.org/10.1109/RadarConf2147009.2021.9455245},
  DOI = {10.1109/radarconf2147009.2021.9455245},
  booktitle = {2021 IEEE Radar Conference (RadarConf21)},
  publisher = {IEEE},
  author = {Jahangir,  Mohammed and Baker,  Chris J and Antoniou,  Michail and Griffin,  Benjamin and Balleri,  Alessio and Money,  David and Harman,  Stephen},
  year = {2021},
  month = may,
  pages = {1–6}
}

@inproceedings{ref49,
  title = {Multistatic and Networked Radar: Principles and Practice},
  url = {http://dx.doi.org/10.1109/RadarConf2147009.2021.9455149},
  DOI = {10.1109/radarconf2147009.2021.9455149},
  booktitle = {2021 IEEE Radar Conference (RadarConf21)},
  publisher = {IEEE},
  author = {Griffiths,  Hugh and Farina,  Alfonso},
  year = {2021},
  month = may,
  pages = {1–5}
}

@INPROCEEDINGS{ref50,
  author={Alaee-Kerharoodi, Mohammad and Bhavani, Shankar M. R. and Mishra, Kumar Vijay and Ottersten, Björn},
  booktitle={ICASSP 2020 - 2020 IEEE International Conference on Acoustics, Speech and Signal Processing (ICASSP)}, 
  title={Information Theoretic Approach for Waveform Design in Coexisting MIMO Radar and MIMO Communications}, 
  year={2020},
  volume={},
  number={},
  pages={1-5},
  doi={10.1109/ICASSP40776.2020.9053048}}

@article{ref51,
  title = {State Space Initiation for Blind Mobile Terminal Position Tracking},
  volume = {2008},
  ISSN = {1687-6180},
  url = {http://dx.doi.org/10.1155/2008/394219},
  DOI = {10.1155/2008/394219},
  number = {1},
  journal = {EURASIP Journal on Advances in Signal Processing},
  publisher = {Springer Science and Business Media LLC},
  author = {Algeier,  Vadim and Demissie,  Bruno and Koch,  Wolfgang and Thom\"{a},  Reiner},
  year = {2007},
  month = oct 
}

@article{ref55,
  title = {High-Accuracy Localization for Assisted Living: 5G systems will turn multipath channels from foe to friend},
  volume = {33},
  ISSN = {1053-5888},
  url = {http://dx.doi.org/10.1109/MSP.2015.2504328},
  DOI = {10.1109/msp.2015.2504328},
  number = {2},
  journal = {IEEE Signal Processing Magazine},
  publisher = {Institute of Electrical and Electronics Engineers (IEEE)},
  author = {Witrisal,  Klaus and Meissner,  Paul and Leitinger,  Erik and Shen,  Yuan and Gustafson,  Carl and Tufvesson,  Fredrik and Haneda,  Katsuyuki and Dardari,  Davide and Molisch,  Andreas F. and Conti,  Andrea and Win,  Moe Z.},
  year = {2016},
  month = mar,
  pages = {59–70}
}

@article{ref56,
  title = {Applying Random Forest and Multipath Fingerprints to Enhance TDOA Localization Systems},
  volume = {18},
  ISSN = {1548-5757},
  url = {http://dx.doi.org/10.1109/LAWP.2019.2934466},
  DOI = {10.1109/lawp.2019.2934466},
  number = {11},
  journal = {IEEE Antennas and Wireless Propagation Letters},
  publisher = {Institute of Electrical and Electronics Engineers (IEEE)},
  author = {de Sousa,  Marcelo Nogueira and Thoma,  Reiner S.},
  year = {2019},
  month = nov,
  pages = {2316–2320}
}

@article{ref57,
  title = {Deep Learning-Based Positioning With Multi-Task Learning and Uncertainty-Based Fusion},
  volume = {2},
  ISSN = {2831-316X},
  url = {http://dx.doi.org/10.1109/TMLCN.2024.3441521},
  DOI = {10.1109/tmlcn.2024.3441521},
  journal = {IEEE Transactions on Machine Learning in Communications and Networking},
  publisher = {Institute of Electrical and Electronics Engineers (IEEE)},
  author = {Foliadis,  Anastasios and Castañeda Garcia,  Mario H. and Stirling-Gallacher,  Richard A. and Thom\"{a},  Reiner S.},
  year = {2024},
  pages = {1127–1141}
}

@INPROCEEDINGS{ref58,
  author={Zea, Antonio and Faion, Florian and Hanebeck, Uwe D.},
  booktitle={2015 18th International Conference on Information Fusion (Fusion)}, 
  title={Exploiting clutter: Negative information for enhanced extended object tracking}, 
  year={2015},
  volume={},
  number={},
  pages={1030-1037},
  doi={}}

@ARTICLE{ref59,
  author={Hirsch, Ole and Janson, Malgorzata and Wiesbeck, Werner and Thomä, Reiner S.},
  journal={IEEE Transactions on Instrumentation and Measurement}, 
  title={Indirect Localization and Imaging of Objects in an UWB Sensor Network}, 
  year={2010},
  volume={59},
  number={11},
  pages={2949-2957},
  doi={10.1109/TIM.2010.2046359}}

@inproceedings{ref60,
  title = {Forward scatter radar detection},
  url = {http://dx.doi.org/10.1109/RADAR.2002.1174649},
  DOI = {10.1109/radar.2002.1174649},
  booktitle = {RADAR 2002},
  publisher = {IEE},
  author = {Gould,  D.M. and Orton,  R.S. and Pollard,  R.J.E.},
  pages = {36–40}
}

@article{ref63,
  title = {Measurement of random time-variant linear channels},
  volume = {15},
  ISSN = {0018-9448},
  url = {http://dx.doi.org/10.1109/TIT.1969.1054332},
  DOI = {10.1109/tit.1969.1054332},
  number = {4},
  journal = {IEEE Transactions on Information Theory},
  publisher = {Institute of Electrical and Electronics Engineers (IEEE)},
  author = {Bello,  P.},
  year = {1969},
  month = jul,
  pages = {469–475}
}

@book{ref64,
  title={Estimation of Radio Channel Parameters: Models and Algorithms},
  author={Richter, A.},
  isbn={9783938843024},
  url={https://books.google.de/books?id=XZEVMQAACAAJ},
  year={2005},
  publisher={ISLE}
}

@ARTICLE{ref66,
  author={Semper, Sebastian and Naviliat, Joël and Gedschold, Jonas and Döbereiner, Michael and Schieler, Steffen and Thomä, Reiner S.},
  journal={IEEE Open Journal of the Communications Society}, 
  title={Distributed Computing and Model-Based Estimation for Integrated Communications and Sensing: A Roadmap}, 
  year={2024},
  volume={5},
  number={},
  pages={6279-6290},
  doi={10.1109/OJCOMS.2024.3467683}}

@article{ref67,
  title = {High-Resolution Parameter Estimation for Wideband Radio Channel Sounding},
  volume = {71},
  ISSN = {1558-2221},
  url = {http://dx.doi.org/10.1109/TAP.2023.3286024},
  DOI = {10.1109/tap.2023.3286024},
  number = {8},
  journal = {IEEE Transactions on Antennas and Propagation},
  publisher = {Institute of Electrical and Electronics Engineers (IEEE)},
  author = {Semper,  Sebastian and D\"{o}bereiner,  Michael and Steinmetz,  Christian and Landmann,  Markus and Thom\"{a},  Reiner S.},
  year = {2023},
  month = aug,
  pages = {6728–6743}
}

@article{ref68,
  title = {60 GHz Ultrawideband Polarimetric MIMO Sensing for Wireless Multi-Gigabit and Radar},
  volume = {61},
  ISSN = {1558-2221},
  url = {http://dx.doi.org/10.1109/TAP.2013.2243398},
  DOI = {10.1109/tap.2013.2243398},
  number = {4},
  journal = {IEEE Transactions on Antennas and Propagation},
  publisher = {Institute of Electrical and Electronics Engineers (IEEE)},
  author = {Garcia Ariza,  Alexis Paolo and Muller,  Robert and Wollenschlager,  Frank and Schulz,  Alexander and Elkhouly,  Mohamed and Sun,  Yaoming and Glisic,  Srdjan and Trautwein,  Uwe and Stephan,  Ralf and Muller,  Jens and Thoma,  Reiner S. and Hein,  MatthiasA.},
  year = {2013},
  month = apr,
  pages = {1631–1641}
}

@INPROCEEDINGS{ref69,
  author={Alves Costa, Heraldo Cesar and James Myint, Saw and Andrich, Carsten and Giehl, Sebastian W. and Novotny, Dieter and Beuster, Julia and Schneider, Christian and Thomä, Reiner S.},
  booktitle={2025 16th German Microwave Conference (GeMiC)}, 
  title={Bistatic Micro-Doppler Analysis of a Vertical Takeoff and Landing (VTOL) Drone in ICAS Framework}, 
  year={2025},
  volume={},
  number={},
  pages={506-509},
  doi={10.23919/GeMiC64734.2025.10979139}
}

@inproceedings{ref70,
  title = {Modelling Micro-Doppler Signature of Drone Propellers in Distributed ISAC},
  url = {http://dx.doi.org/10.1109/RadarConf2458775.2024.10548468},
  DOI = {10.1109/radarconf2458775.2024.10548468},
  booktitle = {2024 IEEE Radar Conference (RadarConf24)},
  publisher = {IEEE},
  author = {Costa,  Heraldo Cesar Alves and Myint,  Saw James and Andrich,  Carsten and Giehl,  Sebastian W. and Schneider,  Christian and Thom\"{a},  Reiner S.},
  year = {2024},
  month = may,
  pages = {1–6}
}

@article{ref74,
  title = {5G PRS-Based Sensing: A Sensing Reference Signal Approach for Joint Sensing and Communication System},
  volume = {72},
  ISSN = {1939-9359},
  url = {http://dx.doi.org/10.1109/TVT.2022.3215159},
  DOI = {10.1109/tvt.2022.3215159},
  number = {3},
  journal = {IEEE Transactions on Vehicular Technology},
  publisher = {Institute of Electrical and Electronics Engineers (IEEE)},
  author = {Wei,  Zhiqing and Wang,  Yuan and Ma,  Liang and Yang,  Shaoshi and Feng,  Zhiyong and Pan,  Chengkang and Zhang,  Qixun and Wang,  Yajuan and Wu,  Huici and Zhang,  Ping},
  year = {2023},
  month = mar,
  pages = {3250–3263}
}

@phdthesis{ref77,
  doi = {10.22032/DBT.49241},
  url = {https://www.db-thueringen.de/receive/dbt_mods_00049241},
  author = {Nogueira de Sousa,  Marcelo},
  language = {en},
  title = {Multipath Exploitation for Emitter Localization using Ray-Tracing Fingerprints and Machine Learning},
  publisher = {Universit\"{a}tsverlag Ilmenau},
  year = {2021},
  copyright = {all rights reserved},
  school = {TU Ilmenau},
}

@article{ref83,
  title = {Sensing-Aided Beamforming: The Impact of Distributed Sensing Network Geometry},
  url = {http://dx.doi.org/10.36227/techrxiv.175687316.68599313/v1},
  DOI = {10.36227/techrxiv.175687316.68599313/v1},
  publisher = {Institute of Electrical and Electronics Engineers (IEEE)},
  author = {Zhao,  Zhixiang and Semper,  Sebastian and Schneider,  Christian and Thom\"{a},  Reiner S},
  year = {2025},
  month = sep 
}

@inproceedings{ref85,
  title = {Excitation Signal Design for THz Channel Sounding and Propagation Parameter Estimation},
  url = {http://dx.doi.org/10.23919/EuCAP60739.2024.10501423},
  DOI = {10.23919/eucap60739.2024.10501423},
  booktitle = {2024 18th European Conference on Antennas and Propagation (EuCAP)},
  publisher = {IEEE},
  author = {Gedschold,  Jonas and Semper,  Sebastian and D\"{o}bereiner,  Michael and Thom\"{a},  Reiner S.},
  year = {2024},
  month = mar,
  pages = {1–5}
}

@article{ref87,
  title = {Time reversal of ultrasonic fields. I. Basic principles},
  volume = {39},
  ISSN = {0885-3010},
  url = {http://dx.doi.org/10.1109/58.156174},
  DOI = {10.1109/58.156174},
  number = {5},
  journal = {IEEE Transactions on Ultrasonics,  Ferroelectrics and Frequency Control},
  publisher = {Institute of Electrical and Electronics Engineers (IEEE)},
  author = {Fink,  M.},
  year = {1992},
  month = sep,
  pages = {555–566}
}

@inproceedings{ref88,
  title = {Time Reversal Precoding at SubTHz Frequencies: Experimental Results on Spatiotemporal Focusing},
  url = {http://dx.doi.org/10.1109/CSCN53733.2021.9686163},
  DOI = {10.1109/cscn53733.2021.9686163},
  booktitle = {2021 IEEE Conference on Standards for Communications and Networking (CSCN)},
  publisher = {IEEE},
  author = {Mokh,  Ali and De Rosny,  Julien and Alexandropoulos,  George C. and Khayatzadeh,  Ramin and Ourir,  Abdelwaheb and Kamoun,  Mohamed and Tourin,  Arnaud and Fink,  Mathias},
  year = {2021},
  month = dec,
  pages = {78–82}
}

@article{ref89,
  title = {Characterization of space-time focusing in time-reversed random fields},
  volume = {53},
  ISSN = {0018-926X},
  url = {http://dx.doi.org/10.1109/TAP.2004.836399},
  DOI = {10.1109/tap.2004.836399},
  number = {1},
  journal = {IEEE Transactions on Antennas and Propagation},
  publisher = {Institute of Electrical and Electronics Engineers (IEEE)},
  author = {Oestges,  C. and Kim,  A.D. and Papanicolaou,  G. and Paulraj,  A.J.},
  year = {2005},
  month = jan,
  pages = {283–293}
}

@article{ref90,
  title = {Pilot-Based SFO Estimation for Bistatic Integrated Sensing and Communication},
  volume = {73},
  ISSN = {1557-9670},
  url = {http://dx.doi.org/10.1109/TMTT.2024.3508018},
  DOI = {10.1109/tmtt.2024.3508018},
  number = {7},
  journal = {IEEE Transactions on Microwave Theory and Techniques},
  publisher = {Institute of Electrical and Electronics Engineers (IEEE)},
  author = {Giroto de Oliveira,  Lucas and Li,  Yueheng and Mandelli,  Silvio and Brunner,  David and Henninger,  Marcus and Wan,  Xiang and Jun Cui,  Tie and Zwick,  Thomas and Nuss,  Benjamin},
  year = {2025},
  month = jul,
  pages = {4143–4161}
}

@article{ref91,
  title = {Synchronization Techniques for Orthogonal Frequency Division Multiple Access (OFDMA): A Tutorial Review},
  volume = {95},
  ISSN = {0018-9219},
  url = {http://dx.doi.org/10.1109/JPROC.2007.897979},
  DOI = {10.1109/jproc.2007.897979},
  number = {7},
  journal = {Proceedings of the IEEE},
  publisher = {Institute of Electrical and Electronics Engineers (IEEE)},
  author = {Morelli,  Michele and Kuo,  C.-C. Jay and Pun,  Man-On},
  year = {2007},
  month = jul,
  pages = {1394–1427}
}

@book{ref92,
  title={5G NR: The Next Generation Wireless Access Technology},
  author={Dahlman, E. and Parkvall, S. and Skold, J.},
  isbn={9780128223208},
  lccn={2022304459},
  url={https://books.google.de/books?id=PZH9DwAAQBAJ},
  year={2020},
  publisher={Elsevier Science}
}

@ARTICLE{ref95,
  author={Bello, P.},
  journal={IEEE Transactions on Communications Systems}, 
  title={Characterization of Randomly Time-Variant Linear Channels}, 
  year={1963},
  volume={11},
  number={4},
  pages={360-393},
  doi={10.1109/TCOM.1963.1088793}}

@article{ref96,
  title = {Frequency Domain System Identification Toolbox for MATLAB: Improvements and New Possibilities},
  volume = {30},
  ISSN = {1474-6670},
  url = {http://dx.doi.org/10.1016/S1474-6670(17)42968-5},
  DOI = {10.1016/s1474-6670(17)42968-5},
  number = {11},
  journal = {IFAC Proceedings Volumes},
  publisher = {Elsevier BV},
  author = {Kollár,  I. and Pintelon,  R. and Schoukens,  J.},
  year = {1997},
  month = jul,
  pages = {943–946}
}

@article{ref97,
  title = {Hybrid Precoder Design for Angle-of-Departure Estimation With Limited-Resolution Phase Shifters},
  volume = {73},
  ISSN = {1558-0857},
  url = {http://dx.doi.org/10.1109/TCOMM.2024.3487517},
  DOI = {10.1109/tcomm.2024.3487517},
  number = {6},
  journal = {IEEE Transactions on Communications},
  publisher = {Institute of Electrical and Electronics Engineers (IEEE)},
  author = {Huang,  Huiping and Furkan Keskin,  Musa and Wymeersch,  Henk and Cai,  Xuesong and Wu,  Linlong and Thunberg,  Johan and Tufvesson,  Fredrik},
  year = {2025},
  month = {6},
  pages = {4439–4453}
}

@ARTICLE{ref98,
  author={Costa, Heraldo C. A. and Myint, Saw J. and Andrich, Carsten and Giehl, Sebastian W. and Engelhardt, Maximilian and Schneider, Christian and Thomä, Reiner S.},
  journal={IEEE Journal of Selected Topics in Electromagnetics, Antennas and Propagation}, 
  title={Modeling Micro-Doppler Signature of Multi-Propeller Drones in Distributed {ISAC}}, 
  year={2025},
  volume={},
  number={},
  pages={1-15},
  doi={10.1109/JSTEAP.2025.3604407}}
\end{document}